\documentclass[12pt]{article}
\usepackage{epsfig}

\def\hybrid{\topmargin -20pt    \oddsidemargin 0pt
        \headheight 0pt \headsep 0pt
        \textwidth 6.25in       
        \textheight 9.5in       
        \marginparwidth .875in
        \parskip 5pt plus 1pt   \jot = 1.5ex}

\hybrid

\def\baselinestretch{1.2}

\catcode`\@=11

\def\marginnote#1{}
\newcount\hour
\newcount\minute
\newtoks\amorpm
\hour=\time\divide\hour by60
\minute=\time{\multiply\hour by60 \global\advance\minute by-\hour}
\edef\standardtime{{\ifnum\hour<12 \global\amorpm={am}%
        \else\global\amorpm={pm}\advance\hour by-12 \fi
        \ifnum\hour=0 \hour=12 \fi
        \number\hour:\ifnum\minute<10 0\fi\number\minute\the\amorpm}}
\edef\militarytime{\number\hour:\ifnum\minute<10 0\fi\number\minute}

\def\draftlabel#1{{\@bsphack\if@filesw {\let\thepage\relax
   \xdef\@gtempa{\write\@auxout{\string
      \newlabel{#1}{{\@currentlabel}{\thepage}}}}}\@gtempa
   \if@nobreak \ifvmode\nobreak\fi\fi\fi\@esphack}
        \gdef\@eqnlabel{#1}}
\def\@eqnlabel{}
\def\@vacuum{}
\def\draftmarginnote#1{\marginpar{\raggedright\scriptsize\tt#1}}

\def\draft{\oddsidemargin -.5truein
        \def\@oddfoot{\sl preliminary draft \hfil
        \rm\thepage\hfil\sl\today\quad\militarytime}
        \let\@evenfoot\@oddfoot \overfullrule 3pt
        \let\label=\draftlabel
        \let\marginnote=\draftmarginnote
   \def\@eqnnum{(\theequation)\rlap{\kern\marginparsep\tt\@eqnlabel}%
\global\let\@eqnlabel\@vacuum}  }

\def\preprint{\twocolumn\sloppy\flushbottom\parindent 2em
        \leftmargini 2em\leftmarginv .5em\leftmarginvi .5em
        \oddsidemargin -.5in    \evensidemargin -.5in
        \columnsep .4in \footheight 0pt
        \textwidth 10.in        \topmargin  -.4in
        \headheight 12pt \topskip .4in
        \textheight 6.9in \footskip 0pt
        \def\@oddhead{\thepage\hfil\addtocounter{page}{1}\thepage}
        \let\@evenhead\@oddhead \def\@oddfoot{} \def\@evenfoot{} }

\def\numberbysection{\@addtoreset{equation}{section}
        \def\theequation{\thesection.\arabic{equation}}}

\def\underline#1{\relax\ifmmode\@@underline#1\else
        $\@@underline{\hbox{#1}}$\relax\fi}

\def\titlepage{\@restonecolfalse\if@twocolumn\@restonecoltrue\onecolumn
     \else \newpage \fi \thispagestyle{empty}\c@page\z@
        \def\thefootnote{\fnsymbol{footnote}} }

\def\endtitlepage{\if@restonecol\twocolumn \else \newpage \fi
        \def\thefootnote{\arabic{footnote}}
        \setcounter{footnote}{0}}  

\catcode`@=12
\relax

\def\figcap{\section*{Figure Captions\markboth
        {FIGURECAPTIONS}{FIGURECAPTIONS}}\list
        {Figure \arabic{enumi}:\hfill}{\settowidth\labelwidth{Figure
999:}
        \leftmargin\labelwidth
        \advance\leftmargin\labelsep\usecounter{enumi}}}
 \relax
\def\tablecap{\section*{Table Captions\markboth
        {TABLECAPTIONS}{TABLECAPTIONS}}\list
        {Table \arabic{enumi}:\hfill}{\settowidth\labelwidth{Table
999:}
        \leftmargin\labelwidth
        \advance\leftmargin\labelsep\usecounter{enumi}}}
 \relax
\def\reflist{\section*{References\markboth
        {REFLIST}{REFLIST}}\list
        {[\arabic{enumi}]\hfill}{\settowidth\labelwidth{[999]}
        \leftmargin\labelwidth
        \advance\leftmargin\labelsep\usecounter{enumi}}}
 \relax
\makeatletter
\newcounter{pubctr}
\def\publist{\@ifnextchar[{\@publist}{\@@publist}}
\def\@publist[#1]{\list
        {[\arabic{pubctr}]\hfill}{\settowidth\labelwidth{[999]}
        \leftmargin\labelwidth
        \advance\leftmargin\labelsep
        \@nmbrlisttrue\def\@listctr{pubctr}
        \setcounter{pubctr}{#1}\addtocounter{pubctr}{-1}}}
\def\@@publist{\list
        {[\arabic{pubctr}]\hfill}{\settowidth\labelwidth{[999]}
        \leftmargin\labelwidth
        \advance\leftmargin\labelsep
        \@nmbrlisttrue\def\@listctr{pubctr}}}
 \relax
\makeatother
\newskip\humongous \humongous=0pt plus 1000pt minus 1000pt

\newif\ifdtup

\relax

\def\be{\begin{equation}}
\def\ee{\end{equation}}
\def\ba{\begin{eqnarray}}
\def\ea{\end{eqnarray}}

\def\no{\noindent}

\def\IR{\relax{\rm I\kern-.18em R}}

\def\IR{\relax{\rm I\kern-.18em R}}
\def\inv{^{\raise.15ex\hbox{${\scriptscriptstyle -}$}\kern-.05em 1}}

\def\tL{{\tilde L}}

\begin{document}

\renewcommand{\theequation}{\thesection.\arabic{equation}}

\newcommand{\beq}{\begin{equation}}
\newcommand{\eeq}[1]{\label{#1}\end{equation}}
\newcommand{\ber}{\begin{eqnarray}}
\newcommand{\eer}[1]{\label{#1}\end{eqnarray}}
\newcommand{\eqn}[1]{(\ref{#1})}
\begin{titlepage}
\begin{center}

\hfill hep--th/0307154\\
\vskip -.1 cm
\hfill July 2003\\

\vskip .6in

{\large \bf Renormalization group flows and continual Lie algebras}

\vskip 0.7in

{\bf Ioannis Bakas}
\vskip 0.2in
{\em Department of Physics, University of Patras \\
GR-26500 Patras, Greece\\
\footnotesize{\tt bakas@ajax.physics.upatras.gr}}\\

\end{center}

\vskip .7in

\centerline{\bf Abstract}

\no
We study the renormalization group flows of two-dimensional metrics in 
sigma models using the one-loop beta functions, 
and demonstrate that they provide 
a continual analogue of the Toda field equations in conformally flat 
coordinates. In this algebraic setting, the logarithm of the world-sheet 
length scale, $t$, is interpreted as Dynkin parameter on the root  
system of a novel continual Lie algebra, denoted by 
${\cal G}(d/dt; 1)$, with anti-symmetric Cartan  
kernel $K(t, t^{\prime}) = \delta^{\prime} (t-t^{\prime})$; as  
such, it coincides with the Cartan matrix of the superalgebra
$sl(N | N+1)$ 
in the large $N$ limit. The resulting Toda field equation is a 
non-linear generalization of the heat equation, which is  
integrable in target space and shares the same dissipative 
properties in time, $t$. We provide 
the general solution of the renormalization group flows in terms of 
free fields, via B\"acklund transformations, and present some simple 
examples that illustrate the validity of their formal power series 
expansion in terms of algebraic data. We study in detail 
the sausage model that arises as geometric deformation of the 
$O(3)$ sigma model, and give a new interpretation to its ultra-violet 
limit by gluing together two copies of Witten's two-dimensional 
black hole in the asymptotic region. We also provide some new solutions 
that describe the renormalization group flow of negatively curved 
spaces in different patches, which look like a cane in the infra-red 
region. Finally, we revisit the transition of a flat cone    
$C/Z_n$ to the plane, as another special solution, and note that   
tachyon condensation in closed string theory exhibits a hidden 
relation to the infinite dimensional algebra ${\cal G}(d/dt; 1)$
in the regime of gravity. Its exponential growth holds the key for the 
construction of conserved currents and their systematic interpretation 
in string theory, but they still remain unknown.  

\vfill
\end{titlepage}
\eject

\def\baselinestretch{1.2}
\baselineskip 16 pt
\noindent

\def\tT{{\tilde T}}
\def\tg{{\tilde g}}
\def\tL{{\tilde L}}


\section{Introduction}
\setcounter{equation}{0}

Vacuum selection in string theory is an important problem that 
remains unsolved to this day. Different vacua are constructed 
by imposing conformal invariance on the world-sheet quantum 
field theories, which yield classical solutions of the 
space-time field equations as fixed points of the two-dimensional
renormalization group equations. Then, in this context, the 
transitions among vacua can be studied by considering the
space of all two-dimensional quantum field theories and use 
the renormalization group flows to interpolate between 
different classical solutions. Non-conformal backgrounds 
provide a way to continue off-shell in string theory, in which case  
the world-sheet renormalization group equations hold the key for 
exploring the structure of the infinite dimensional space 
of all possible string configurations. One may also assume 
that there is an evolution in physical time, which can 
be identified with the world-sheet renormalization group time, 
but this on-shell formulation has not been made
systematic in the general case. There are compelling 
reasons to expect that the transitions can be addressed
equally well in these two different frames, since fixed 
points of the renormalization group flows are stationary 
solutions for classical strings and there is no known 
decay process for which there is no corresponding 
renormalization group flow. We will adopt the renormalization
group approach in the gravity regime  
and present some new ideas for the 
reformulation of string dynamics in terms of a novel infinite
dimensional Lie algebra with exponential growth. 
This framework will also prove useful for exploring 
the integrability properties of the renormalization group
flows in target space. 

The renormalization group flows lead to 
transitions from unstable extrema towards more stable 
ones by the condensation of tachyons in various string 
models. 
It is also important to realize in this context that a 
deeper understanding of non-supersymmetric string dynamics 
may help to connect string theory with the real world. 
The problem of tachyon condensation is fairly 
well understood in open string theory, following the 
seminal work by Sen \cite{sen1, sen2}, Witten \cite{ed1}, 
and many others, but in 
closed string theory it is more difficult and remains 
largely unexplored. The main complication is
the presence of gravity, which makes it difficult to 
isolate the condensation phenomena and their 
properties. Closed string tachyon 
condensation has been recently addressed in a number 
of special situations, where tachyons are localized 
at a defect, such as branes or an orbifold fixed 
point \cite{polch, minwa, harv1, emil1, vafa, david, emil2}. 
Some of these results will be revisited in the
present work in order to compare our new algebraic 
framework with what is already known about them.       
From this point of view, 
the framework we are proposing here is only the beginning for
the algebraic reformulation of vacuum selection in 
closed string theory, where gravity plays a prominent 
role.   
However, further work is definitely required in order to make
progress on the general problem of tachyon 
condensation, using the hidden relation to 
infinite dimensional Lie algebras. 

We will study geometric transitions induced by the 
renormalization group flows of two-dimensional sigma models in 
the gravity regime by limiting our attention  
to the simplest case of two-dimensional target spaces and taking into
account only the one-loop contribution to the beta function equations. 
We will be able to systematize the action of the 
renormalization group flows, as well as the construction of 
their general solution, using an integrable Toda field 
equation. This equation arises for purely metric backgrounds in  
conformally flat coordinates and provides a non-linear 
generalization of the heat equation in target space. The 
algebraic reformulation is made possible using an 
infinite dimensional Lie algebra with generalized Cartan kernel 
$K (t, t^{\prime}) = \delta^{\prime} (t-t^{\prime})$, 
which falls into the general class of the so-called {\em continual 
Lie algebras} \cite{misha1, misha2, vershik}. 
The Cartan variable $t$ (or $t^{\prime}$)  
plays the role of the logarithmic world-sheet length scale that drives
the renormalization group flows, whereas the flow equations 
admit a zero curvature formulation as integrable two-dimensional  
system in target space based on this infinite dimensional algebra.  

Although infinite dimensional Lie algebras have been used on
several occasions in physics and mathematics, the particular  
algebra that arises in the theory of renormalization group flows has    
been studied only very little in the mathematics literature. 
This bosonic algebra is rather special 
for two main reasons. First, the first order dependence 
of the beta function on the renormalization group time, $t$, 
implies that the algebra has an anti-symmetric Cartan kernel,
which happens to    
coincide with the Cartan matrix of 
the Lie superalgebra $sl(N|N+1)$ for a certain choice of the 
simple odd root system when   
$N \rightarrow \infty$. As we will see later, this large $N$ limit 
is naturally defined by taking
a continuous analogue of the Dynkin diagram, which 
justifies the use of the term ``continual" for the 
corresponding Lie algebra. 
Second, this infinite dimensional Lie algebra exhibits 
exponential growth, which can be 
seen by taking successive commutators of the 
basic system of its Cartan-Weyl generators. 
This special property makes it 
rather difficult but worth studying in the context of string 
theory. It is natural to expect that higher modes of the string 
can be accommodated in the exponential growth of the Lie 
algebra elements at higher levels, and they can be 
subsequently used to describe beta functions beyond the metric  
deformations that are only addressed in this paper.    

The ability to apply this new algebraic framework to the 
problem of tachyon condensation, in all generality, appears to be 
intimately related to the knowledge of the complete structure of
the algebra, which remains unsolved problem. After all, 
other classes of generalized Kac-Moody algebras, 
like hyperbolic algebras, are conjectured symmetries of string 
theory via the vertex operator construction of higher string 
states, but their complete algebraic structure is also still 
lacking in all generality (see, for 
instance, \cite{kac1, olive1, nico, peter} and references 
therein).     
At the same time, the exponential growth of the 
renormalization group algebra holds the
key for the construction of conserved currents in target
space, and their systematic interpretation in string theory. Finally,
it could be used as a framework for the algebraic definition of 
an entropy function that determines the dynamics of string 
theory in renormalization group time, and for the selection of 
stable vacua in the space of all two-dimensional quantum field
theories. The intuitive expectation that the energy of space-time
decreases along the renormalization group flow when there are 
transitions to more stable vacua (see, for instance, \cite{minwa}), 
as well as Zamolodchikov's 
$c$-theorem that provides a monotonic function on  
the trajectories as they run from the ultra-violet to the infra-red fixed 
points \cite{zamo1} (but see also \cite{tsey1}, \cite{joe2}), 
might also admit a systematic interpretation in 
the framework of this particular infinite dimensional Lie algebra. 

Thus, it becomes clear that the algebraic framework we are 
proposing in this paper sets up the stage for future developments 
in the renormalization group approach to off-shell string 
theory. Here, it is predominately used to prove the integrability 
of the metric beta function equations in target space, at least
to lowest order in the perturbative renormalization of the 
world-sheet sigma models, and parametrize their general solution 
in terms of free fields via B\"acklund transformations. 
We examine several special solutions in this context by 
considering geometric deformations that depend on a finite 
number of moduli for compact, as well as non-compact 
two-dimensional  
target spaces. These examples are used further to deepen our
understanding of the geometric and algebraic aspects of the
renormalization group equations, and their interrelation. 
Solutions with conical 
curvature singularities are also considered in detail, since 
their resolution in the infra-red region of the renormalization 
group flow serves as prototype for studying the problem of tachyon
condensation in orbifold models. 

It is natural to expect that
the behavior of the renormalization group flows in the 
vicinity of singular geometries, and other unstable vacua, will 
admit a systematic description in the context of 
Toda field equations. The resulting integrable equation 
provides a non-linear generalization of the heat equation, but 
there is no general proof that it exhibits the same dissipative
properties by diffusing in space any initial singular data after 
infinite long time. At this stage, one may only use the 
heat equation to approximate the Toda field equation for all 
times by 
studying the transition of a two-dimensional cone $C/Z_n$ 
to $C/Z_m$ with $n<m$ when both $n$ and $m$ are very 
large \cite{polch}.    
In all other cases the approximation is valid only asymptotically 
in the infra-red region, which is very far away from any initial 
singular data, but there are also exact results that are indicative 
of the dissipative properties of the non-linear equation. 
In any case, we expect to be able to provide
a general proof of the dissipative properties of 
the renormalization group equations elsewhere, using the 
algebraic structure of the corresponding Toda system. This 
is also bound to have profound applications to the algebraic 
description of tachyon condensation based on the properties of  
the underlying infinite dimensional Lie algebra.   

Most of the applications we will consider in the present 
work are not directly related to string theory, but they 
are included as testing ground for the algebraic  
formulation of the renormalization group flows as 
Toda system. Geometric 
deformations of constant curvature spaces, which can be either 
positive or negative, do not interpolate between an ultra-violet 
and an infra-red fixed point, as it is usually required 
for addressing  
the problem of vacuum selection in string theory. Positive 
curvature spaces become asymptotically free in the ultra-violet
region and they reach a state of big crunch at some finite 
time, where strong curvature singularities are present and 
the lowest order approximation to the beta function equations 
breaks down. Negative curvature spaces, on the other hand,   
may start from a state of infinite negative curvature and 
flow towards a free theory in the infra-red region, which is 
well defined within the lowest order approximation to their
beta function equations. Thus, apart from the special 
solutions that describe the decay of a cone $C/Z_n$ to 
$C/Z_m$,  
it will be very interesting to know other 
explicit solutions that interpolate between different string 
vacua. Unfortunately, there are no general solutions 
with this property in our disposal, as we do not yet have 
a systematic algebraic prescription for
selecting the relevant configurations of the continual 
Toda field equation. 

A few clarifying remarks are also put in order to avoid 
other possible 
sources of confusion. It appears as if the integrable 
structure we are advocating for the renormalization group
flows is in contradiction with the dissipative properties
of the time evolution, which resolve singularities
and lead to 
the decay of unstable vacua to more stable ones. 
Dissipation is certainly not a 
property of integrable systems, but in our case space and 
time are treated differently. The zero curvature formulation
of the flows leads to a relativistic system of second order 
equations  
in two-dimensional target space, whereas the time dependence is
first order and it is encoded in the defining relations of the
continual Lie algebra used in the Lax pair. Thus,
it is still possible to have integrability in target space 
and dissipation in time. Put it differently, if we were 
able to formulate the transition among any two vacua as a  
dynamical process in real time, and not just as a flow in 
renormalization group time, we would have devised an ansatz
for time dependent string backgrounds using a certain frame 
with specific $2+1$ (or more generally $9+1$) decomposition 
of space and time. The selection of a preferred time direction
breaks general reparametrization invariance in space-time
and the resulting equations may look asymmetric in the 
space and time variables in this particular frame.        
What is more surprising, however, is the apparent ability to model 
dissipative phenomena and decay processes in terms of 
infinite dimensional algebras in a systematic way. We will  
develop this formalism for the case of renormalization group
flows of two-dimensional sigma models, but we suspect
that it may also have far reaching consequences to other 
physical problems of this kind, including the systematic investigation
of time dependent backgrounds in general relativity and 
in string theory.      

The best we can presently do in order to relate the renormalization 
group flows with space-time dynamics is to employ the light-front 
evolution. It is known that any non-conformal 
gravitational background in $d$ 
dimensions can be used to define a conformally 
invariant theory in $d+2$ space-time dimensions using 
the embedding \cite{tsey2} 
\be
ds^2 = -2 du dv + G_{ij} (u, x) dx^i dx^j ~, ~~~~  
\Phi^{\prime} = -v + \Phi (u, x)  
\ee
and identify the null coordinate $u$ with the 
renormalization group time $t$, as $u=t$. This embedding, which has 
been written here in all generality for non-conformal 
backgrounds with metric and dilaton fields $(G, \Phi)$
implies that the renormalization group flow becomes the 
profile of a gravitational wave in $(d+2)$-dimensional space-time, 
and it can be applied to all solutions that will be 
encountered in subsequent sections. Here, we also see that
the coordinates of transverse space and the light-cone 
coordinate $u$ are treated differently to account for    
the different order of the differential equations in 
space and time variables. We also note for completeness 
that other deformations of two-dimensional conformal field
theories can be made dynamical
by constructing time dependent solutions of string 
theory in $d+1$ space-time dimensions, which describe 
changes of topology and other interesting 
transitions (see, for instance, \cite{elias, enriq}), but 
it is not known how to achieve this embedding for arbitrary 
deformations including the ones we are considering here (see,
however, some recent results described in \cite{harv2}).

It is also worth emphasizing that the integrability of the 
renormalization group flows in the space of all 
target metrics looks very different in nature from the 
integrable deformations of conformal field theories away from 
criticality that were studied extensively some years ago.
In all cases, one may describe the breaking of conformal 
invariance by adding appropriate operators to the 
world-sheet action, which are schematically represented by
perturbing the initial conformal background  
\be
S_{\rm t} = S_0 + g(t) \int d^2 w ~ {\cal O}(w, \bar{w}) ~. 
\ee
There are special operators ${\cal O}$ that lead to deformations
of the two-dimensional conformal field theory and which are 
integrable with respect to the world-sheet coordinates, as 
those initially studied by Zamolodchikov \cite{zamo2}. There are other
deformations that correspond to adding relevant operators 
on the world-sheet, which are tachyonic in space-time, 
and which describe tachyon condensation in string theory by
flowing towards the infra-red region of the renormalization
group flow. The latter are not integrable on the world-sheet, 
but they provide special solutions of an integrable system 
in target space within the context of our discussion. 
In fact, it will be interesting to know whether all different 
kind of perturbations away from criticality, apart from 
the metric perturbations we are considering in the 
present work, can be viewed as special solutions of a 
big integrable structure in target space that casts the
beta function equations of all fields in zero curvature 
form. Thus, we may have integrability on the world-sheet 
only for some special choices of ${\cal O}$, whereas integrability
in target space appears to be a property of the superspace of all 
two-dimensional quantum field theories. As noted before, 
the complete structure of the renormalization group algebra
and its systematic interpretation in string theory hold
the key to the general formulation of this proposal. 
      
The remaining sections of this paper are organized as follows. 
In section 2, we include some background material 
and formulate the renormalization group equations 
for metric backgrounds in 
different frames. The equations assume rather 
simple form for two-dimensional target spaces in 
conformally flat coordinates, where they can be formally 
viewed as Toda system. As such, they provide a 
non-linear generalization of the heat flow equation 
whose basic properties, as well as its fundamental 
solutions, are also reviewed. In section 3, we 
provide a systematic description of the renormalization
group flows as integrable Toda system, using an 
infinite dimensional Lie algebra with exponential 
growth. This algebra is defined in the framework 
of continual Lie algebras by identifying the 
Dynkin parameter of its root system with the renormalization
group time. Then, using the zero curvature formulation  
of the beta function equations, we present the general 
solution in terms of arbitrary families of free fields
via B\"acklund transformations. The general solution is also 
expressed as formal power series expansion about the corresponding 
free field configurations by making use of the Lie algebra 
commutation relations and its formal highest weight 
representations, in close analogy with finite dimensional Toda
systems. We also formulate some generalized Toda field 
equations, which could be relevant for the systematic 
description of other renormalization group equations 
beyond the metric deformations. In section 4, we    
revisit the sausage model which represents special 
deformations of the $O(3)$ sigma model under the 
renormalization group flows. We also 
describe the deformation in proper coordinates and give a new
interpretation to the ultra-violet limit of model
by gluing together two copies of Witten's two-dimensional
black hole in the asymptotic region. This example proves
particularly useful for illustrating the validity of
the formal power series expansion of the general solution
to Toda field equation in terms of free fields. 
In section 5, we extend the discussion to special geometric 
deformations of constant negative curvature metrics in two 
dimensions, using different patches that correspond to 
hyperbolic, parabolic or elliptic configurations. 
We find that the target space looks like a cane in the infra-red 
region, apart from the elliptic case which has a 
conical curvature singularity. In section 6, we   
consider more complicated solutions of the continual Toda 
field equation that describe the decay of conical singularities
under the renormalization group flows. 
The transition from the  
cone $C/Z_n$ to $C/Z_m$ with $m<n$, and eventually to 
a two-dimensional plane, serves as prototype for studying the 
behavior of the general solution close to curvature 
singularities. The end-point of the flow is reached by diffusing
the curvature singularity all over the space and, thus, the 
existence of these special solutions indicates the dissipative 
nature of the non-linear evolution, as in the linearized heat equation.    
Applications to the problem of tachyon condensation are also briefly 
discussed in this context. Finally, in section 7, we 
present our conclusions and indicate some directions 
for future work. 

\section{Renormalization group flows of sigma models}
\setcounter{equation}{0}

The renormalization group properties of two-dimensional 
sigma models have attracted considerable attention following the
seminal work by Polyakov on the $O(N)$ sigma model,
\be
S = {1 \over g^2} \int d^2 w \left(\partial \vec{n} 
\right)^2 
\ee
with $\vec{n}$ being an $N$-dimensional unit vector, 
$\vec{n} \cdot \vec{n} = 1$. It was
found that the theory is perturbatively renormalizable and 
the beta function $\beta (g^2)$ is negative 
for all non-abelian groups with $N \geq 3$, 
due to the positive curvature of the target space 
manifold $S^{N-1} = SO(N)/SO(N-1)$. More precisely, introducing 
a momentum cut-off $\Lambda$, the coupling
$g^2$ changes as follows, \cite{poly, zinn},  
\be
{1 \over {\tilde{g}}^2} = {1 \over g^2} + {N-2 \over 4\pi} 
{\rm log} {\tilde{\Lambda} \over \Lambda} \label{mena}  
\ee
and, therefore, $\beta (g^2) = -(N-2) g^4 / 4\pi$ to lowest 
order. Consequently, the theory   
becomes asymptotically free in the ultra-violet region, in
close analogy with the more complicated example of non-abelian
gauge theories in four dimensions. 
For $N=2$, the theory is conformal and describes
the physics of a single compact scalar field in two 
dimensions. 

Further generalizations were also considered
later by studying the renormalization group equations for 
arbitrary background metrics in target space with far 
reaching consequences   
to the development of string theory (see, for instance,  
\cite{gsw, joe1} and references therein). 
Here, we consider the structure of the renormalization group
equations away from fixed points and present some new ideas
that set up the basis for their integrability in 
target space for the 
simplest case of two-dimensional manifolds.
   
\subsection{General considerations}

Generalized two-dimensional sigma models are well studied 
both classically and 
quantum mechanically. Classically, they are described by the 
action
\be
S = {1 \over 2} \int d^2 w \sqrt{h} h^{ij} (\partial_i X^{\mu})  
(\partial_j X^{\nu}) G_{\mu \nu} ~,  
\ee
where $\{w^i ; i = 1, 2 \}$ are coordinates on the world-sheet 
(base space) with
metric $h_{ij} (w)$ and $\{X^{\mu} ; \mu = 1, 2, \cdots , n \}$ are 
coordinates in target space with metric $G_{\mu \nu} (X)$.  
The standard approach to quantization is via perturbation theory.
When the curvature of the target space geometry is small, the 
sigma model action is perturbatively renormalizable and it can 
be treated by adding counter terms of the form  
\be
S_{\rm c.t.} = {1 \over 2} \int d^2 w \sqrt{h} h^{ij} 
(\partial_i X^{\mu})  
(\partial_j X^{\nu}) {\cal R}_{\mu \nu} ~,    
\ee
which cancel the ultra-violet divergencies of the theory.  
${\cal R}_{\mu \nu}$ is a symmetric tensor that depends on the curvature
and it can be computed systematically at one-loop or 
higher orders \cite{honer, fried} (but see also \cite{luis}, 
\cite{zachos}, \cite{shore} among many other relevant references). 

The renormalization group equations of two-dimensional 
sigma models are
\be
\Lambda^{-1} {\partial \over \partial \Lambda^{-1}} 
G_{\mu \nu} \equiv 
- \beta (G_{\mu \nu}) = - R_{\mu \nu} - {1 \over 2} 
R_{\mu \rho \sigma \tau} {R_{\nu}}^{\rho \sigma \tau} + \cdots ~,  
\ee
where $R_{\mu \nu}$ is the Ricci curvature tensor of the target
space metric that provides the one-loop contribution to 
${\cal R}_{\mu \nu}$ in appropriate units, 
$2\pi \alpha^{\prime} = 1$. The quadratic terms  
arise at two loops and they involve contributions 
from the Riemann curvature 
tensor, whereas higher order terms, which are 
denoted by dots, account for higher order curvature terms 
at higher loops. $\Lambda^{-1}$ sets the world-sheet 
renormalization scale parameter, which can be traded with 
the logarithmic length scale defined by $t = {\rm log} 
\Lambda^{-1}$. Using this variable, which is usually called
the renormalization group time, the ultra-violet limit is
reached as $t \rightarrow -\infty$, whereas the infra-red 
limit lies at $t \rightarrow +\infty$. 
Then, to lowest order in perturbation theory, we have the equations
\be
{\partial \over \partial t} G_{\mu \nu} = - 
R_{\mu \nu} ~, \label{loia}  
\ee
which describe changes of the target space geometry induced
by changes of the world-sheet logarithmic length scale. 

Their integration yields trajectories in {\em superspace}, 
which consists of all 
Riemannian metrics\footnote{It should not
be confused with the notion of superspace used in the formulation of
supersymmetric field theories.}. The problem is quite complicated
in all generality, as superspace is an infinite dimensional manifold  
and the renormalization group equations act as dynamical system
in the time variable, $t$. We do not know how to treat 
this system of equations for all 
target space manifolds, but many simplifications 
occur when the dimension  
of target space is two and the equations become more tractable. 
We will in fact show that 
the system becomes integrable in target space when $n=2$, 
using a novel infinite 
dimensional Lie algebra which is introduced later to rewrite the 
renormalization group equations in zero curvature form. This is 
made possible to lowest order in perturbation theory by omitting 
all higher order curvature terms. 
Inclusion of higher order terms, which apparently become 
important when the geometry develops strong curvature, may  
change our conclusions but their effect will not be addressed
in the present work. We are also excluding the effect of 
other fields which might be present in the given class of string 
backgrounds, and 
leave their treatment to future work.       

There are a few general results one can derive for the action of the
renormalization group flows on geometric collective parameters 
of two-dimensional sigma models. We first note 
that the volume $V$ of the target space manifold $M$,  
\be
V = \int_{M} \sqrt{{\rm det} G} ~ d^n X ~, 
\ee
changes under the renormalization group flow as follows,  
\be
{\partial V \over \partial t} = 
{1 \over 2} \int_M d^n X \sqrt{{\rm det} G} ~ G^{\mu \nu} 
{\partial  G_{\mu \nu} \over \partial t} = 
-{1 \over 2} \int_M d^n X \sqrt{{\rm det} G} 
~ R[G] ~,  
\ee
using the identity $2 \partial (\sqrt{{\rm det} G}) / \partial t = 
\sqrt{{\rm det} G} G^{\mu \nu} \partial G_{\mu \nu} / \partial t$.  
The rate of change is given by the 
Einstein-Hilbert action of the target space metric and the result 
is certainly valid in any number of dimensions.  
In two dimensions, in particular, $n=2$, which is our main concern, the 
Einstein-Hilbert action is topological and it provides the  
Euler number of the target space manifold, namely 
\be
\chi(M) = {1 \over 4 \pi} \int_M d^2 X \sqrt{{\rm det} G} ~ R[G] ~.  
\ee
Therefore, we deduce the linear dependence of the two-dimensional  
target space volume
\be
V(t) = V_0 - 2 \pi t \chi(M) ~, 
\ee
where $V_0$ is an arbitrary integration constant equal to the volume 
of the space at $t=0$. 

For compact spaces with positive Euler number, 
$V_0$ is finite and it can be normalized
to $V_0 = 2\pi t_0 \chi (M)$, by introducing a positive 
definite time scale $t_0$. Then, we may write  
\be
V(t) = 2 \pi (t_0 - t) \chi(M) ~,  
\ee
where $t_0$ corresponds to the renormalization group scale for which  
$V(t_0) = 0$ (see also \cite{sausa}). The structure of the renormalization
group equations predicts in this case that the manifold shrinks to zero size 
at some given instant of time, $t=t_0$.  
Strong curvature singularities will develop close to the big crunch, which may  
invalidate the lowest order description of the renormalization group
equations. In view of this behavior, it only makes sense to consider 
flows from the
ultra-violet region to $t_0$. The $O(3)$ sigma model falls   
precisely in this class, as it has positive curvature and $\chi (S^2) = 2$,  
and exhibits ultra-violet asymptotic freedom.  
On the other hand, when the space is compact with Euler number $\chi (M) <0$,  
as in the case of all Riemann surfaces with genus $g \geq 2$, which also have
negative curvature, it only
makes sense to consider the action of the renormalization group flows
from some finite time to the infra-red region.  
For the torus, $S^1 \times S^1$, which is a 
Riemann surface with genus $1$ and Euler number $0$, the volume 
remains unchanged, as the two-dimensional field theory
is conformal and the metric does not run. 
Finally, $V_0$ is infinite for non-compact manifolds, in which case the   
renormalization group flows may end up at infra-red geometries with
variable volume, depending on the parameters of specific 
trajectories. This behavior will be encountered in section 
5 by considering geometric deformations of constant negative
curvature spaces in different patches. 
Strictly speaking, however, it is appropriate
to regularize the volume of non-compact target spaces before 
considering the evolution under the renormalization group flow.  

It is also natural to inquire at this point whether the topology of the 
target space $M$ can change under the renormalization group flows,   
given the drastic geometric deformations that
$M$ undergoes in time. This question can be examined in all 
target space dimensions, but in two dimensions, in particular,  
we may only consider the flow of the Euler number. It turns out  
that $\partial \chi(M) / \partial t = 0$,   
because in two dimensions we have identically 
$2R_{\mu \nu} = G_{\mu \nu} R$. Therefore, we conclude that 
the Euler number of the target space 
manifold remains invariant under the 
flow, as expected.

The results we have obtained so far show that there  
exist topological as well as geometrical integrals of the renormalization
group flow in two dimensions, 
which are given by the quantities $I_1 = \chi(M)$ and 
$I_2 = V + 2\pi t I_1$ and satisfy conservation laws 
in time, $\partial I_i / \partial t = 0$. 
Actually, these might be simple 
examples from an extended list of higher integrals of motion, which are
currently unknown.    
Although this problem will not be investigated in 
the present work, we note 
that there is a way to proceed systematically by embedding 
the two-dimensional manifold $M$ in three-dimensional Euclidean space.
Such an embedding 
may also be used in order to provide a concrete visualization of the 
geometric deformations of $M$ under the renormalization 
group flows. Since  
any embedding problem requires knowledge of the metric $G_{\mu \nu}$,  
as well as the extrinsic curvature tensor $K_{\mu \nu}$, it is natural 
to expect that higher integrals of motion, 
if they exist at all in the time variable $t$, 
will involve
components of the extrinsic curvature tensor.
Integrals of the extrinsic curvature tensor may also be used to
characterize the rigidity of the space and the response of its shape
under the renormalization group flow via the associated system of
Gauss-Codazzi equations.  
This provides a well-posed problem in differential geometry, which 
has not been investigated so far for the special  
geometric deformations \eqn{loia}, to the best of our knowledge.

\subsection{Renormalization group equations in different frames}

The renormalization group equations \eqn{loia} generalize 
when arbitrary reparametrizations of the target space coordinates take 
place along the flows. The reparametrizations are elements of the
diffeomorphism group of the manifold $M$ and they are generated by 
vector fields  
$\xi_{\mu}$ that may depend on all target
space coordinates, as well as on time. Then, the beta function equations
of a purely metric background assume the generalized form \cite{fried} 
\be
- \beta (G_{\mu \nu}) \equiv {\partial \over \partial t} G_{\mu \nu} = 
- R_{\mu \nu} + \nabla_{\mu} \xi_{\nu} + \nabla_{\nu} 
\xi_{\mu} ~. \label{tomar}      
\ee
They incorporate the effect of all possible field redefinitions 
$\delta X_{\mu} = - \xi_{\mu}$, which is a poor man's way to
describe diffeomorphisms, but they can be excluded if we are 
only interested in changes of the geometry itself (see also 
\cite{luis, shore}).  
Also, no additional terms arise for reparametrizations associated
to Killing vector fields, as they satisfy $\nabla_{\mu} \xi_{\nu} 
+ \nabla_{\nu} \xi_{\mu} = 0$ by definition.  

In order to further understand the meaning of such additional terms, it is 
useful to write down the renormalization group equations 
for both metric and dilaton fields, 
$G_{\mu \nu}$ and $\tilde{\Phi}$,
which assume the following form in the sigma model frame:
\ba
& & {\partial \over \partial t} G_{\mu \nu} 
= -R_{\mu \nu} - 2 \nabla_{\mu} \nabla_{\nu} \tilde{\Phi} + 
\nabla_{\mu} \xi_{\nu} + \nabla_{\nu} \xi_{\mu} ~, \nonumber\\
& & {\partial \over \partial t} \tilde{\Phi} 
= - (\nabla \tilde{\Phi})^2 + {1 \over 2} \nabla^2 \tilde{\Phi} +
\xi_{\mu} \nabla^{\mu} \tilde{\Phi} ~. 
\ea  
Then, it is always possible to choose a frame by appropriate 
reparametrizations generated by the vector field 
$\xi_{\mu} = \partial_{\mu} 
\tilde{\Phi}$, so that the system of the renormalization group equations 
simplifies to
\be
{\partial \over \partial t} G_{\mu \nu} 
= -R_{\mu \nu} ~, ~~~~~  
{\partial \over \partial t} \tilde{\Phi} 
= {1 \over 2} \nabla^2 \tilde{\Phi} ~, \label{loib}  
\ee 
and provides a simple extension of the metric beta function 
equation \eqn{loia}.   
According to this simple observation, the effect of the dilaton is similar
to the effect of an arbitrary reparametrization generated by the vector
field $\xi_{\mu}$. 

We also recall at this point that the one-loop
effective action of two-dimensional sigma models exhibits    
conformal invariance only on-shell under the naive 
conformal transformation laws 
\be
\delta_{\epsilon} h_{ij} = \epsilon h_{ij} ~, ~~~~~ \delta_{\epsilon}
X^{\mu} = 0 ~. \label{naive}  
\ee  
Conformal invariance can be achieved off-shell by 
writing down the modified Weyl transformation laws
\be
\delta_{\epsilon} h_{ij} = \epsilon h_{ij} ~, ~~~~~ \delta_{\epsilon}
X^{\mu} = \epsilon G^{\mu \nu} \partial_{\nu} 
\tilde{\Phi} ~,  \label{anoma} 
\ee
which are applicable to both cases: $\tilde{\Phi}$ can be either 
the dilaton or
the ``would be" dilaton that acts as potential for the vector field 
$\xi_{\mu}$.
Then, the meaning of the simplified system of equations \eqn{loib} is that 
conformal invariance can be reinstated under the transformation 
\eqn{naive} by arranging the reparametrizations to cancel the effect 
of the dilaton on the modified Weyl transformations \eqn{anoma}.
Thus, {\em reparametrizations assign non-trivial Weyl transformation 
laws to the target space coordinates and act as non-trivial dilaton 
gradient in the renormalization group flows of two-dimensional metric 
sigma models}. We will make extensive use of this 
well known fact in subsequent 
sections by changing frames from conformal to proper 
coordinates.          

Let us now restrict attention to two-dimensional target
spaces in the conformally flat frame
\be
ds_{\rm t}^2 = 2 e^{\Phi(z_+, z_-; t)} dz_+ dz_- = 
{1 \over 2} e^{\Phi(X,Y; t)} (dX^2 + dY^2) ~, 
\ee
using the Cartesian coordinates $X$, $Y$ or the complex conjugate 
variables $2z_{\pm} = Y \pm i X$. 
It is well-known that any two-dimensional metric can always be written 
in this form, at least locally, for appropriately chosen conformal
factor $\Phi$, in which case  
the non-vanishing components of the Ricci curvature tensor are  
\be
R_{+-} = - \partial_{+} \left( e^{-\Phi} \partial_- e^{\Phi} \right) 
\equiv - \nabla^2 \Phi ~.   
\ee
Here, for notational purposes, we write $\nabla^2 = \partial^2 / 
\partial z_{+} \partial z_{-} = \partial^2 / \partial X^2 + 
\partial^2 / \partial Y^2$. 
Then, the renormalization group equations \eqn{loia} simplify to  
the following non-linear differential equation for $\Phi (X,Y; t)$, 
\be
{\partial \over \partial t} e^{\Phi(X,Y; t)} = \nabla^2  
\Phi (X,Y; t) ~, \label{basist} 
\ee
which has to be solved for all $t$. 

It proves advantageous for the purposes of the present work 
to rewrite this equation in
the form
\be
\nabla^2 \Phi (X,Y; t) =  
\int dt^{\prime} K (t, t^{\prime}) e^{\Phi (X,Y; t^{\prime})} ~,  
\ee
where $K(t, t^{\prime})$ is the kernel 
\be
K (t, t^{\prime}) = {\partial \over \partial t} 
\delta (t - t^{\prime}) ~. 
\ee
This equation admits a natural algebraic interpretation in the  
framework of Toda field theory, which is an 
integrable system of non-linear differential equations in two 
dimensions with local coordinates $(z_+, z_-)$. 
In Toda theory we consider a  
collection of non-linear fields $\{ \phi_i (z_+, z_-)\}$, which are 
are labeled by the simple roots of a given Lie algebra and  
interact via the Cartan matrix $K_{ij}$ of the Lie algebra, as
follows, 
\be
\nabla^2 \phi_i (X,Y) =  
\sum_{j} K_{ij} e^{\phi_j (X,Y)} ~.   
\ee
The indices $i$ and $j$ are typically discrete, but there are also 
generalizations to continuous variables for 
some infinite dimensional Lie algebras, which are obtained by
replacing the collection of two-dimensional 
Toda fields $\{\phi_i \}$ with a 
``master" field $\Phi (z_+, z_-; t)$, the Cartan matrix $K_{ij}$  
with a Cartan kernel $K (t, t^{\prime})$, and the summation
over $j$ by an integral over $t^{\prime}$.     
Then, in this framework, the renormalization group 
equation \eqn{basist}   
can be viewed as limiting case of a Toda system with Cartan matrix 
\be
K_{ij} = \delta_{i+1, j} - \delta_{i, j+1}
\ee
when the discrete indices are replaced by the  
continuous variable $t$.   
This formal description of the non-linear equation \eqn{basist} is 
actually very precise in the context of infinite dimensional 
continual Lie algebras. 
The details will be given in section 3 together with all 
necessary background material. 

The general solution of the renormalization group equation 
\eqn{basist} can be 
written, at least formally, by appealing to the Toda field theory 
interpretation. Here, as in all 
integrable Toda theories, non-linear field configurations 
can be constructed group theoretically, 
via B\"acklund transformations, using 
arbitrary families of two-dimensional free 
fields. This algebraic 
method is outlined in section 3 in all generality and 
allows to integrate the evolution 
of two-dimensional sigma models in 
superspace using free field coordinates. 
However, it is convenient for practical purposes
to develop a {\em mini-superspace} approach to the renormalization
group equations by considering geometric deformations of the 
target space metric that depend only on a finite number of degrees 
of freedom, say $\{a_i (t)\}$.  
For consistent mini-superspace reductions of the field equation 
\eqn{basist}, the moduli $a_i (t)$ satisfy a system of first order
differential equations that describe specific geometric deformations 
of the target space geometry, which are much easier to understand 
algebraically as well as geometrically. All explicit solutions 
that are worked out in subsequent sections are precisely of  
this form, and they can be used to check the validity of the 
general solution in terms of free fields.  

There are situations, however, where the mini-superspace 
approximation is mathematically consistent, but the resulting
solutions of the renormalization group equation are not 
physically sensible. Moreover, there can be instabilities 
in some mini-superspace models, in that the  
trajectories become very sensitive to relevant geometric  
deformations which were ignored at first sight.  
Although it is not easy to analyze these issues 
in all generality at this moment, using a stability 
analysis in the space of all 
possible metrics, we expect that the behavior of some rather
simple geometries with conical singularities will  
help to understand the problem in general terms. 
Besides, the fate of singularities is also an issue of paramount 
importance in applications of the renormalization group
flows to the problem of 
tachyon condensation in closed string theory. Examples
of all different types will also be given later.     

\subsection{The heat equation}

The renormalization group equation \eqn{basist} for conformally  
flat two-dimensional metrics provides a 
non-linear generalization of the well-known heat equation in 
flat space, 
\be
{\partial \over \partial t} \Theta (X, Y; t) = 
\nabla^2 \Theta (X,Y; t) ~. \label{hote} 
\ee
Indeed, the renormalization group equation can be  
approximated by the heat equation when $\Phi(X,Y; t)$ becomes
very small, i.e., when the target space metric 
becomes approximately flat and smooth, 
\be
G_{\mu \nu} \simeq \delta_{\mu \nu} ~, ~~~~~ {\rm with} ~~ 
h_{\mu \nu} <<1 ~~ {\rm and} ~~ \partial h_{\mu \nu} << 1 ~.   
\ee
This is typically valid in the infra-red region of the renormalization 
group flow, where $\Theta (X,Y; t)$ also
becomes very small and smooth, and we may set  
\be
\Phi(X,Y;t) \simeq \Theta (X,Y;t) ~, ~~~~~ {\rm as} ~~ 
t \rightarrow +\infty . 
\ee
Thus, under appropriate 
conditions, the heat equation provides a linearization of the
renormalization group equation in the infra-red limit, 
where small fluctuations around the corresponding conformal field theory 
are well defined in the weak field approximation. 
We will present examples of this
behavior in subsequent sections by 
considering theories with positive definite
beta functions, which exhibit trivial infra-red fixed points.    

The fundamental solution of the heat
equation in one spatial dimension is a Gaussian pulse with height
$1/\sqrt{t}$ and width $\sqrt{t}$, whereas 
in two dimensions it is given explicitly by
\be
\Theta (X,Y;t) = {1 \over 4 \pi t} e^{-(X^2 + Y^2)/4t} ~, 
\ee
up to normalization. 
It starts as singular function at the initial time $t=0$, 
where the Gaussian 
function becomes delta function, and it spreads in time 
with diminishing height as  $t \rightarrow +\infty$. 
As such, it encodes
the dissipative properties of the heat equation that resolves in time
the initial singularity. 
This elementary, but rather fundamental, 
decay process is presented 
schematically in figure 1 below.       

\vspace{-40pt}
\begin{figure}[h] \centering
\epsfxsize=8cm
\epsfbox{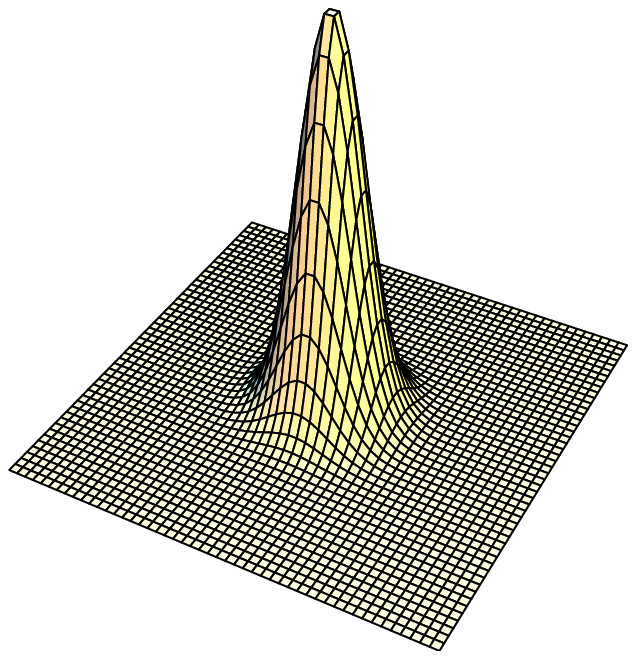}\mbox{\hspace{-0.8cm}}\nolinebreak
\epsfxsize=8cm
\epsfbox{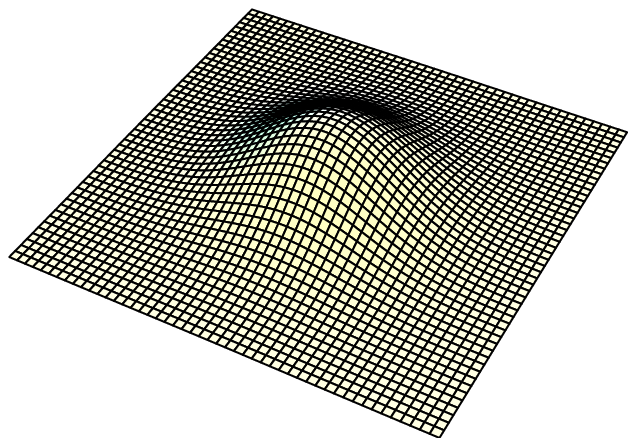} \caption{A Gaussian distribution spreading in time.} 
\end{figure}

Since the weak field approximation is not valid for finite
values of $t$, one may wonder whether the most characteristic
property of the heat equation to dissipate singularities  
is also a property in the non-linear Toda field equation. Quite 
remarkably, the answer is affirmative with 
many important consequences
to the problem of tachyon condensation in closed string theory.  
The most elementary example, which is the analogue of the
fundamental solution to the heat equation, is provided by the 
two-dimensional cone $C/Z_n$. This geometry is everywhere flat
apart from a point, the tip of the cone, where the curvature has
delta function singularity. There is a special exact solution of 
the renormalization group equations in this case, 
which is valid for all $t$ 
and describes the decay of the initial conical singularity into
nothing by spreading smoothly its curvature in space 
\cite{polch, minwa}. The
end-point of this process is a two-dimensional 
plane that arises as infra-red fixed point, 
while the curvature singularity
is washed away in close analogy with the evolution of
the Gaussian distribution depicted in figure 1. This characteristic 
solution of the continual Toda field equations will be examined 
closely in section 6 by expanding on earlier work by other 
authors.  
 
More complicated solutions of the heat equation can be easily
obtained by taking appropriate superposition of elementary 
Gaussian functions located at different points in space, thanks to
its linearity. In one dimension, for example, 
one may consider the superposition of 
infinite many fundamental solutions of the heat equation on the 
line, $1/ \sqrt{4\pi t}~ {\rm exp}(-z^2/ 4t)$, with initial data   
being a sum of delta functions for all integer values $z=n$.  
The function that results in this case is   
\be
\Theta (z; t) = {1 \over \sqrt{4 \pi t}} \sum_{n \in Z} 
{\rm exp}\left(-{(z-n)^2 \over 4t} \right)  
\ee
and provides a periodic solution of the heat equation with 
$\Theta (z+1 ; t) = \Theta (z; t)$. Actually, this particular
construction defines the Jacobi theta function, as it can be
seen using the remarkable identity \cite{watson, mumf}  
\be
\Theta (z; t) = \sum_{n \in Z} {\rm exp}(-4 \pi^2 n^2 t + 2 \pi inz) 
= {1 \over \sqrt{4 \pi t}} \sum_{n \in Z} 
{\rm exp}\left(-{(z-n)^2 \over 4t} \right) . 
\ee
The Jacobi theta function $\Theta (u; \tau)$ follows from above
by setting $u = 2zK(k)$ and $\tau = 4\pi i t$.   
All other theta functions, namely
$H(u; \tau)$, $\Theta_1 (u; \tau) = \Theta (u + K(k); \tau)$ and 
$H_1 (u; \tau) = H(u + K(k)); \tau)$ also satisfy the
same heat equation\footnote{In  
more standard notation, $H$, $H_1$, $\Theta_1$ and $\Theta$ correspond 
to the theta functions ${\cal \theta}_1$, 
${\cal \theta}_2$, ${\cal \theta}_3$ and 
${\cal \theta}_4$ respectively. This sets up the notation that will be
used in subsequent sections. For a detailed account of their 
properties, we refer the reader to standard texts on elliptic
functions (see, for instance, \cite{watson}).}.    

Parametrizing the complex modulus of the theta functions as 
$\tau = i K(k^{\prime})/K(k)$, where $K(k)$ denotes the complete
elliptic integral of the first kind with modulus $0 \leq k \leq 1$ and
$k^{\prime} = \sqrt{1 - k^2}$, we
see immediately that $\Theta (Y; t)$ describes a train of 
delta functions at $t=0$, or equivalently at $k=1$, 
which corresponds to the 
degeneration limit of the underlying Riemann surface when 
the $b$-cycle vanishes. On the other hand, the value $k=0$ 
is attained in the infra-red limit $t \rightarrow +\infty$, 
which corresponds to the degeneration 
limit of the underlying Riemann surface when the $a$-cycle   
vanishes. In this limit we have $\Theta (Y; 0) = 1$ and, likewise,  
the other theta functions yield 
$\Theta_1 (Y; 0) = 0$, $H(Y; 0) = 0$ and $H_1 (Y; 0) = 1$. 
Thus, all theta functions share the same dissipative property with   
the simple Gaussian function on the real line. 
An important feature of these periodic solutions is the interpretation 
of the complex modulus $\tau$ as Wick rotation of the real time 
variable $t$. Then, the time evolution is 
solely described by changes of the complex structure of the underlying 
Riemann surface between the two different degeneration limits. 

We do not know any elliptic solutions of the renormalization group  
equation in conformal coordinates, but  
as we will see later, 
there are simple algebraic solutions of equation \eqn{basist} 
which are expressed in terms of theta functions in another frame
by employing 
proper coordinates. Then, the heat equation for the theta functions
can be used to prove that these configurations satisfy the 
renormalization group equations for all times, and not only 
asymptotically; $t$ is also assigned to the shape modulus  
of a Riemann surface with $k=0$ corresponding to the  
infra-red limit. Put it differently, there
are special ansatz for embedding solution of the heat equation 
into the generalized system of renormalization group equations 
\eqn{tomar}, which are valid at all times. 
A more systematic understanding 
of these ansatz is definitely required in order to construct
other interesting solutions in the future. This embedding  
might also be closely related to the linearization of 
the renormalization 
group equations, which is proved in the next section, and could 
be used further to explain why both equations share the same
dissipative properties that lead to the resolution of 
initial singularities.

\section{Toda systems and Lie algebras}
\setcounter{equation}{0}
In this section we review the basic properties of Toda field theories and
outline 
the systematic description of their solutions in terms of two-dimensional
free field configurations. We start with some preliminary material on
Toda theories associated with finite dimensional Lie algebras 
with Cartan matrix $K_{ij}$, and then 
extend the discussion to infinite dimensional algebras with generalized 
Cartan matrix given by a kernel $K(t, t^{\prime})$ that depends on a 
continuous parameter rather than a discrete Dynkin index $i$. 
These are the so 
called continual Lie algebras, which provide a general class of 
infinite dimensional Lie algebras 
that include, among other cases, the large $N$ limit of the finite 
dimensional simple Lie algebras $sl(N)$. 
For the purposes of the present work,
we focus on the case of a rather odd continual Lie algebra with 
anti-symmetric Cartan kernel 
$K(t, t^{\prime}) = \delta^{\prime} (t-t^{\prime})$, which provides
the renormalization group flow for the conformal factor of a two-dimensional
metric as its Toda theory. In this new algebraic setting, the logarithm of
the world-sheet length 
scale associated with the renormalization group time is interpreted
as a continuous Dynkin parameter on the root system of this infinite
dimensional Lie algebra; at the same time, its Cartan matrix happens to be 
identical to the continuous large $N$ limit of the superalgebra 
$sl(N | N+1)$ for a certain choice of its odd root system. We will 
present the general solution of the associated Toda field theory as a 
formal series in terms of two-dimensional free fields and also 
discuss some characteristic algebraic properties of this particular
continual Lie algebra that prove its exponential growth.
This algebraic framework will be applied in subsequent sections to study
a variety of geometric transitions under the renormalization group 
flow.   

\subsection{Toda equations for simple Lie algebras}

Recall first the essential facts about Toda field theories
associated with a simple Lie algebra ${\cal G}$ of finite rank $n$. 
Let $\phi_i(z_+ , z_-)$ be a collection of 
two-dimensional fields with $i = 1, 2, \cdots , n$
and let $K_{ij}$ be the Cartan matrix of ${\cal G}$, which serves as
the coupling matrix among them.  Toda 
field theory is described by the following relativistic
system of non-linear differential equations with exponential interactions,
\be
\partial_+ \partial_- \phi_i (z_+ , z_-) = \lambda^2 \sum_{j=1}^{n} K_{ji} 
e^{\phi_j (z_+ , z_-)} ~, 
\ee
which turns out to be integrable (see, for instance, 
\cite{toda, leznov1, leznov2, manf}). 

The integrability properties of the theory follow 
easily from the zero curvature formulation of these equations,  
\be
[\partial_+ +  A_+(z_+ , z_-) ~, ~ \partial_- + 
A_-(z_+ , z_-) ] 
= 0 ~, \label{toda1} 
\ee
where $A_{\pm} (z_+, z_-)$ are gauge connections with values in  
${\cal G}$,
\be
A_+ = \sum_{i=1}^{n} \left( 
\left(\sum_{j=1}^{n} K_{ij}^{-1} \partial_+ \phi_j 
\right) H_i + \lambda X_i^+ \right) , ~~~~ 
A_- = \lambda \sum_{i=1}^{n} e^{\phi_i} X_i^- ~.  
\ee
The elements $H_i$, $X_i^{\pm}$ with $i = 1, 2, \cdots , n$
form the system of Weyl generators of ${\cal G}$ and they  
satisfy the defining relations 
\be
[H_i , X_j^{\pm}] = \pm K_{ij} X_j^{\pm} ~, ~~~~ 
[X_i^+ , X_j^-] = \delta_{ij} H_j ~, ~~~~ 
[H_i , H_j] = 0 ~.  
\ee
The ``coupling constant" $\lambda^2$ that appears in the classical equations 
of motion can be 
set equal to $1$ (or $-1$) by rescaling the coordinates $z_{\pm}$, 
or shifting all $\phi_i$ 
by $2{\rm log}\lambda$, provided that $\lambda \neq 0$. For $\lambda =0$, 
however, the Toda field equations become linear,  
as it can be readily seen. We will keep $\lambda$ as 
book keeping
parameter for future reference, in order to gain better insight into the 
free field realization of the general solution, but later we may set
$\lambda^2 = \pm 1$, depending on the sign of the coupling, 
without loss of generality.     

The complete structure of the algebra ${\cal G}$ is not described 
by the system of Weyl generators alone, but they are sufficient to 
generate the remaining elements. In particular,  
all other generators of ${\cal G}$ can be obtained by taking successive 
commutators of these basic elements, namely 
$[\cdots [[X_{i_1}^+ , X_{i_2}^+] , X_{i_3}^+] , \cdots , X_{i_m}^+]$, and
likewise for the $X_i^-$'s. Put it differently, $H_i$ and $X_i^{\pm}$ span the 
subspaces of the Lie algebra ${\cal G}_0$ and ${\cal G}_{\pm 1}$ respectively,
which form the so called local part of the Lie algebra,  
${\cal G}_{-1} \oplus {\cal G}_0 \oplus {\cal G}_{+1}$, 
whereas the other elements belong in the subspaces 
${\cal G}_{\pm 2} = [{\cal G}_{\pm 1} , {\cal G}_{\pm 1}]$,
${\cal G}_{\pm 3} = [{\cal G}_{\pm 1} , {\cal G}_{\pm 2}]$,
and so on.     
Of course, we also have the Serre conditions, which are imposed 
on the various successive commutators and they 
effectively provide the number of all independent generators
of the simple Lie algebra ${\cal G}$.  
The system of Weyl generators is not only important for establishing the 
integrability properties of the underlying Toda field equations, but it may also
be used for describing their general solution. For this, we also need to introduce
representations of the Lie algebra ${\cal G}$ based on a highest weight state
$|j>$, so that 
\be
X_i^+ |j> = 0 ~, ~~~~ <j| X_i^- = 0 ~, ~~~~ H_i |j> = \delta_{ij} |j> 
\ee
with the normalization $<j|j> =1$. 

It is well known that the Toda field equations admit B\"acklund transformations 
that relate its general configuration to two-dimensional free fields.
A particularly simple way to describe the free 
field realization of $\phi_i(z_+ , z_-)$ 
requires the introduction of two matrices $M_{\pm}(z_{\pm})$, which satisfy 
the first order equations  
\be
\partial_{\pm} M_{\pm}(z_{\pm}) = M_{\pm} (z_{\pm}) \left(\lambda \sum_{i=1}^{n} 
e^{f_i^{\pm} (z_{\pm})} X_i^{\pm} \right) .  
\ee
Here, $f_i^{\pm}(z_{\pm})$ are $n$ arbitrary holomorphic and anti-holomorphic 
functions so that the fields 
\be
\phi_i^{(0)} (z_+ , z_-) = f_i^+(z_+) + f_i^-(z_-)
\ee
satisfy the 
two-dimensional free field equation, $\partial_+ \partial_- 
\phi_i^{(0)} = 0$, for all 
values of $i$. The matrices $M_{\pm}(z_{\pm})$ can be obtained by simple 
integration, using the path-ordered exponentials 
\be
M_{\pm}(z_{\pm}) = {\cal P} {\rm exp}\left(\lambda 
\int^{z_{\pm}} dz_{\pm}^{\prime} 
\sum_{i = 1}^{n} e^{f_i^{\pm}(z_{\pm}^{\prime})} X_i^{\pm} \right) ,  
\ee
where, as usual, we have the ordering prescription 
\be
{\cal P} {\rm exp} \left(\int^{z} dz^{\prime} A(z^{\prime}) \right) = 
\sum_{m=0}^{\infty} \int^{z} dz_1 \int^{z_1} dz_2 \cdots \int^{z_{m-1}} dz_m 
A(z_m) \cdots A(z_2) A(z_1) \label{rikos}    
\ee
with $z_m \leq \cdots \leq z_2 \leq z_1 \leq z$. 
Here, the lower value in the integration range
is left arbitrary to account for the integration constants in $M_{\pm}$ that arise  
by solving the corresponding first order equations in terms of path-ordered
exponentials. Also, since $\partial M = M A$ implies $\partial M^{-1} = -AM$ for
all $M$ and $A$, the inverse of \eqn{rikos} is defined 
by taking the opposite
ordering of paths and changing $A$ to $-A$.   

Then, according to the literature, the general solution of the Toda field equations 
is given by the formula \cite{leznov1, leznov2, manf}  
\be
\phi_i(z_+ , z_-) = \phi_i^{(0)} (z_+ , z_-) - \sum_{j=1}^{n} K_{ji} 
{\rm log} <j|M_+^{-1}(z_+) M_- (z_-) |j> ~, 
\ee
which thus provides the required free field realization of the non-linear Toda field 
theory configurations in closed form. 

The parametrization of the general solution in terms of free fields 
$\phi_i^{(0)} (z_+ , z_-)$  
allows to express the non-linear Toda fields $\phi_i(z_+ , z_-)$ as a power 
series expansion, using the book keeping parameter $\lambda$. 
The individual terms
can be computed by first expanding the path-ordered exponentials $M_{\pm}$ in power series  
in $\lambda$ and then expressing the expectation values 
$<j|M_+^{-1}(z_+) M_-(z_-)|j>$  
as superposition of the Lie algebra matrix elements  
\be
D_j^{\{i_1, i_2 , \cdots , i_m; i_1^{\prime} , i_2^{\prime} , \cdots , i_m^{\prime}\}} = 
<j| X_{i_1}^+ X_{i_2}^+ \cdots X_{i_m}^+ X_{i_m^{\prime}}^- \cdots 
X_{i_2^{\prime}}^- X_{i_1^{\prime}}^- |j> \label{toda2}  
\ee
for all $j$. Their coefficients depend on 
the components of the free fields $f_i^{\pm}(z_{\pm})$, as
they come out from the expansion of the exponentials. All such matrix elements can be
computed recursively by shifting the $X_i^+$'s to the right, using the commutation 
relations of the Weyl generators, and the resulting expressions are polynomials of the 
Cartan matrix elements $K_{ij}$. For instance, we find  
\ba
& & D_j^{\{i; i^{\prime}\}} = \delta_{i i^{\prime}} \delta_{ij} ~, \nonumber\\
& & D_j^{\{i_1, i_2; i_1^{\prime} , i_2^{\prime}\}} = \delta_{i_1 i_1^{\prime}} 
\delta_{i_2 i_2^{\prime}} \delta_{i_1^{\prime} j} \left(2 \delta_{i_2^{\prime} j} 
- K_{i_2^{\prime} i_1^{\prime}} \right) ,   
\ea
and so on for all higher order terms. 
We also note, for completeness, that more general matrix elements of the 
form  
$D_j^{\{i_1 , \cdots , i_m ; i_1^{\prime} , \cdots , i_{m^{\prime}}^{\prime}\}}$
arise as cross-terms from the expansion of $M_+^{-1}(z_+) M_-(z_-)$, 
but they are 
all zero, as it can be easily seen by direct computation, unless  
$m=m^{\prime}$, as in equation \eqn{toda2} above.  
Thus, all terms of the expansion 
can be determined entirely by the Lie algebra ${\cal G}$ and its highest 
weight representations, using free fields.

We summarize the results from a slightly different perspective
that shows the power of integrability.  
Since $\lambda = 0$ yields free field
equations for the Toda variables, we may think of the power series expansion 
of the non-linear Toda fields as in perturbation theory,  
\be
\phi_i(z_+, z_-) = \phi_i^{(0)} (z_+, z_-) + \lambda^2 \phi_i^{(1)} (z_+, z_-) 
+ \lambda^4 \phi_i^{(2)}(z_+, z_-) + \cdots ~, 
\ee
where $\phi_i^{(0)}(z_+, z_-)$ is a collection of two-dimensional free
fields that arise for $\lambda = 0$. The higher order 
correction terms 
$\phi_i^{(1)}(z_+, z_-), ~ \phi_i^{(2)}(z_+, z_-), ~ \cdots$ can be 
determined recursively in terms of the free fields 
$\phi_i^{(0)} (z_+, z_-)$, as it is 
usually done order by order in
the parameter $\lambda^2$ using the Toda field equations with $\lambda \neq 0$.
This is another way to say that the Toda field equations admit a free field 
realization, using the book keeping parameter $\lambda$.
However, $\lambda$ is not a small parameter, as it can be rescaled at will. 
Therefore, computing only the first few 
terms in the perturbative expansion is not
sufficient for extracting an approximate general solution. One has to determine
all higher order terms in the Taylor series expansion, 
which is rather cumbersome 
algorithm to carry out in all detail. It is right here that 
the integrability of the Toda
field equations saves the day, since all terms in the expansion can be computed 
systematically in terms of algebraic data, as above, and moreover, the entire series
can be summed up in close form with the aid of the path-ordered exponential 
functions $M_{\pm}(z_{\pm})$. 

This remark concludes our brief discussion of the Toda field 
equations for simple Lie algebras ${\cal G}$ and makes us appreciate even more
the algebraic description of their general solutions. 
Clearly, in all considerations,  
it is implicitly assumed that 
there are no non-perturbative corrections to the general 
configuration, which may then give rise to more exotic solutions 
of Toda theory; this is 
certainly true when the rank of ${\cal G}$ is finite, and we will 
assume that it is so 
in the infinite dimensional generalizations that are introduced next.

\subsection{Toda equations for continual Lie algebras}

The notion of continual Lie algebras arises naturally by considering a system of
Weyl generators $\{H(t), X^{\pm}(t)\}$, which depend on a continuous variable $t$
instead of a discrete Dynkin label, so that the commutation relations 
of the local part of the algebra assume
the form \cite{misha1, misha2, vershik}  
\ba
& & [X^+(t) , X^-(t^{\prime})] = \delta(t-t^{\prime}) H(t^{\prime}) ~, ~~~~ 
[H(t) , H(t^{\prime})] = 0 ~, \nonumber\\  
& & [H(t) , X^{\pm}(t^{\prime})] = \pm K(t, t^{\prime}) X^{\pm}(t^{\prime}) ~.  
\ea
Here, $K(t, t^{\prime})$ is a continual analogue of the Cartan matrix, which 
can be a distribution. The resulting algebras are infinite dimensional and one 
may also formally define their highest weight representations by introducing a 
normalized state $|t>$, so that
\be
X^+(t^{\prime})|t> = 0 ~, ~~~~ <t|X^-(t^{\prime}) = 0 ~, ~~~~ 
H(t^{\prime}) |t> = \delta(t-t^{\prime}) |t>  
\ee
with $<t|t> = 1$. This algebraic data will play an 
important role for the formulation of Toda field theories
for all continual Lie algebras and the systematic 
construction of their general solution in 
terms of free fields, as in the finite dimensional case.  

The general theory of continual Lie algebras involves the theory of distributions 
with values in Lie algebras, and therefore it is more natural to define them 
using the smeared form of the Weyl generators $\{H(\varphi), X^{\pm}(\varphi)\}$
with respect to arbitrary functions $\varphi(t)$; we set, for notational 
purposes,  
\be
A(\varphi) = \int dt \varphi(t) A(t) ~. 
\ee
Then, the Cartan operator $K$
is a bilinear map on the vector space of all functions $\varphi$ and the 
commutation relations of the local part of the algebra,  
${\cal G}_{-1} \oplus {\cal G}_0 \oplus {\cal G}_{+1}$, assume the form 
\ba
& & [X^+(\varphi) , X^-(\psi)] = H(S(\varphi , \psi)) , ~~~~
[H(\varphi) , H(\psi)] = 0 ~, \nonumber\\  
& & [H(\varphi) , X^{\pm} (\psi)] = \pm X^{\pm}(K(\varphi , \psi)) ~, 
\ea
where, $S(\varphi , \psi) = \varphi \cdot \psi$. More general continual Lie
algebras can be constructed for other choices of the bilinear form $S$, 
which may differ from the identity; however, all pairs $(K, S)$ must satisfy 
the relation 
\be
S(\varphi , K(\psi , \chi)) = S(K(\psi, \phi), \chi) 
\ee
for consistency with the Jacobi identities. 
Then, if the operator $S$ is invertible, the 
substitution $H(S\varphi) \rightarrow H(\varphi)$ 
will bring the commutation relations    
of the algebra into a standard form,  
\be
\tilde{K} = KS ~, ~~~~~  \tilde{S} = I ~.
\ee

Continual Lie algebras are denoted in general by ${\cal G}(K, S)$, by specifying 
the two bilinear maps $K$, $S$, and they are 
$Z$-graded algebras in the sense that
\be
{\cal G}(K, S) = \oplus_{n \in Z} {\cal G}_n ~, 
\ee
where apart from the local part corresponding to $n=0$ and $\pm1$, 
the remaining elements of the algebra are  
constructed as usual, by taking successive commutators, so that 
${\cal G}_n = [{\cal G}_{n-1} , {\cal G}_{+1}]$ for all $n>0$ and 
${\cal G}_n = [{\cal G}_{n+1} , {\cal G}_{-1}]$ for all $n<0$. Finally, an 
important concept is the dimension of the subspaces ${\cal G}_n$ relative to the
dimension of the Cartan subalgebra ${\cal G}_0$, 
\be
d_n = {\rm dim}_{{\cal G}_0} {\cal G}_n 
\ee
which varies from one example to another. 
We always have $d_0 = d_{\pm 1} = 1$ by definition, but
there are cases with $d_n = 1$ for all $n$, i.e., algebras of constant growth, 
like ordinary simple Lie algebras including $sl(\infty)$, 
and cases that exhibit polynomial growth in $n$; finally, there can be  
algebras with exponential growth, like the algebra that will be used
for the systematic description of the renormalization group flows. 

Using this abstract notion of continual Lie algebras, it is 
fairly easy to see, as a simple example, that 
the $sl(2)$-current algebra, 
\ba
& & [X^+(\varphi) , X^-(\psi)] = H(\varphi \cdot \psi) , ~~~~
[H(\varphi) , H(\psi)] = 0 ~, \nonumber\\  
& & [H(\varphi) , X^{\pm} (\psi)] = \pm 2 X^{\pm}(\varphi \cdot \psi) ~, 
\ea
arises for the choice $K = 2I$ and $S=I$; all other current algebras 
can also be described in this fashion.

Another standard example is provided by the choice $K(t, t^{\prime}) = 
-\delta^{\prime \prime} (t-t^{\prime})$, which can be thought as the continuous
large $N$ limit of the Cartan matrix of the simple Lie algebras $sl(N)$, 
with $K_{ij} = 2\delta_{i,j} - \delta_{i+1,j} - \delta_{i,j+1}$. Thus,
the Dynkin diagram of the continual Lie algebra $sl(\infty)$ 
is a continuous 
line parametrized by the variable $t$, whereas for finite $N$ the  
skeletonization of $t$ yields the standard discrete Dynkin diagram of $sl(N)$.
The resulting infinite dimensional algebra $sl(\infty)$ 
turns out to be isomorphic to the 
algebra of area preserving diffeomorphisms of a two-dimensional 
surface. Working with smeared generators, we choose the bilinear maps  
$K(\varphi , \psi) = i \varphi^{\prime} \cdot \psi$ and 
$S(\varphi, \psi) = -i(\varphi \cdot \psi)^{\prime}$, which is same as having 
$\tilde{K} (\varphi, \psi) = - \varphi^{\prime \prime} \cdot \psi$ and 
$\tilde{S}(\varphi , \psi) = \varphi \cdot \psi$, so that 
$\tilde{K}(t, t^{\prime}) = - \delta^{\prime \prime} (t - t^{\prime})$ and 
$\tilde{S}(t,t^{\prime}) = \delta(t-t^{\prime})$ as above. Then, the  
complete structure of this continual Lie algebra can be worked out and
it is described by the following system of commutation relations  
\be
[X_n(\varphi) , X_m(\psi)] = i X_{n+m} (m \varphi^{\prime} \psi - n 
\varphi \psi^{\prime})  
~; ~~~~  {\rm for ~ all} ~ ~X_n \in {\cal G}_n ~. 
\ee
Associating to each $X_n(\varphi)$ the mode functions $\varphi_n(t) e^{ins}$, we 
see immediately that this algebra is isomorphic to the Poisson bracket algebra
of all functions $\varphi(t, s)$ on a two-dimensional surface (e.g., torus) 
with coordinates $(t, s)$, \cite{misha2, vershik, flora}. 
Thus, $sl(\infty)$ is an infinite dimensional
Lie algebra with constant growth, since $d_n = 1$ for all $n$; 
this is not always true, however, 
for other continual Lie algebras.

There is a natural generalization of Toda field theory to continual Lie algebras
provided that one introduces a ``master" field $\Phi(z_+, z_-; t)$ instead of a
discrete collection of two-dimensional Toda fields $\phi_i(z_+, z_-)$. Then, 
the continual analogue of the Toda equations is given by 
\cite{misha1, misha2, vershik, misha3, misha4}   
\be
\partial_+ \partial_- \Phi(z_+, z_-; t) = 
\lambda^2 \int dt^{\prime} K(t^{\prime}, t) 
e^{\Phi(z_+, z_-; t^{\prime})} ~,
\ee
which defines an integrable equation, as in the ordinary case, 
but lives in higher
dimensions; however, this equation is not relativistic in higher dimensions, 
but only in the two dimensional space with light-cone coordinates $(z_+, z_-)$.        
Then, in analogy with Toda theories for finite dimensional Lie algebras, the 
continual analogue of the field equations follow from the two-dimensional zero 
curvature condition, 
\be
[\partial_+ + A_+ , \partial_- + A_-] = 0 ~, 
\ee
where the gauge connections $A_{\pm}(z_+, z_-)$ take values in the local
part of the Lie algebra, ${\cal G}_{-1} \oplus {\cal G}_0 \oplus {\cal G}_{+1}$, 
with
\be
A_+ = H(\Psi) + \lambda X^+(1) ~, ~~~~~ A_- = 
\lambda X^-(e^{\Phi}) ~. \label{toda3}  
\ee
These choices are straightforward generalizations of the ansatz \eqn{toda1}  
used for the case of finite dimensional algebras. 

For any continual Lie algebra ${\cal G}(K, S)$, the resulting equations 
take the form 
\be
\partial_- \Psi = \lambda^2 S(1, e^{\Phi}) ~, ~~~~~
\partial_+ e^{\Phi} = K(\Psi, e^{\Phi}) ~.   
\ee
Then, for the special choice of bilinear maps, which will be implicitly assumed in
all subsequent cases,  
\be
K(\varphi, \psi) = (K \varphi) \cdot \psi ~, ~~~~~ 
S(\varphi, \psi) = S(\varphi \cdot \psi) ~, 
\ee
and which describe $K$ and $S$ in terms of a much simpler 
set of linear maps, the system  
reduces to the continual Toda field equation 
\be
\partial_+ \partial_- \Phi = \lambda^2 \tilde{K} \left({\rm exp} \Phi \right) 
\ee
by elimination of the 
variable $\Psi$, or equivalently by making the consistent choice 
$\Psi = K^{-1} (\partial_t \Phi)$, as in the finite dimensional case. 
Here, $\tilde{K} = KS$
as before, but the tilde can be dropped for all practical purposes.     

A particularly interesting list of equations, which are closely related to each 
other in various ways, is provided by the following choices of Cartan kernel
$K(t, t^{\prime})$, having $S(t, t^{\prime}) = \delta(t-t^{\prime})$ in all cases:
\ba
& & K(t, t^{\prime}) = 2\delta(t - t^{\prime}): 
~~~~~~~ \partial_+ \partial_- \Phi 
(z_+, z_-; t) =2 \lambda^2 e^{\Phi(z_+, z_-; t)} ~, \\
& & K(t, t^{\prime}) = \delta^{\prime} (t - t^{\prime}): ~~~~~~~~ 
\partial_+ \partial_- \Phi 
(z_+, z_-; t) = -\lambda^2 {\partial \over \partial t} 
e^{\Phi(z_+, z_-; t)} ~, \\
& & K(t, t^{\prime}) = - \delta^{\prime \prime} (t - t^{\prime}): ~~~~~ 
\partial_+ \partial_- \Phi 
(z_+, z_-; t) = - \lambda^2 {\partial^2 \over \partial t^2} 
e^{\Phi(z_+, z_-; t)} ~. \label{toda4} 
\ea

The first equation, which is associated to the $sl(2)$-current algebra, is 
just the Liouville equation for which $t$ acts as
spectator; its geometric interpretation is well known from the 19th century
and provides the condition for constant curvature two-dimensional metrics in
the conformal gauge, 
\be
ds^2 = e^{\Phi(z_+ , z_-)} dz_+ dz_- ~. 
\ee
Positive or negative curvature spaces correspond to $\lambda^2 = -1$ or
$+1$ respectively. 
The third equation, which is associated to the algebra $sl(\infty)$, 
is the so called {\em heavenly
equation} and arose in more recent years as the 
hyper-K\"ahler condition for four-dimensional metrics
with a rotational (i.e., non-triholomorphic) isometry
generated by a vector field $\partial_{\tau}$, using the  
system of adapted coordinates \cite{geo, park, misha5, bak},
\be
ds^2 = {1 \over \partial_t \Phi} \left(d \tau \pm i (\partial_+ \Phi) 
dz_+ \mp i (\partial_- \Phi) dz_- \right)^2 + (\partial_t \Phi) 
\left(e^{\Phi} dz_+ dz_- + dt^2 \right)   
\ee 
for $\lambda^2 = 1$. Finally, the second equation is the one that describes  
the geometric deformations of 
two-dimensional metrics under the renormalization group flow in the 
conformal gauge for $\lambda^2 = -1$. 
It will be investigated separately in all detail in section 3.3.    

Straightforward generalization of the group theoretical framework that 
describes the solutions of Toda field equations in terms of arbitrary 
free field configurations yields the expression \cite{misha1}  
\be
\Phi(z_+, z_-; t) = \Phi_0 (z_+, z_-; t) - 
\int dt^{\prime} K(t^{\prime}, t) 
{\rm log} <t^{\prime}| M_+^{-1}(z_+) M_-(z_-) 
|t^{\prime}> ~, \label{toda5}  
\ee
where 
\be
\Phi_0 (z_+, z_-; t) = f^+(z_+; t) + f^-(z_-; t) \label{toda6} 
\ee
is a ``master" free field 
in two dimensions,
i.e., $\partial_+ \partial_- \Phi_0 (z_+, z_-; t) = 0$, 
which depends on the continuous
variable $t$ rather than a discrete index $i$, as in ordinary Toda theory. 
Furthermore, 
\be
M_{\pm}(z_{\pm}; t) = {\cal P} {\rm exp}
\left(\lambda \int^{z_{\pm}} dz_{\pm}^{\prime} 
\int^t dt^{\prime} e^{f^{\pm}(z_{\pm}^{\prime}; t^{\prime})} X^{\pm}(t^{\prime})
\right) \label{toda7} 
\ee
are path-ordered exponentials that generalize the corresponding expressions 
of Toda field theories to the case of continual Lie algebras 
in an obvious way, and  
the matrix elements $<t|M_+^{-1}(z_+) M_-(z_-)|t>$ 
can be computed using the continual (but formal) 
generalization of highest weight representations to the corresponding 
infinite dimensional Lie algebras.  

The path-ordered exponentials can be expanded in infinite power series
and each term can be determined algebraically in terms of the special
matrix elements, as before, 
\be
D_t^{\{t_1, t_2, \cdots , t_m; t_1^{\prime}, t_2^{\prime}, \cdots , t_m^{\prime}\}}
= <t|X^+(t_1) X^+(t_2) \cdots X^+(t_m) X^-(t_m^{\prime}) \cdots 
X^-(t_2^{\prime}) X^-(t_1^{\prime})|t> , 
\ee
which involve an equal number of $X^+$'s and $X^-$'s. Then, using the 
commutation relations among the Weyl generators of the Lie algebra, one may
easily derive recursive relations for these matrix elements, namely  
\be
D_t^{\{t_1, t_2, \cdots , t_m; t_1^{\prime}, t_2^{\prime}, \cdots , t_m^{\prime}\}}
= \sum_{j=1}^{m} \delta(t_m - t_j^{\prime}) \left(\delta(t-t_j^{\prime}) - 
\sum_{k=1}^{j-1} K(t_j^{\prime}, t_k^{\prime}) \right) 
D_t^{\{t_1, \cdots , t_{m-1}; t_1^{\prime}, \cdots , \hat{t}_j^{\prime}, 
\cdots , t_m^{\prime}\}} ~, \label{toda8}  
\ee
with $D_t^{\{t_1; t_1^{\prime}\}} = \delta (t-t_1) \delta (t-t_1^{\prime})$. 
Here, $\hat{t}_j^{\prime}$ is used to denote that $t_j^{\prime}$ has been omitted 
by contraction with the index $t_m$. Although it is not possible to 
iterate the recursive relations in favor of producing a closed formula for the 
general matrix elements involved in the expansion, even in the 
simplest possible cases where the  
Cartan kernel has derivatives of the delta function, the 
first few terms of the expansion can be obtained 
explicitly starting with the normalization 
$<t|t> = 1$. More details will be given later for the special Lie  
algebra with $K(t, t^{\prime}) = \delta^{\prime}(t-t^{\prime})$. In any case,
this procedure provides a systematic description of the general solutions to 
Toda field equations for any continual Lie algebra in terms of free fields.

A practical question concerning the validity of the general solution 
is the convergence of the formal power series expansion, or else
what is the rigorous meaning of the group elements 
$M_{\pm}(z_{\pm})$ that sum up the infinite many 
terms into the matrix elements
$<t|M_{+}^{-1}(z_+) M_-(z_-)|t>$.     
Note that the path-ordered exponentials 
$M_{\pm}(z_{\pm})$ correspond to 
specific group elements obtained by exponentiation
of the Lie algebra elements in ${\cal G}_{\pm 1}$, with the free fields  
${\rm exp} f_i^{\pm}$ playing the role of their canonical parameters.  
They are all well defined for finite dimensional simple Lie algebras, but 
for infinite dimensional algebras some appropriate regularization  
might be needed. Also, there might be solutions that admit 
free field realizations in patches, as one single patch $(f^+, f^-)$ 
might not be sufficient for their construction\footnote{A characteristic
example of this type is provided by the Atiyah-Hitchin metric, which  
is a hyper-K\"ahler metric with rotational isometry that 
solves equation \eqn{toda4} for the algebra $sl(\infty)$ with 
$K(t, t^{\prime}) = - \delta^{\prime \prime} (t-t^{\prime})$ (see, 
for instance, \cite{bak} and references therein).}. 
A more rigorous definition   
of the group elements 
and their action on the highest weight states $|t>$, which are only 
formally described, is certainly required from a mathematical 
point of view.   
In general, the investigation of the convergence properties of the 
power series expansion in terms of free fields may  
depend crucially on the properties of the Cartan kernel 
$K(t, t^{\prime})$. It is natural 
to expect that algebras of finite growth, i.e., $d_n < \infty$, will  
exhibit the necessary convergence properties, but it 
is fair to say that   
there is no systematic understanding of this issue at the moment.  

\subsection{Renormalization group flows as Toda system}
 
Next, we specialize the general discussion to the continual Lie algebra 
with $K(t, t^{\prime}) = \delta^{\prime}(t-t^{\prime})$ and 
$S(t, t^{\prime}) = \delta(t-t^{\prime})$, which is appropriate for
the systematic description of the renormalization group flows of 
two-dimensional metrics as Toda field equations.  
According to the general theory, 
the local part of the algebra, ${\cal G}_{-1} 
\oplus {\cal G}_0 \oplus {\cal G}_{+1}$, is described by the 
system of Weyl 
relations that are conveniently written here in smeared form,
\ba
& & [X_{+1}(\varphi), X_{-1}(\psi)] = H (\varphi \psi) ~, ~~~ 
[H(\varphi), H(\psi)] = 0 ~, \nonumber\\ 
& & [H(\varphi), X_{\pm 1}(\psi)] = \mp X_{\pm 1} (\varphi^{\prime} \psi) ~.  
\ea
The remaining part of the full $Z$-graded algebra ${\cal G} = 
\oplus_{n \in Z} {\cal G}_n$ is generated recursively, as usual, by 
taking successive commutators of $X_{\pm 1}$. We do not need to 
know it for the purposes of the proposed association, but we
will collect, nevertheless, everything that is known about 
the structure of the full algebra at the end of this
section, focusing on the proof of its exponential growth.

Consider the zero curvature condition $[\partial_+ + A_+(z_+, z_-) , 
\partial_- + A_-(z_+ , z_-)] = 0$ and make the most general ansatz
for the gauge connections $A_{\pm} (z_+, z_-) \in {\cal G}_0 \oplus 
{\cal G}_{\pm 1}$, 
\be
A_{\pm} (z_+, z_-) = H(u_{\pm}) + \lambda X_{\pm 1} (f_{\pm}) ~, 
\ee
where $u_{\pm}$ and $f_{\pm}$ are all functions of $(z_+, z_-; t)$.
Then, by comparing terms of the Lie algebra elements $H$ and 
$X_{\pm 1}$ that arise in this case, we find the following 
system of equations 
\ba
& & {\partial u_+ \over \partial t} = - \partial_+ ({\rm log} f_-) ~, 
~~~~~ 
{\partial u_- \over \partial t} = \partial_- ({\rm log} f_+) ~, 
\nonumber\\
& & \partial_+ u_- - \partial_- u_+ + \lambda^2 f_+ f_- = 0 ~.  
\ea
Obviously, $u_{\pm}$ can be easily eliminated by taking appropriate
derivatives of these equations to yield
\be
\partial_+ \partial_- \Phi (z_+, z_-; t) = - \lambda^2  
{\partial \over \partial t} e^{\Phi(z_+, z_-; t)} ~,  ~~~~~ 
{\rm with} ~~
\Phi = {\rm log}(f_+f_-) ~,  
\ee
as the continual Toda field equation associated with the Lie algebra
${\cal G}(d/dt; 1)$. It describes the renormalization group flow
of two-dimensional sigma models in a conformally flat system of 
target space coordinates $(z_+, z_-)$, as advertised, with 
$\lambda^2 = -1$.  
 
Note that our ansatz appears to be slightly more general than the ansatz 
\eqn{toda3} used in section 3.2, 
but this is not really so, as it reflects the
gauge arbitrariness of the zero curvature formulation of our equations.
In fact, it can be easily seen that one can set $u_- =0$ by performing 
a gauge transformation $A_{\pm} \rightarrow g^{-1}(\partial_{\pm} 
+ A_{\pm}) g$ with $g = {\rm exp} (H(\varphi))$ and 
$\partial_- \varphi = -u_-$. The function $\Phi = {\rm log}(f_+f_-)$    
is gauge invariant, and, as a result, we may set $f_+ =1$, 
$f_- = {\rm exp} \Phi$ in order to bring the general ansatz into a 
more simple and standard form. In any case, it casts the
renormalization group flows into a zero curvature form with the aid
of the infinite dimensional algebra ${\cal G}(d/dt; 1)$, which is
thus an integrable system\footnote{However, the notion of integrability
should be used with care when the gauge connections $A_{\pm}$ 
take values in infinite dimensional Lie algebras, because in some  
cases the growth of the algebra can be large and there might 
not be a straightforward way to construct and count the integrals
of motion; we will make a few extra comments about it later, while 
discussing the exponential growth of ${\cal G}(d/dt; 1)$.}. 

We may take advantage of the reformulation of these equations as 
Toda system in order 
to apply the algorithm for the systematic description of the
general solution in terms of a much simpler family of two-dimensional
free field configurations $f^{\pm} (z_{\pm}; t)$, given by 
\be
\Phi (z_+, z_-; t) = \Phi_0 (z_+ , z_-; t) + \partial_t 
\left( {\rm log} <t| M_{+}^{-1} (z_+; t) M_- (z_-; t) 
|t> \right) ~, \label{toda133}  
\ee
by specializing equations \eqn{toda5}, \eqn{toda6} and \eqn{toda7} 
to ${\cal G}(d/dt; 1)$ with $\lambda = i$, so that $\lambda^2 = -1$.  
Then, expanding the path ordered exponentials in $M_{\pm}(z_{\pm}; t)$,  
we obtain the series
\ba
& & <t| M_+^{-1} M_- |t>  = 1 + \sum_{m=1}^{\infty} 
\int^{z_+} dz_1^+ \cdots \int^{z_{m-1}^+}dz_m^+ 
\int^{z_-} dz_1^- \cdots \int^{z_{m-1}^-}dz_m^- \times  
\nonumber\\ 
& & ~~~~~~~~~ \times \int \prod_{i=1}^m dt_i \int 
\prod_{i=1}^m dt_i^{\prime} ~  
{\rm exp}{f^+ (z_i^+ ; t_i)}  
{\rm exp}{f^- (z_i^- ; t_i^{\prime})}   
D_t^{\{t_1, \cdots , t_m ; t_1^{\prime} , \cdots , 
t_m^{\prime}\}} ~, \label{toda9}  
\ea
with the ordering $z_{\pm} \geq z_1^{\pm} \cdots \geq z_{m-1}^{\pm} 
\geq z_m^{\pm}$. The elements  
$D_t^{\{t_1, t_2, \cdots , t_m; t_1^{\prime}, t_2^{\prime}, \cdots , 
t_m^{\prime}\}}$ follow from the recursive relations \eqn{toda8} for 
$K(t, t^{\prime}) = \delta^{\prime}(t-t^{\prime})$, and the 
result turns out to be  
\ba
& & D_t^{\{t_1; t_1^{\prime}\}} = \delta (t-t_1) 
\delta(t_1 - t_1^{\prime}) ~, \nonumber\\
& & D_t^{\{t_1, t_2; t_1^{\prime}, t_2^{\prime}\}} = 
\delta(t-t_1) \delta(t_1-t_1^{\prime}) \delta(t_2-t_2^{\prime}) 
\left(2\delta(t-t_2) + \partial_t \delta(t-t_2) \right) , \nonumber\\
& & D_t^{\{t_1, t_2, t_3; t_1^{\prime}, t_2^{\prime}, t_3^{\prime}\}}
= \delta(t-t_1) \delta(t-t_3) \delta(t_1-t_1^{\prime}) 
\delta(t_2-t_3^{\prime}) \delta(t_3-t_2^{\prime}) \times \nonumber\\ 
& & ~~~~~~~~~~ \times 
\left(2\delta(t-t_2) + \partial_t \delta(t -t_2) \right) + \nonumber\\
& & ~~~~~~ + \delta(t-t_1) \delta(t_1-t_1^{\prime}) 
\delta(t_2 - t_3^{\prime}) \delta(t_3 -t_2^{\prime}) \times \nonumber\\ 
& & ~~~~~~~~~~ \times 
\left(\delta(t-t_3) + \partial_t \delta(t-t_3) \right) 
\left(2\delta(t-t_2) + \partial_t \delta (t-t_2) \right) + \nonumber\\
& & ~~~~~~ + \delta(t-t_1) \delta(t_1-t_1^{\prime}) 
\delta(t_2-t_2^{\prime}) \delta(t_3-t_3^{\prime}) \times \nonumber\\ 
& & ~~~~~~~~~~ \times  
\left(2\delta(t-t_2) + \partial_t \delta(t-t_2) \right) 
\left(\delta(t-t_3) + \partial_t \delta(t-t_3) +  
\partial_{t_2} \delta(t_2 -t_3) \right) ~~ 
\ea
and so on. Thus, we obtain  
a power series expansion of the non-linear field 
$\Phi(z_+, z_-;t)$ in terms of free fields, as advertised before, 
but its structure becomes messy rather quickly. 

The families of the two-dimensional chiral fields $f^{\pm}$,
with $\partial_{\mp} f^{\pm}(z_{\pm}; t) = 0$, 
can be thought as providing a natural system of flat 
coordinates in the infinite dimensional space of all 
two-dimensional metrics, where the renormalization group flows act. 
Every physical trajectory in superspace corresponds to all possible 
geometric deformations  
of a given sigma model, which, according to our formalism, 
it can be solely described by the dependence of the parameters  
$f^{\pm}(z_{\pm}; t)$ on 
$t$. However, the convergence of the infinite series \eqn{toda9}  
cannot be easily proved in the strict mathematical sense, and one 
might question its validity, as well as its practical use, for
solving the renormalization group equations.    
We will encounter some simple algebraic solutions of the
continual Toda field equation 
later, while studying the renormalization 
group flows of constant curvature metrics (with positive or 
negative curvature), which describe axially symmetric deformations
of their volume and shape. These special examples, which are  
mini-superspace approximations of the more general trajectories in
superspace, will be used as a guide for testing the validity of 
the formal power series expansion \eqn{toda9} that describes    
$\Phi(z_+, z_-; t)$ 
in free field coordinates. Explicit comparison will be made  
in those cases by first extracting the form of the corresponding 
functions $f^{\pm}(z_{\pm}; t)$, and then checking (order by order) 
the terms that arise in the power series expansion of the  given 
solution. A similar investigation has been carried out before for
the continual Lie algebra ${\cal G} (-d^2 / dt^2 ; 1)$ that arises
in problems of four-dimensional hyper-K\"ahler geometry.   

In view of future applications, we
describe the simplifications that occur 
in the power series expansion \eqn{toda9}  
for the special class of free field configurations 
\be
\Phi_0 (z_+, z_-; t) = c \cdot (z_+ + z_-) + d(t) 
\equiv cY + d(t) ~, \label{toda56}  
\ee
which are used to parametrize solutions of the Toda field  
equations with an isometry, i.e., solutions which are 
independent of $z_+ - z_-$.
Here, $c$ is taken to be an arbitrary constant and $d(t)$ is
an arbitrary function of $t$. Then, the first few terms in the
power series expansion \eqn{toda9} can be easily computed and
the result reads
\ba
& & <t| M_+^{-1} M_- |t> ~ = ~ 1 + {1 \over c^2} e^{\Phi_0} + 
{1 \over 4c^4} e^{\Phi_0} \left(2e^{\Phi_0} + \partial_t 
e^{\Phi_0} \right) + \nonumber\\
& & ~~~~~~~~ + {1 \over 36 c^6} e^{\Phi_0} \left(6e^{2\Phi_0} 
+ 9 e^{\Phi_0} \partial_t e^{\Phi_0} + 
3\left(\partial_t e^{\Phi_0}\right)^2 + e^{\Phi_0} 
{\partial_t}^2 e^{\Phi_0} \right) + \cdots ~.   
\ea
Subsequently, using the identity ${\rm log}(1+x) = 
x - x^2 / 2 + x^3 / 3 - \cdots$,   
it can be shown that $\Phi(z_+, z_-; t)$ 
takes the form  
\be
\Phi = \Phi_0 + {1 \over c^2}  \partial_t e^{\Phi_0} +  
{1 \over 4c^4} \partial_t 
\left(e^{\Phi_0} \partial_t e^{\Phi_0} \right) +  
{1 \over 36 c^6} \partial_t \left( 
3 e^{\Phi_0} \left(\partial_t e^{\Phi_0}\right)^2 
+ e^{2\Phi_0} \partial_t^2 e^{\Phi_0} \right) 
+ \cdots \label{toda10}  
\ee
and provides a perturbative expansion of the Toda field around 
${\rm exp}\Phi \simeq 0$, i.e., for $\Phi_0 \rightarrow -\infty$.

We note finally that there other special solutions of the continual Toda
field equation that arise by suitable embedding the general solution of 
the Liouville equation,
\be
e^{\Phi_{\rm L}(z_+, z_-)} = {\partial_+ F^+(z_+) \partial_- F^-(z_-) \over
\left(1 \pm F^+(z_+) F^-(z_-) \right)^2} ~~~~~ 
{\rm with} ~~ 
\partial_+ \partial_- \Phi_{\rm L} = \mp 2 e^{\Phi_{\rm L}} ~,  
\ee
where the different signs correspond to positive or negative curvature 
spaces respectively. 
Here, to convey with the notation used earlier for the free 
field realization of Toda theories, we have chosen a pair of arbitrary  
holomorphic and anti-holomorphic functions $F^{\pm}(z_{\pm})$ so that 
$F^{\pm}(z_{\pm}) = \int^{z_{\pm}} dz_{\pm}^{\prime} {\rm exp} f^{\pm} 
(z_{\pm}^{\prime})$.  
Then, we have, in particular, the special class of solutions
\be
e^{\Phi(z_+, z_-; t)} = \pm 2(t_0 - t) 
e^{\Phi_{\rm L}(z_+, z_-)} \label{toda11}   
\ee
with $t_0$ being an arbitrary constant, 
which admit a very simple free field realization. These special solutions 
have an interesting geometric interpretation, as we will see 
later; we will call 
{\em Liouville lines} all the  
trajectories of the renormalization group flow that correspond to
these special solutions, provided that $\pm (t_0 -t) \geq 0$ 
for real $\Phi$.

\subsection{Novelties of the Lie algebra ${\cal G}(d/dt ; 1)$}

We end this section by discussing two novel features of the bosonic Lie
algebra ${\cal G}(d/dt ; 1)$, which make it rather unique from the  
mathematical and physical point of view. 
   
\underline{\em Supersymmetric Toda field equations}

First, there is a remarkable connection between the continual Toda field theories 
associated to the Cartan kernels $K(t, t^{\prime}) = \delta^{\prime} 
(t-t^{\prime})$ and $K(t, t^{\prime}) = - \delta^{\prime \prime}(t-t^{\prime})$,
which can be high-lighted with the aid of supersymmetric generalizations of
Toda systems. This connection also shows that $\delta^{\prime} (t-t^{\prime})$,
which is anti-symmetric, has a natural interpretation as the Cartan 
kernel of a continual superalgebra, but it is rather peculiar for 
bosonic Lie algebra like ${\cal G}(d/dt; 1)$.  

Consider the simple Lie superalgebra $sl(N|N+1)$ for which the
system of simple roots is odd and corresponds to the super-principal 
embedding of $osp(1|2)$ in $sl(N|N+1)$, as depicted in figure 2 below that
represents its Dynkin diagram in Kac's notation \cite{kac2} (but see also 
\cite{sorba1} for a comprehensive exposition).

\vspace{15pt}
\begin{figure}[h] \centering
\epsfxsize=8cm \epsfbox{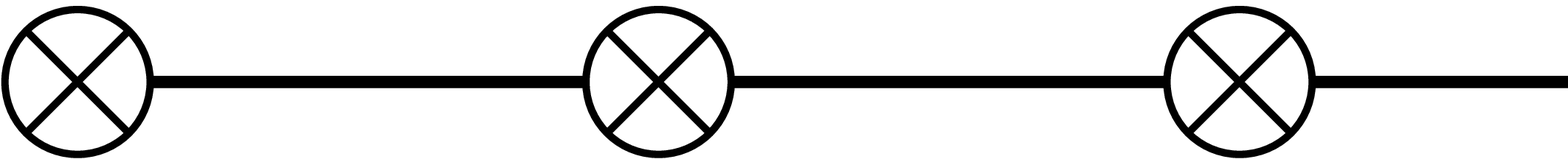} \caption{Simple odd roots of the 
Lie superalgebra
$sl(N|N+1)$.} 
\end{figure}

\noindent
Then, the continuous large $N$ limit of the Lie superalgebra $sl(N|N+1)$ is 
naturally defined by the following relations among its Weyl generators, 
\ba
& & [H(\varphi), X_{\pm 1}(\psi)]_- = \mp X_{\pm 1}(\varphi^{\prime} \psi) ~, 
~~~~ [H(\varphi), H(\psi)]_- = 0 ~, \nonumber\\
& & [X_{+1}(\varphi) , X_{-1}(\psi)]_+ = H(\varphi \cdot \psi) ~. 
\ea
Here, $H(\varphi)$ is an even element, $X_{\pm 1}(\varphi)$ are odd elements, 
whereas the Cartan kernel, which arises as the continuous limit of the Cartan
matrix of $sl(N|N+1)$, $K_{ij} = \delta_{i+1, j} - \delta_{i,j+1}$, is
chosen anti-symmetric and coincides with  
$K(t, t^{\prime}) = \delta^{\prime}
(t-t^{\prime})$. As a result, the local part of this superalgebra looks the 
same as for the contragradient continual Lie algebra ${\cal G}(d/dt; 1)$, but
now the elements $X_{\pm 1}(\varphi)$ are odd and 
satisfy anti-commutation (rather than commutation) relations.
 
The Toda field equation associated to the continual Lie superalgebra
$sl(\infty|\infty +1)$ is best described using a superfield 
$\hat{\Phi}(z_{\pm}; \theta_{\pm}; t)$ which is defined in $2/2$-dimensional 
superspace with two even ($z_{\pm}$) and two odd ($\theta_{\pm}$) coordinates.
We have the supersymmetric Toda field equation \cite{sorba2}   
\be
D_+ D_- \hat{\Phi}(z_{\pm}; \theta_{\pm}; t) = {\partial \over \partial t} 
e^{\hat{\Phi}(z_{\pm}; \theta_{\pm}; t)} ~,  
\ee
where the supersymmetric covariant derivatives in superspace are defined,  
as usual,
\be
D_{\pm} = {\partial \over \partial \theta_{\pm}} + i \theta_{\pm} 
{\partial \over \partial z_{\pm}} ~. 
\ee
In analogy with the non-supersymmetric case, the Toda field equation for the 
superfield $\hat{\Phi}(z_{\pm}; \theta_{\pm}; t)$ can be obtained from the
zero curvature condition
\be
[D_+ + A_+ , D_- + A_-]_+ = 0 ~, 
\ee
using the odd gauge connections
\be
A_+ = H(D_+ \hat{\Psi}) + X_+(1) ~, ~~~~ A_- = X_-(e^{\hat{\Phi}}) ~, ~~~~ 
{\rm with} ~~ \hat{\Phi} = {\partial \hat{\Psi} \over \partial t} ~.  
\ee
   
Parametrizing the continual Toda superfield as 
\be
\hat{\Phi}(z_{\pm}; \theta_{\pm}; t) = \Phi(z_{\pm}; t) + 
\theta_+ \Omega_+(z_{\pm}; t) + \theta_- \Omega_-(z_{\pm}; t) + 
\theta_+ \theta_- F(z_{\pm}; t) ~,   
\ee
we may write down the field equations in terms of components. This is achieved
by eliminating the auxiliary field $F(z_{\pm}; t)$, 
as usual, using the identification
$F = \partial ({\rm exp} \Phi) / \partial t$, in which case the resulting 
equations read  
\ba
& & \partial_+ \Omega_+ = -i {\partial \over \partial t} \left(e^{\Phi} 
\Omega_- \right) , ~~~~ 
\partial_- \Omega_- = i {\partial \over \partial t} \left(e^{\Phi} 
\Omega_+ \right) , \nonumber\\
& & \partial_+ \partial_- (2\Phi) = {\partial^2 \over \partial t^2} 
e^{2\Phi} - {\partial \over \partial t} \left(e^{\Phi} \Omega_+ \Omega_-
\right) .  
\ea
Setting $\Omega_+ = 0 = \Omega_-$, we arrive at the continual Toda field
equation \eqn{toda4}  
for the infinite dimensional 
Lie algebra ${\cal G}(-d^2/dt^2; 1) \simeq sl(\infty)$, which 
is just the bosonic part of the superalgebra $sl(\infty | \infty +1)$.   

Thus, we find that the anti-symmetric 
Cartan kernel $K(t, t^{\prime}) = \delta^{\prime} (t-t^{\prime})$
is naturally associated to the superalgebra of area preserving 
diffeomorphisms by giving a suitable geometric interpretation to 
$sl(\infty |\infty +1)$, as for $sl(\infty)$. We do not know whether this
fact plays a deeper role in understanding the structure of the 
bosonic Lie algebra ${\cal G}(d/dt; 1)$ that describes the renormalization 
group flows in two-dimensional sigma models. For, if we were defining 
the elements of the Cartan matrix as inner product of the simple 
roots, $K_{ij} = \vec{a}_i \cdot \vec{a}_j$, the roots 
$\vec{a}_i$ would have 
been fermionic by the anti-symmetry of $K$. In the continual case,
in particular, $\vec{a}(t)$ would correspond to a one-dimensional fermion.  
Bosonic Lie algebras with such root systems have not been studied 
before in the mathematics literature, to the best of our knowledge,
and they certainly lead to rather exotic structures. 

\underline{\em Exponential growth and related topics}

The bosonic Lie algebra ${\cal G}(d/dt; 1)$ exhibits another remarkable
property, which can only be seen by going beyond the local part  
${\cal G}_{-1} 
\oplus {\cal G}_0 \oplus {\cal G}_{+1}$ associated to the 
system of Weyl generators
\ba
& & [X_{+1}(\varphi), X_{-1}(\psi)] = H (\varphi \psi) ~, ~~~ 
[H(\varphi), H(\psi)] = 0 ~, \nonumber\\  
& & [H(\varphi), X_{\pm 1}(\psi)] = \mp X_{\pm 1} 
(\varphi^{\prime} \psi) ~.  
\ea
The full algebra is a $Z$-graded algebra ${\cal G} = 
\oplus_{n \in Z} {\cal G}_n$, which according to the general theory can be 
generated recursively by 
taking successive commutators of $X_{\pm 1}$. 
The generators of ${\cal G}_{\pm 2}$ are defined, as usual, by $X_{\pm 2} 
\in [{\cal G}_{\pm 1}, {\cal G}_{\pm 1}]$, 
and their commutation relations assume
the form
\ba
& & [X_{\pm 1}(\varphi), X_{\pm 1}(\psi)] = 
\pm X_{\pm 2} (\varphi^{\prime} \psi
- \varphi \psi^{\prime}) ~, ~~~~ 
[H(\varphi), X_{\pm 2}(\psi)] = \pm 2 X_{\pm 2} (\varphi^{\prime} 
\psi) ~, 
\nonumber\\ 
& & [X_{\pm 2}(\varphi), X_{\mp 1} (\psi)] = \pm X_{\pm 1}(\varphi \psi) ~, 
~~~~ [X_{+2}(\varphi), X_{-2}(\psi)] = 0 ~. 
\ea
They can be easily derived by requiring consistency with the 
Jacobi identities \cite{vershik}. 

Proceeding further to ${\cal G}_{\pm 3}$, one finds that  
there are two independent generators that parametrize  
each one of the Lie algebra elements 
$[{\cal G}_{\pm 1}, {\cal G}_{\pm 2}]$, which turn to be 
\be
X_{\pm 3}^{(s)}(\varphi) = 2^{s-3}  
[X_{\pm 1}(1) , X_{\pm 2}(\varphi)] + (-1)^s [X_{\pm 1}(\varphi), 
X_{\pm 2} (1)] ~, ~~~~ {\rm for} ~~ s=1, 2 ~.  
\ee
It can be verified that these combinations satisfy the 
commutation relations
\ba
& & [X_{\pm 3}^{(1)} (\varphi) , 
X_{\mp 1} (\psi)] = \mp X_{\pm 2} (\varphi \psi) ~, ~~~~~   
[X_{\pm 3}^{(2)} (\varphi) ~ ,  
X_{\mp 1} (\psi)] = \pm  X_{\pm 2} 
((\varphi \psi)^{\prime}) ~, \nonumber\\  
& & [H(\varphi) , X_{\pm 3}^{(s)}(\psi)] = \pm 3 X_{\pm 3}^{(s)}
(\varphi^{\prime} \psi) ~, 
\ea
which, thus, prove the linear independence of the elements 
$X_{\pm 3}^{(1)}$ and $X_{\pm 3}^{(2)}$ \cite{vershik}.  

More generally, for any positive integer $n \geq 2$, 
it can be shown by induction that if ${\cal G}_{\pm n}$ is 
spanned by the independent elements $X_{\pm n}^{(1)}, \cdots , 
X_{\pm n}^{(d_{n})}$, the subspace ${\cal G}_{\pm (n+1)} = 
[{\cal G}_{\pm 1} , {\cal G}_{\pm n}]$ will be spanned by the following 
$2d_n$ independent elements \cite{vershik}  
\ba
X_{\pm (n+1)}^{(s)} (\varphi) & = & \alpha_n^{(s)}  
[X_{\pm 1} (1) , X_{\pm n}^{(s)}(\varphi)] - [X_{\pm 1} (\varphi) , 
X_{\pm n}^{(s)} (1)] ~; ~~~ 1 \leq s \leq d_n ~, \\ 
X_{\pm (n+1)}^{(s)} (\varphi) & = & \beta_n^{(s)}  
[X_{\pm 1} (1) , X_{\pm n}^{(s)}(\varphi)] + [X_{\pm 1} (\varphi) , 
X_{\pm n}^{(s)} (1)] ~; ~~~d_n +1 \leq s \leq 2d_n \nonumber   
\ea
for some appropriately chosen constants $\alpha_n^{(s)}$ and 
$\beta_n^{(s)}$.  
Thus, the dimension of the subspaces ${\cal G}_n$ (relative to the dimension
of ${\cal G}_0$, which is taken to be 1) is given by
\be
d_0 = d_{\pm 1} = 1 ~, ~~~~~ d_n = 2^{|n| -2} ~~~~ {\rm for} ~~ |n| \geq 2  
\ee
and therefore, this particular continual algebra exhibits 
{\em exponential growth} 
that makes it rather exotic, but worth studying in great detail. 
It can also be shown that
\be
[{\cal G}_{\pm 2} , 
X_{\mp (n+1)}^{(s)}] = 0
\ee  
for all $n \geq 2$ and $1 \leq s \leq d_n$. 

The complete set of commutation relations of this infinite dimensional 
algebra have not been derived systematically to this day and there has 
been no analogue of the corresponding Serre relations, to the best of our 
knowledge, due to its exponential growth. 
In fact, this is a typical behavior of indefinite 
Lie algebras, with hyperbolic algebras being the simplest, yet quite  
intractable, examples. Recall that 
hyperbolic algebras are usually defined using 
a symmetric Cartan matrix with  Lorentzian structure 
in their root system \cite{kac1}, but here, 
the algebra ${\cal G}(d/dt; 1)$, also has 
the additional feature that its Cartan kernel is anti-symmetric. Then, 
in analogy with hyperbolic algebras, one 
expects that the structure of the full algebra ${\cal G}(d/dt; 1)$ 
can only be accommodated in string theory, but there is no clear 
way to proceed in this direction at the moment.  
Fortunately, we do not need to know the structure of the full algebra
in order to describe the renormalization group
flows of two-dimensional sigma models as Toda field equations, because 
we are only considering the simplest case of Toda field equations
with both gauge connections $A_{\pm}(z_+, z_-)$
taking their values in the local part of the algebra,  
${\cal G}_{-1} \oplus {\cal G}_0 \oplus {\cal G}_{+1}$. 
The structure of the 
full algebra is only needed to 
describe the form of the conserved currents, which in ordinary Toda 
theory are constructed in terms of the Casimir operators \cite{jean}, 
but they are not known yet in the present case. For this  
reason we are not in the position to provide a systematic list of
all integrals of the renormalization group flows, despite their   
integrability. We postpone the construction
of the conserved currents 
to future work with the note that unexpected results might
arise, as for some Toda field theories associated to hyperbolic
algebras \cite{inami}. 
A few more comments will be made at the end of this paper.

It is worth mentioning now  
that generalized versions of the Toda field 
equations can be considered by introducing gauge connections $A_{\pm}$
taking values in the subspace ${\cal G}_{-N} \oplus {\cal G}_{-N+1} 
\oplus \cdots \oplus {\cal G}_0 \oplus \cdots 
\oplus {\cal G}_{N-1} \oplus {\cal G}_N$ 
of the complete algebra for any 
finite, but fixed, value $N \geq 2$, i.e., 
\be
A_{\pm} (z_+, z_-) = H(\varphi_0^{\pm}) + \sum_{n=1}^N \sum_{s=1}^{d_n} 
X_{\pm n}^{(s)} (\varphi_n^{\pm; s}) ~.  
\ee  
Here, the gauge connection are parametrized by arbitrary functions 
$\varphi_n^{\pm ; s}(z_+, z_-; t)$ for each independent element that
generates the subspaces ${\cal G}_{\pm n}$. Thus, taking into account 
the exponential growth of the continual algebra 
${\cal G}(d/dt; 1)$, we see that $A_+$ 
depends on $2^{N-1} +1$ such arbitrary functions, and likewise for 
the gauge connection $A_-$. Of course, one can always make a gauge 
transformation to set $\varphi_0^- = 0$ in the zero curvature condition,
in which case the generalized system of Toda field equations 
involves $2^N + 1$ functions in total that satisfy the same number
of non-linear equations. These equations follow by comparing terms
for the generators $H$ and 
$\{X_{\pm n}^{(s)} ; 1\leq n \leq N, 
~ 1\leq s \leq d_n \}$ that arise by computing the commutator 
$[A_+ , A_-] \in {\cal G}_{-N} \oplus \cdots \oplus {\cal G}_0 \oplus 
\cdots \oplus {\cal G}_N$ in the zero 
curvature condition. 

This is precisely the way that the equation 
$\partial_+ \partial_- \Phi = \partial_t({\rm exp} \Phi)$ arose earlier
for $N=1$, but for $N \geq 2$ the details of the generalized Toda system
of field equations depend crucially on the structure  
of the commutation relations of the algebra, which is still lacking
in all generality. For $N=2$, it is indeed possible to work out the
details of the generalized system of Toda field equations using the 
explicit form of the commutation relations in the subspace 
${\cal G}_{-2} \oplus {\cal G}_{-1} \oplus {\cal G}_0 \oplus {\cal G}_{+1} 
\oplus {\cal G}_{+2}$ given above, but for $N \geq 3$ we cannot write
down a systematic answer at this moment. 
In any case, it will be interesting to
examine whether such generalized continual Toda field equations have 
any relevance to the beta function equations and the 
renormalization group flows 
of sigma models, where, apart from the metric, couplings to other fields 
are also allowed (see, for instance, \cite{tsey5}, and references therein). 
We expect that this line of thought will help to clarify
the full structure of the continual Lie algebra ${\cal G}(d/dt; 1)$ 
and provide a string theoretic realization of it in two dimensions.       

\section{The sausage model}
\setcounter{equation}{0}

There is a well studied solution of the continual Toda field equation 
associated with the renormalization group flow of two-dimensional 
spaces with spherical topology, $S^2$. Consider the 
class of axially symmetric sigma models with    
two-dimensional action
\be
S_{\rm t} = {1 \over 2} \int e^{\Phi(Y; t)} 
\left((\partial_\mu X)^2 + (\partial_\mu Y)^2 
\right) d^2 w,  
\ee
where the conformal factor of the target space metric is given by the
one-parameter family of functions $\Phi(Y; t)$, \cite{sausa},   
\be
e^{\Phi (Y; t)} = {2 \over a(t) + b(t) {\rm cosh}2Y} ~.  
\ee
Here, $X$ is an angular coordinate, $ 0 \leq X \leq 2 \pi$, which accounts 
for the axial symmetry of the metric that is independent of $X$, whereas 
$Y$ ranges over the whole real axis, $-\infty < Y < +\infty$. Furthermore, 
the end points $Y = \pm \infty$ are used to parametrize the north and the 
south poles of the manifold with spherical topology, 
where the boundary conditions
\be
\Phi(Y; t) \sim -2 |Y| ~~~~~ {\rm for} ~ Y \rightarrow \pm \infty ~,  
\ee
are imposed for regularity. 

In general, the metric of the two-dimensional target space is real  
provided that $a \geq -b$ and $b \geq 0$, 
which is implicitly assumed here, but later we will impose the stronger 
condition $a \geq b \geq 0$ as more appropriate for the 
geometric deformations of the $O(3)$ sigma model.  
The target space is a deformed sphere with axial symmetry, rather
than round sphere, and so the geometry looks like   
a sausage, in general, which accounts for the term ``sausage model" used in 
the literature.  Further justification will be given shortly by studying
the geometric deformation of this class of models under the 
action of the renormalization group flows.    

There are two special configurations that arise in the general class
of sausage models and deserve special reference.
First, the case $a(t) = b(t) = 1$ corresponds to the 
usual $O(3)$ sigma model with the geometry of a round sphere associated to  
the conformal factor 
\be
e^{\Phi(Y)} = {1 \over {\rm cosh}^2 Y} ~.  
\ee
Actually, to make contact with the standard description of the $O(3)$  
sigma model, it is convenient to introduce two angular coordinates
$(\theta, \varphi)$ with 
\be
{\rm cos} \theta = {\rm tanh} Y ~, ~~~~~  
\varphi = X ~, 
\ee
so that $0 \leq \theta \leq \pi$, as $Y$ ranges 
from $+\infty$ to $-\infty$, whereas $0 \leq \varphi \leq 2 \pi$. They 
provide a more natural parametrization of the two-dimensional 
sphere using the geographical chart $(\theta , \varphi)$ with the 
north and south poles located  
at $\theta = 0$ and $\pi$ respectively. Then, the target space metric
assumes the equivalent form
\be
ds^2 = {1 \over 2} \left( d\theta^2 + {\rm sin}^2 \theta d\varphi^2 
\right)   
\ee
and describes a round sphere of radius $1/\sqrt{2}$; 
clearly, the beta function is 
non-vanishing in this case. Another special configuration in the  
one-parameter family of sausage  
metrics arises for the choice $a(t) = \gamma$, where $\gamma$ is an 
arbitrary positive  
constant, and $b(t) = 0$. Then, the target space metric takes the form
\be
ds^2 = {1 \over \gamma} (dX^2 + dY^2)   
\ee
and the model describes two free scalar fields $Y$ and $X$ 
taking values in the spaces $R$ and $U(1)$, respectively. The resulting
theory, which is denoted by $RU(1)_{\gamma}$, 
defines a two-dimensional quantum conformal field theory with 
vanishing beta function. The radius of the compact space $U(1)_{\gamma}$  
is $1/\sqrt{\gamma}$ and assumes all 
possible values as $\gamma$ ranges from $0$ to $+\infty$. For $\gamma = 0$, 
the coordinate $X$ decompactifies and the geometry describes a plane, 
which can also be thought at the infinite radius limit of a round 
sphere. For $\gamma = + \infty$, the coordinate $X$ shrinks to zero size
and the geometry degenerates to a one-dimensional line. Finally, for 
all intermediate values of $\gamma$, we obtain the geometry of 
a two-dimensional cylinder. Thus, it follows that  
the one-parameter family of sausage metrics encompasses
both conformal as well as non-conformal backgrounds.  
 
One expects that for generic choices of the functions $a(t)$ and $b(t)$ 
the resulting two-dimensional quantum field theory will violate 
scale invariance and, hence, have non-vanishing beta function. 
It turns out that the renormalization group flows, which result 
by changing the world-sheet length scale of the model, act within
the above one-parameter class of metrics in a consistent way and provide
a set of first order non-linear differential equations for the two 
yet unknown functions. They read as follows,       
\be
a^{\prime}(t) = 2 b^2 (t) ~, ~~~~~ b^{\prime}(t) = 2 a(t) b(t) ~,  
\ee
and they can be easily integrated by noting that the quantity  
\be
\gamma^2 = a^2(t) - b^2(t) 
\ee
is conserved along the flow. Hence, choosing $\gamma$ to be any arbitrary real 
constant, so that $a(t) \geq -b(t)$ in all generality, 
we obtain the solution \cite{sausa}  
\be
a(t) = \gamma {\rm coth}2\gamma(t_0 -t) ~, ~~~~~ 
b(t) = {\gamma \over {\rm sinh} 2\gamma (t_0 -t)} ~.  
\ee
This solution does not depend on the sign of the integration 
constant $\gamma$, and so we 
may choose $\gamma \geq 0$ for all
practical purposes. Moreover, the restriction $b(t) \geq 0$ 
implies that $t \leq t_0$, in which case the model exhibits a
well defined ultra-violet limit for $t \rightarrow -\infty$.
This behavior is natural for all compact spaces with 
positive curvature 
and, therefore, the physically relevant region of the sausage  
model is restricted to the wedge $a(t) \geq b(t) \geq 0$ of the  
parameter space.   

There is another arbitrary 
integration constant $t_0$ appearing in the general solution, where
both $a(t)$ and $b(t)$ become infinite. In fact, we have the following
leading behavior near $t_0$,
\be
a(t) \simeq b(t) \simeq {1 \over 2(t_0 -t)} ~, ~~~~~ 
e^{\Phi(Y;t)} \simeq 2 {t_0 -t \over {\rm cosh}^2 Y} ~,  
\ee
which is independent of the parameter $\gamma$, and hence universal.  
It means that all renormalization group trajectories in the two-dimensional 
parameter space $(a, b)$ approach the diagonal line $a (t)= b (t)$ as 
$t \rightarrow t_0$, thus converging to the point $(\infty, \infty)$.
We also note, by comparing the resulting metric with the $O(3)$ sigma model, 
that all solutions tend to the geometry of a round sphere, but with 
vanishing volume as $t \rightarrow t_0^-$. Then,  
$t_0$ provides a limiting upper bound in the allowed range of the renormalization 
group time $t$, and it makes no sense to continue the flow 
beyond $t = t_0$. The  
curvature also becomes infinite at the degeneration point $t=t_0$, 
thus invalidating the
lowest order approximation to the beta function equations that was 
obtained by disregarding all higher order curvature terms. On the 
other hand, $t$ can vary without restriction all the way to $-\infty$, in
which case the parameters reach the fixed values  
\be
a(t) = \gamma ~, ~~~~~ b(t) = 0 ~, ~~~~ {\rm for} ~~ t = -\infty ~.  
\ee
These constant values define the line of {\em ultra-violet points} of the  
trajectories for all $\gamma \geq 0$, and they correspond to the 
quantum conformal field theories $RU(1)_{\gamma}$.        
 
The renormalization group trajectories are hyperbola in the $(a, b)$ parameter
space, and they are described by the first integral of the flow $a^2(t) - b^2(t) = 
\gamma^2$. As such, they fill up the entire physical 
wedge $a(t) \geq b(t) \geq 0$ by starting from the ultra-violet fixed points 
$(a=\gamma, b=0)$ at $t= -\infty$, and they evolve continuously without
intersecting each other towards the diagonal boundary line of the wedge,
$a=b \geq 0$,  
which they only reach asymptotically, as $t \rightarrow t_0$. The diagonal 
boundary line is also an allowed trajectory for $\gamma = 0$. 
The renormalization 
group trajectories are depicted in figure 3 below. 

\vspace{10pt}
\begin{figure}[h] \centering
\epsfxsize=10cm
\epsfysize=8cm
\epsfbox
{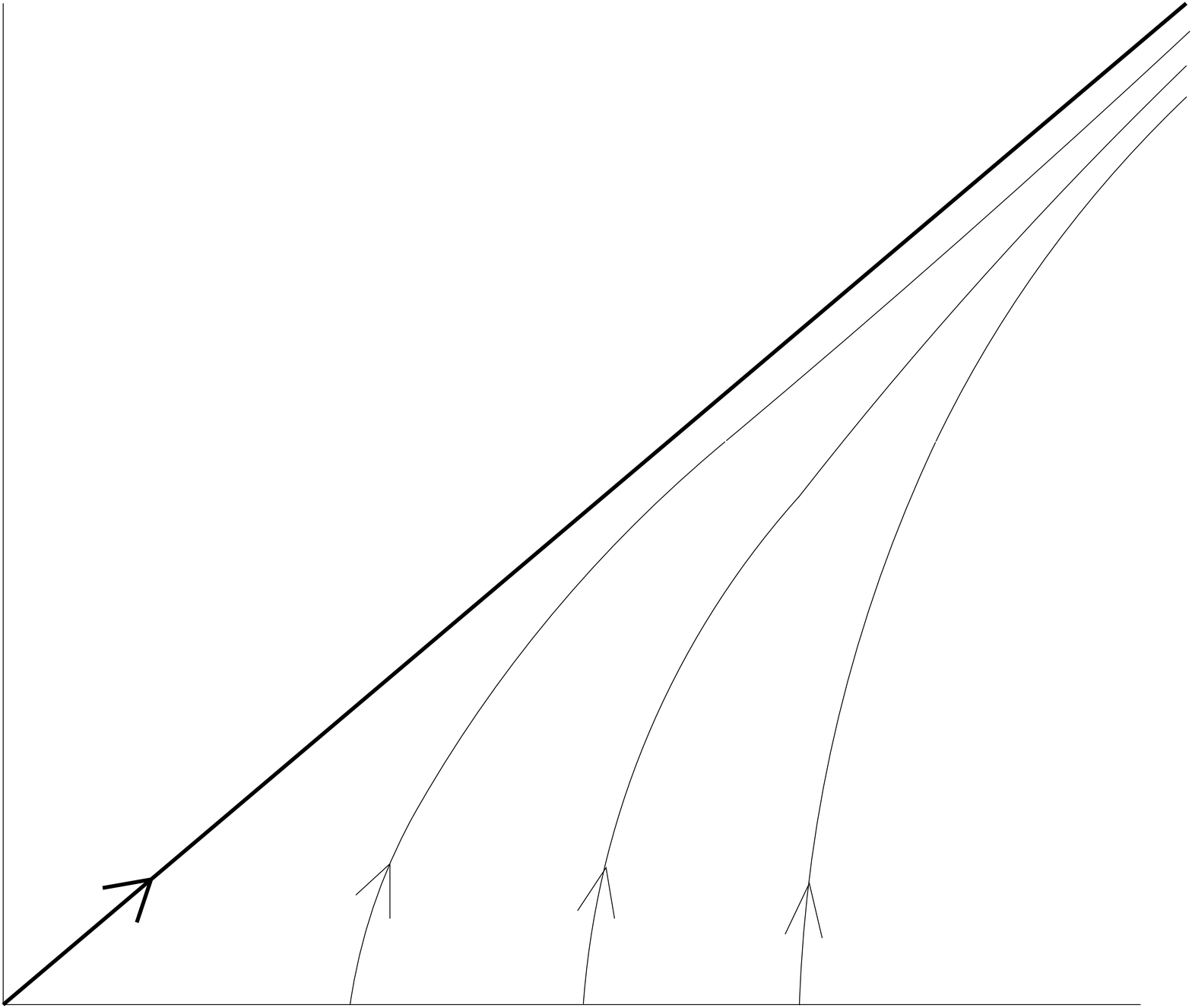} \put(-20,-15){$a(t)$} \put(-305,230){$b(t)$}
\put(-290,-15){0}
\caption{Renormalization group trajectories of the sausage model.}
\end{figure}

The geometry of the space changes along the trajectories in a way that
is reminiscent of a sausage deformation of the round sphere.
A way to visualize this deformation is provided 
by the embedding of the two-dimensional surface in three-dimensional
Euclidean space, in analogy with the standard description of the 
$O(3)$ sigma model. For this, we introduce of an $O(3)$ unit vector 
field 
$\{n_i; ~i = 1, 2, 3\}$, with $\vec{n} \cdot \vec{n} = 1$, 
\be
\vec{n} = \left({\rm sin} \theta {\rm cos} \varphi , ~ 
{\rm sin} \theta {\rm sin} \varphi , ~  
{\rm cos} \theta \right) . 
\ee
Then, in terms of these variables, the action  
$S_{\rm t}$ is cast in
the form \cite{sausa} 
\be
S_{\rm t} = {1 \over g^2(t)} \int {(\partial \vec{n})^2 \over
1 - {\gamma}^2 n_3^2/g^2(t)} d^2w ~; ~~~~~ {\rm with} ~~ 
g^2(t) = \gamma {\rm coth} \gamma (t_0 -t) ~,  
\ee
and describes an axially symmetric sausage configuration 
embedded in three-dimensional Euclidean space spanned by $n_1$, $n_2$ and
$n_3$.
The sausage is elongated in the direction $n_3$, which  
acts as the axis of symmetry in three-dimensional 
Euclidean space, as depicted in figure 4 below. 

\vspace{-15pt}
\begin{figure}[h] \centering
\epsfxsize=8cm
\epsfbox{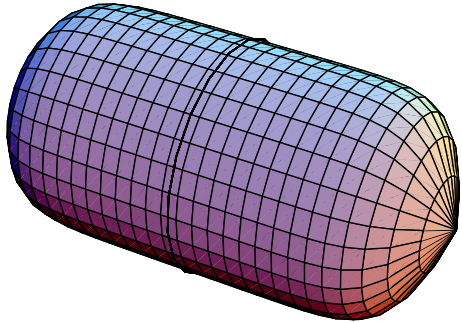} 
\\[-20pt] \caption{A sphere deforming to a sausage configuration.}
\end{figure}

The sausage becomes infinite long at the ultra-violet points,  
where it looks like a cylinder $R \times S^1$ with radius
equal to $1/\sqrt{\gamma}$. The theory is asymptotically free in the
ultra-violet region, as expected on general grounds, 
and corresponds to the quantum conformal 
field theories $RU(1)_{\gamma}$ for all $\gamma$. 
As $t$ flows towards $t_0$, the configuration starts to become 
round with diminishing size, and it shrinks 
asymptotically to zero as $t \rightarrow t_0$, as in a big crunch. 
For $\gamma = 0$, we 
have the usual spherical geometry of the $O(3)$ sigma model for all  
times, since the model follows the diagonal trajectory by starting from 
a sphere of infinite radius and evolving towards zero size; 
we will say more about this 
special solution later, in connection with the reduction of the continual
Toda equation to Liouville theory.  

An alternative way to visualize the geometry of the resulting configuration
is described by changing coordinates to the proper variable  
\be
\tilde{Y} = \sqrt{2} {\rm cosh} \gamma (t_0 - t) 
\int {d Y \over \sqrt{{\rm cosh} 2Y + {\rm cosh} 2 \gamma 
(t_0 - t)}} = 
F(\psi; k) ~,  
\ee
where $F(\psi; k)$ denotes the incomplete elliptic integral of the first kind,
\be
F(\psi; k) = \int_0^{\psi} {d \chi \over \sqrt{1 - k^2 {\rm sin}^2 \chi}} =
\int_0^{{\rm sin} \psi} {dx \over \sqrt{(1-x^2)(1-k^2x^2)}} ~, 
\ee
with parameters
\be
{\rm sin} \psi = {\rm tanh} Y ~, ~~~~~ k = {\rm tanh} \gamma (t_0 - t) ~.  
\ee
By the defining relations of the elliptic functions, we also have
${\rm sin} \psi = {\rm sn} 
( \tilde{Y} ; k )$, 
where ${\rm sn}(u; k)$ is the standard sine-amplitude Jacobi elliptic function 
with modulus $k$. 
The metric of the deformed configuration assumes the
following form in proper coordinates $(X, \tilde{Y})$, 
\be
ds_{\rm t}^2 = {k \over \gamma} 
\left(d{\tilde{Y}}^2 + {\rm sn}^2 
\left( \tilde{Y} + K(k); k \right) dX^2 \right) , \label{guya} 
\ee
where we also made use of the identity 
\be
{\rm sn}^2 (u + K(k); k) = {{\rm cn}^2 (u; k) \over 
{\rm dn}^2 (u; k)} = {1 - {\rm sn}^2 (u; k) \over 1 - k^2 {\rm sn}^2 (u; k)} ~.
\ee
Then, the shape of the space can be easily drawn, as in figure 4  
above, and it  
represents the deformation of a sphere into an axially symmetric sausage-like  
configuration for all $\gamma \neq 0$, with $\tilde{Y}$ ranging from 
$-K(k)$ to $K(k)$.     

The change of coordinates to $(X, \tilde{Y})$ has to be compensated by 
a vector field $\xi_{\mu} = \partial_{\mu} \tilde{\Phi}$ that describes  
reparametrization along the renormalization group flows. Recall that 
in this case
one has to solve the more general system of equations 
\be
{\partial \over \partial t} G_{\mu \nu} = -R_{\mu \nu} + 
2 \nabla_{\mu} \nabla_{\nu} \tilde{\Phi} ~,  
\ee
which determine the
``would be" dilaton field $\tilde{\Phi}$ of the sausage configuration 
in proper coordinates, following the
general framework of section 2.3.  We find that
the relevant configuration is given by
\ba
ds_{\rm t}^2 & = & {1 \over \gamma} 
\left(k d{\tilde{Y}}^2 + \left({H_1(\tilde{Y}) \over 
\Theta_1 (\tilde{Y})}\right)^2 dX^2 \right) , \nonumber\\
\tilde{\Phi}(\tilde{Y}) & = & {\rm log} \Theta_1 (\tilde{Y}) + {1 \over 2} 
\left({E(k) \over K(k)} - {1 \over 2}{k^{\prime}}^2 \left(1 + 
{1 \over \gamma} \right) \right)  
{\tilde{Y}}^2 ~, \label{funny1} 
\ea
where $\Theta_1 (u) = \Theta (u + K(k))$, $H_1 (u) = H(u + K(k))$, as
usual, and $K(k)$, $E(k)$ denote the complete elliptic integrals of the
first and second kind, respectively. Here, we have written 
the metric \eqn{guya} and the corresponding dilaton field 
in terms of Jacobi theta functions, choosing the value of 
$\tilde{\Phi} (\tilde{Y})$ equal to ${\rm log} \Theta (K(k))$ at 
$\tilde{Y} = 0$, rather than zero. Then, 
in this frame, the solution of 
the renormalization group flow follows immediately from the 
heat equation \eqn{hote} for the theta functions. It is also useful to
know in the course of the calculations that 
\be
{d\tau \over dk} = -{i \pi \over 2}{1 \over k{k^{\prime}}^2 K^2 (k)} ~, 
~~~~~ {\rm where} ~~  
\tau = i{K(k^{\prime}) \over K(k)} ~. 
\ee
Thus, we encounter a novel embedding of periodic solutions of the 
heat equation into the non-linear system of the renormalization group
flows, which is applicable for all times; more examples of 
this embedding will be discussed in section 5.2.  
 
The description of the sausage configuration in proper coordinates, using 
elliptic functions, proves advantageous for studying some special 
features of its geometry. 
The elliptic modulus is $k = {\rm tanh} \gamma 
(t_0 -t)$, thus ranging from $1$ to $0$ as the renormalization group time $t$ 
flows from the ultra-violet region $t \rightarrow -\infty$ to the curvature
singularity that occurs at $t=t_0$. The function ${\rm sn}(u; k)$ is periodic 
in $u$, with two consecutive zeros occurring at a distance $2K(k)$ 
apart from each other,    
and so the characteristic length of the sausage in the 
$\tilde{Y}$-direction is given by $2\sqrt{k}K(k)$ for all $0 \leq k \leq 1$. 
For $k \simeq 1$, in particular, 
we have ${\rm cn}(u; k) \simeq {\rm dn}(u; k)$
and the geometry is an infinite long cylinder with metric  
$ds^2 = (d{\tilde{Y}}^2 + dX^2)/ \gamma$, as seen from any finite distance 
away from the origin of the proper coordinate, $\tilde{Y} = 0$. However, it
is also interesting to examine the geometry of the sausage close to its tips, 
which are located infinite far away from $\tilde{Y} = 0$  
when $k = 1$. It is convenient for this purpose to introduce the 
change of variables 
\be
\rho = \tilde{Y} + K(k) ~,  
\ee
by first zooming close to one of its tips, and then take the ultra-violet limit 
$k \rightarrow 1$. Since, all Jacobi 
elliptic functions become ordinary hyperbolic functions in the limit $k=1$,
in particular ${\rm sn}(u; 1) = {\rm tanh}u$, we obtain the following form 
of the metric
\be
ds_{\infty}^2 = {1 \over \gamma} \left(d \rho^2 + 
{\rm tanh}^2 \rho ~ dX^2 \right) .  
\ee

Remarkably, this is the geometry of a Euclidean two-dimensional 
black-hole \cite{ed2, others}, which looks like an infinite long cigar, 
and which asymptotically becomes flat, 
$R \times S^1$,  
as $\rho \rightarrow + \infty$\footnote{There is another interesting
occurrence of the two-dimensional black-hole geometry as the {\em infra-red} 
fixed point of all two-dimensional cigar-like geometries, which has been
studied before \cite{hori1} (but see also \cite{tong}). Deforming cigars 
provide yet another solution of the continual Toda field equation 
for non-compact spaces, which,  
however, will not be revisited in the present work. Here, in contrast with
the renormalization group flow of a single cigar, we have the  
occurrence of two semi-infinite cigars glued together in the {\em ultra-violet} 
limit of compact sigma models.}.  Likewise, in the vicinity of the other tip
of the sausage, $\tilde{Y} \rightarrow - K(k)$, 
which can be reached by first performing the change of 
coordinates $\rho = \tilde{Y} - K(k)$ and then taking
the limit $k \rightarrow 1$, the geometry also looks like a two-dimensional
black-hole, 
since ${\rm sn}^2(u+2K(k); k) = {\rm sn}^2(u; k) \simeq {\rm tanh} u$ 
for $k \simeq 1$. Thus,
the infinite long cylinder that describes the conformal field theory
at the ultra-violet point, $RU(1)_{\gamma}$, can be thought as being
constructed by sewing
together two infinite long cigars, which connect to each other 
in the asymptotic region at $\tilde{Y} = 0$. Each of the two Euclidean  
black holes describes the semi-classical geometry of the $SL(2,R)_l / U(1)$ 
Wess-Zumino-Witten model, which is exact in the large level $l$ limit of 
the $SL(2,R)$ current algebra. In this context, we have $l = 2/\gamma$ 
and, therefore, the interpretation we are proposing is 
more appropriate when $\gamma$ 
is close to $0$.  

The dilaton field of Witten's two-dimensional 
black-hole is equal to \cite{ed2}  
\be
\tilde{\Phi} (\rho) = {\rm log} \left({\rm cosh} \rho \right) ~ . 
\ee
It is zero at $\rho = 0$, and it becomes infinite at $\rho = \infty$, where
the string coupling constant ${\rm exp}(-\tilde{\Phi})$ 
vanishes. Of course, we can also shift the
dilaton by an arbitrary constant\footnote{Adding an 
arbitrary constant to the dilaton 
changes the mass of the black hole, however, which comes out to be
$M = \sqrt{\gamma} {\rm exp} (2 \tilde{\Phi} (0))$, and it represents 
a family of objects with variable mass inserted in a fixed 
background. For $\tilde{\Phi}(0) = 0$, the mass is inverse to the 
radius of the compact coordinate $X$.}, 
but this has no effect in the asymptotic
region where $\tilde{\Phi}$ grows linearly, $\tilde{\Phi} \simeq \rho$, 
and becomes infinite as $\rho \rightarrow \infty$ (or, else, as
$\tilde{Y}$ becomes finite). In our case, following equation 
\eqn{funny1}, we find that the dilaton field is constant,  
\be
\tilde{\Phi} (\tilde{Y}) \simeq {\rm log} \Theta (K(k)) = 
{1 \over 2} {\rm log} 
\left({2K(k) \over \pi}\right) ,
\ee
for all $\tilde{Y}$ when $k \rightarrow 1$, or when 
$k \rightarrow 0$. In the first case\footnote{Here, $\tilde{\Phi} 
(\tilde{Y})$ is equal to ${\rm log} \Theta (K(k))$ at any finite distance
away from the origin $\tilde{Y} = 0$, as it can be readily seen 
from equation \eqn{funny1} when $k \rightarrow 1$. Likewise, we 
observe that at the two tips located at $\tilde{Y} = \pm K(k)$, or any
finite distance away from them, the dilaton field behaves as 
$\tilde{\Phi} \simeq {\rm log} \Theta (0) + K(k)/2$, since 
$E(k) \rightarrow 1$ and $k^{\prime} K(k) \rightarrow 0$ as 
$k \rightarrow 1$. Moreover, since ${\rm log} \Theta (0) = 
{\rm log} \Theta (K(k)) + {\rm log} \sqrt{k^{\prime}}$, we arrive 
immediately at the desired expression using the fact that 
$K(k) + {\rm log} k^{\prime}$ is finite ($= {\rm log}4$) for 
$k=1$.}, $k=1$, $\tilde{\Phi}$ becomes infinite
everywhere by assuming the same value as in the cylindrical asymptotic  
region where the two black holes are 
glued together by matching 
their metric and dilaton fields. Then, the resulting
``two body" problem has a dilaton which is infinite everywhere by simple
superposition of the two individual ``one body" problems. 
In the second case\footnote{Here, $\tilde{\Phi} (\tilde{Y} = 0) = 
{\rm log} \Theta (K(0)) = 0$, since $K(0) = \pi / 2$, and it does 
not make sense to have $\tilde{Y} \neq 0$; in this case, the distance
shrinks to zero size, although $-\pi / 2 \leq \tilde{Y} \leq \pi/2$, 
because the metric \eqn{funny1} scales with $k$ and tends to zero.},
$k=0$, the two black holes ``eat" each other and the dilaton field 
becomes zero, which is consistent with the
creation of a curvature singularity.   
According to this picture, the two black holes are initially far away 
from each other and they start ``attracting" each other by forming a 
sausage configuration as intermediate state that ends with a big 
crunch at $t=t_0$. Actually, it might be interesting to put this 
intuitive mechanical analogue on more firm basis, but we will not 
attempt to do so in the present work.   
 
Summarizing, the sausage model corresponds to the following algebraic solution
of the dimensionally reduced continual Toda equation in conformally flat 
coordinates,
\be
ds_{\rm t}^2  =  
{dX^2 + dY^2 \over a(t) + b(t) {\rm cosh}2Y} =  
{1 \over \gamma} {\rm sinh}2\gamma (t_0 -t)  
{ d \omega_+ d \omega_- \over 
1 + 2 \omega_+ \omega_- {\rm cosh}2\gamma(t_0 -t) + 
(\omega_+ \omega_-)^2} ~,   
\ee
using the parametrization $X \in U(1)$,  
$Y \in R$, or,   
equivalently, using a system of complex coordinates $\omega_{\pm}$,
\be
\omega_{\pm} = {\rm tan} {\theta \over 2} e^{\pm i \varphi} = 
\sqrt{{1-{\rm tanh} Y \over 1 + {\rm tanh} Y}} ~ e^{\pm iX} ~.  
\ee

For $\gamma = 0$, we note separately that the target space metric has
the exact form  
\be
e^{\Phi (\omega_+ , \omega_- ; t)} = 
2{(t_0 -t) \over (1+\omega_+ \omega_-)^2} ~ 
\ee
for all $-\infty < t \leq t_0$, 
which corresponds to the geometry of a collapsing round sphere, as 
noted before. Actually,  
in this case, the radius of the sphere 
equals to $\sqrt{t_0 -t}$ and so, its volume,  
\be
V(t) = 4\pi (t_0 -t) ~, 
\ee
depends linearly on $t$, as expected on general grounds 
given in section 2.1.
The collapsing 
round sphere is an exact solution of the continual Toda field equation, which
is obtained by embedding solutions of the Liouville equation with  
factorized linear dependence in $t$, as in equation \eqn{toda11} (see also
equation \eqn{mena}).  
Thus, in the nomenclature used earlier, 
the diagonal line $a(t) = b(t)$ is a Liouville line  
in the two-dimensional parameter space of the sausage model,  
which also has a very simple free field realization. 
All other hyperbolic trajectories
in the parameter space with $\gamma \neq 0$ are lying 
beyond this simple Liouville embedding and their free field realizations 
require full use of the algebraic methods of section 3.3 based on B\"acklund 
transformation for the continual Lie algebra ${\cal G}(d/dt ; 1)$. 

\underline{\em Free field realization}
       
We are turning now to the free field realization of the  
sausage configuration based on the algebraic construction of the 
most general solution of the continual Toda field equations.
We choose to work in the conformally flat system of coordinates $(X, Y)$, 
and note that in the asymptotic region of $Y$, 
i.e., $Y \rightarrow \pm \infty$, 
the conformal factor of the sausage metric goes to zero as follows: 
\be
e^{\Phi(Y; t)} \simeq {4 \over b(t)} e^{-2|Y|} ~, ~~~~ 
Y \rightarrow \pm \infty ~.  
\ee
Since $\Phi (Y; t)$ becomes asymptotically a free field, i.e., 
$\partial_Y^2 \Phi \simeq 0$, we may select the free field configuration
\be
\Phi_0(Y;t) = -2Y + {\rm log}\left({4 \over \gamma} {\rm sinh}2\gamma (t_0 -t)
\right) 
\ee
and try to reconstruct the full solution of the 
continual Toda field equation for all $Y$ as a power series 
expansion of free fields based on the Lie algebra 
${\cal G}(d/dt; 1)$. We will actually 
verify that the algebraic expansion introduced in section 3.3 
coincides order by order with  
the Taylor series expansion of $\Phi(Y;t)$ in powers of   
${\rm exp}(-2Y)$. 

First, using the exact form of the solution, we expand 
${\rm exp} \Phi$ in the vicinity of 
$Y = +\infty$, 
\ba
e^{\Phi(Y; t)} \equiv & & {2 \over a(t) + b(t) {\rm cosh}2Y} = 
{4 \over b(t)} e^{-2Y} \left(1 - 2{a(t) \over b(t)} e^{-2Y} + 
\left(4{a^2(t) \over b^2(t)} -1\right) e^{-4Y} \right. \nonumber\\
& & - 
\left. 4{a(t) \over b(t)} \left(2{a^2(t) \over b^2(t)} - 1 \right) 
e^{-6Y} + {\cal O}(e^{-8Y}) \right) ~,  
\ea
which in turn leads to the power series expansion
\ba
\Phi(Y; t) \equiv & & {\rm log} {2 \over a(t) + b(t) {\rm cosh 2Y}} = 
\Phi_0(Y; t) - 2{a(t) \over b(t)} e^{-2Y} + \left(2{a^2(t) \over b^2(t)} -1 
\right) e^{-4Y} \nonumber\\
& &- {2a(t) \over 3b(t)} \left(4 {a^2(t) \over b^2(t)} - 3 \right) 
e^{-6Y} + {\cal O}(e^{-8Y}) ~,  
\ea
using the identity ${\rm log}(1+x) = x - x^2/2 
+ x^3/3 - x^4/4 + \cdots$. Similarly, we may also  
work with the Taylor series expansion around $Y = - \infty$.

On the other hand, the class of sausage metrics fits 
precisely into the special class of 
solutions of the continual Toda field equation for   
$\Phi_0(Y; t) = c Y + d(t)$ with $c= -2$. Thus, we may employ the expansion 
\eqn{toda10} in 
order to verify our assertion by matching terms in the two
series. Straightforward calculation shows in this case that 
\ba
\underline{{\cal O}(e^{-2Y})}: & & {1 \over 4} \partial_t e^{\Phi_0(Y;t)} = 
-2{\rm cosh} 2\gamma (t_0 -t) e^{-2Y} \equiv -2{a(t) \over b(t)} e^{-2Y} ~, 
\nonumber\\
\underline{{\cal O}(e^{-4Y})}: & & {1 \over 64} 
\partial_t \left(e^{\Phi_0(Y;t)} \partial_t e^{\Phi_0(Y; t)} \right) = 
{\rm cosh} 4\gamma (t_0 -t) e^{-4Y} \equiv \left(2{a^2(t) \over b^2(t)} -1 
\right) e^{-4Y} ~, 
\nonumber\\
\underline{{\cal O}(e^{-6Y})}: & & {1 \over 2,304 } \partial_t \left(  
3 e^{\Phi_0(Y;t)} \left(\partial_t e^{\Phi_0(Y;t)}\right)^2 
+ e^{2\Phi_0(Y;t)} \partial_t^2 e^{\Phi_0(Y; t)} 
\right) = 
\nonumber\\ 
& & -{2 \over 3}{\rm cosh} 2\gamma (t_0 -t) 
\left(4{\rm cosh}^2 2\gamma(t_0 -t) 
- 3 \right) e^{-6Y} \equiv \nonumber\\
& & -{2a(t) \over 3 b(t)} 
\left(4{a^2(t) \over b^2(t)} - 3  
\right) e^{-6Y} ~, 
\ea
and so on. Thus, the sausage metric yields a concrete example for testing  
the validity of the formal algebraic 
power series expansion of the continual Toda field configurations 
in term of free fields.        
It also helps to get a better grasp
of the interplay between the geometric deformations induced by the 
renormalization group flow of two-dimensional metrics and the algebraic 
structure that underlies their systematic description 
as continual Toda system for ${\cal G}(d/dt; 1)$.

\section{The cane and related models}
\setcounter{equation}{0}

We turn next to the geometric deformations of constant negative curvature 
metrics under the renormalization group flows. Contrary to the case of 
positive curvature metrics, which exhibit asymptotic freedom by becoming 
free field theories in the ultraviolet region, negative curvature metrics
are expected to flow towards free field theories in the infra-red region by 
appropriate changes of their target space geometry. We first review the 
classification of constant negative curvature metrics in terms of the 
monodromy of the free field configurations 
of Liouville theory, and then we investigate the 
geometric deformations induced by the renormalization group flow in each case
separately. In all cases we will only consider deformations that preserve the 
axial symmetry of the geometric configurations, thus simplifying the dynamics, 
as in the sausage model. We will use again two degrees of freedom $a(t)$ and 
$b(t)$ to parametrize the mini-superspace models 
in a consistent way and derive 
a system of first order differential equations for their 
evolution. It turns out that these geometric deformations also follow hyperbolic 
trajectories in the parameter space, in close analogy with the renormalization
group flows of the sausage model, but the physical regions 
differ from one model to another; likewise, the shapes of
the interpolating geometries are also different. Thus, we will be 
able to obtain some new, yet simple, solutions of the continual Toda field 
equation that describe the renormalization
group flow of negatively curved spaces in different patches, which look
like a cane in the infra-red limit and account for the name ``cane model". 

\subsection{Constant negative curvature metrics}

We consider two dimensional spaces with Euclidean signature
and metric 
\be
ds^2 = 2e^{\Phi(z_+, z_-)} dz_+ dz_- = 
{8 \over \Lambda^2} {\partial_+ F^+(z_+) 
\partial_- F^-(z_-) \over (1 - F^+(z_+) F^-(z_-))^2} dz_+ dz_- 
\ee
where $F^{\pm}(z_{\pm})$ are functions of the variable $z_+$ and its complex conjugate 
$z_-$. Using this parametrization of the metric 
in terms of the general solution
of the Liouville equation with coupling constant $\Lambda^2 /2$, 
the Ricci scalar curvature 
of the metric is equal to $-\Lambda^2$, i.e.,  
\be
R \equiv -2 e^{-\Phi} \partial_+ \partial_- \Phi = - \Lambda^2 ~,  
\ee
modulo delta-function singularities that depend on the choice of functions 
$F^{\pm}(z_{\pm})$.
As we will see shortly, there are three different type of geometries with constant 
negative curvature whose classification depends solely  
on the monodromy of the arbitrary functions $F^+(z_+)$ and $F^-(z_-)$; likewise,
there can be curvature singularities at $z_+ = 0 = z_-$, which also 
depend on the class 
of the corresponding solutions.

Let us parametrize the complex coordinates $z_{\pm}$ as follows,
\be
z_{\pm} = e^{Y \pm iX} ~; ~~~~ -\infty < Y < +\infty ~, ~~~~ 0 \leq X \leq 2\pi ~, 
\ee
which describe the standard map of an infinite long cylinder onto the plane.  
One end of the cylinder located at $Y= -\infty$ is mapped to the origin 
$z_+ = 0 =z_-$,
whereas the other end with $Y= +\infty$ is mapped to infinity in the  
$z$-plane; this map is also familiar 
from the radial quantization of two-dimensional field theories. Next, we consider
$X$-independent functions for the conformal factor $\Phi(Y)$, which can be 
either hyperbolic or parabolic or elliptic, depending on the behavior of 
$F^+(z_+)$ under  
$z_+ \rightarrow e^{2\pi i} z_+$ or, equivalently, under the winding 
$X \rightarrow X + 2\pi$. 
Hyperbolic solutions have monodromy of the 
form $F^+(z_+) \rightarrow e^{-2\pi \zeta} F^+(z_+)$ with $\zeta$ real, 
parabolic solutions have monodromy $F^+(z_+) \rightarrow F^+(z_+) - 2\pi$, 
and elliptic solutions have monodromy $F^+(z_+) \rightarrow e^{2\pi i \zeta} 
F^+(z_+)$ with $\zeta$ also real. 
Clearly, hyperbolic solutions can be viewed as arising from elliptic 
solutions by simple analytic continuation in $\zeta$, whereas parabolic solutions 
are obtained in the limit $\zeta \rightarrow 0$ from the other two classes. 
Also, since the solutions are independent from the sign of $\zeta$, we may
choose $\zeta$ to be positive definite without loss of generality. 

More explicitly, we have the following three different 
classes of $X$-independent solutions, 
up to dilation (see, for instance, \cite{nati}):

(i) \underline{\em Hyperbolic solutions}: Choosing $F^+(z_+) = {z_+}^{i\zeta}$ and 
$F^-(z_-) = {z_-}^{i\zeta}$, we have
\be
ds^2 = {2 \over \Lambda^2} {\zeta^2 \over z_+ z_- {\rm sin}^2 
\left({\zeta \over 2} {\rm log} z_+ z_- \right)} dz_+ dz_- = 
{2 \over \Lambda^2} {\zeta^2 \over {\rm sin}^2 \zeta Y} (dX^2 + dY^2) ~, 
\ee
with conformal factor that satisfies the equation
\be
\partial_+ \partial_- \Phi - {\Lambda^2 \over 2} e^{\Phi}  = 0 ~.
\ee
Clearly, this metric describes a space of constant negative curvature 
$-\Lambda^2$, which is regular everywhere 
without curvature singularities. The conformal factor is periodic in 
the variable $Y$ and, therefore, we may restrict the values of $Y$
in the fundamental domain $-\pi / \zeta \leq Y \leq 0$, rather than 
taking values on the entire line $-\infty < Y < +\infty$.  

Passing from the conformal variable 
$Y$ to the proper coordinate $\tilde{Y}$, which is defined here via 
$\tilde{Y} = -{\rm log}({\rm tan}(-\zeta Y/2))$, 
we may write the metric in the form
\be
ds^2 = {2 \over \Lambda^2} \left(d{\tilde{Y}}^2 + \zeta^2 {\rm cosh}^2 \tilde{Y} 
dX^2 \right) ,  
\ee
with $\tilde{Y}$ ranging from $-\infty$ to $+\infty$ as $Y$ ranges 
from $-\pi / \zeta$ to $0$. It describes 
a two-dimensional space with constant negative curvature, which  
corresponds to the so called plumbing fixture metric.
The $X$-independent hyperbolic solution is depicted in figure 
5 below.

\vspace{-15pt}
\begin{figure}[h] \centering
\epsfxsize=8cm
\epsfbox{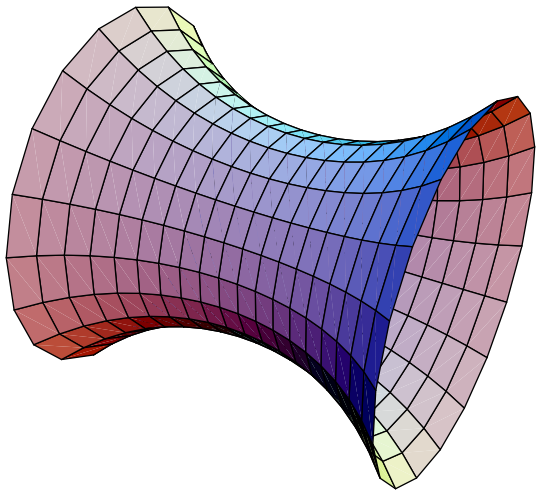}
\\[-25pt] \caption{An axially symmetric hyperbolic solution.} 
\end{figure}

(ii) \underline{\em Parabolic solutions}: Choosing $F^+(z_+) = i {\rm log}z_+$ and 
$F^-(z_-) = i/{\rm log}z_-$, we have
\be
ds^2 = {8 \over \Lambda^2} {1 \over z_+ z_- ({\rm log}z_+ z_-)^2} 
dz_+ dz_- = {2 \over \Lambda^2}{1 \over Y^2} (dX^2 + dY^2) ~,  
\ee
with conformal factor that satifies the equation
\be
\partial_+ \partial_- \Phi - {\Lambda^2 \over 2} e^{\Phi} + 
2\pi \delta^{(2)}(z) = 0 ~.  
\ee
This metric also describes a space of constant negative curvature $-\Lambda^2$, but 
in this case there is a curvature singularity at $z_+ = 0 =z_-$, 
which is half the curvature of the
sphere and corresponds to a puncture. In the vicinity of the singularity, 
however, the metric
does not become flat because of the logarithmic correction.  
We will also restrict $Y$ in the fundamental domain 
$-\infty < Y \leq 0$, rather than taking values on the entire line
$-\infty < Y < +\infty$.

As before, we may pass from the conformal  
variable $Y$ to the proper coordinate $\tilde{Y}$, which is now defined via 
$\tilde{Y} = -{\rm log} (-Y)$. 
Then, in terms of this variable the metric assumes the familiar form  
\be
ds^2 = {2 \over \Lambda^2} \left(d{\tilde{Y}}^2 + e^{2\tilde{Y}} dX^2 \right) ,   
\ee
with $\tilde{Y}$ ranging from $-\infty$ to $+\infty$ as $Y$ ranges from 
$-\infty$ to $0$.
The two-dimensional space exhibits rotational symmetry and can it be 
foliated by circles with variable size, 
starting from zero size at $\tilde{Y}=-\infty$
and growing exponentially to infinite size as $\tilde{Y} \rightarrow \infty$.
The $X$-independent parabolic solution is depicted in figure 
6 below. 

\vspace{-50pt}
\begin{figure}[h] \centering
\epsfxsize=8cm
\epsfbox{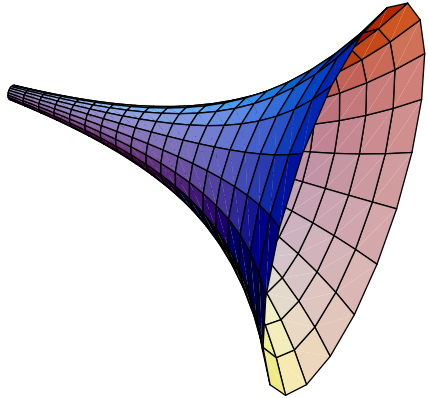}
\\[-25pt] \caption{An axially symmetric parabolic solution.}
\end{figure}

(iii) \underline{\em Elliptic solutions}: Choosing $F^+(z_+) = z_+^{\zeta}$ and 
$F^-(z_-) = z_-^{\zeta}$, with $0 \leq \zeta \leq 1$, we have the metric 
\be
ds^2 = {8 \over \Lambda^2}{\zeta^2 \over (z_+ z_-)^{1-\zeta} 
(1 - (z_+ z_-)^{\zeta})^2} dz_+ dz_- = {2 \over \Lambda^2} {\zeta^2 \over
{\rm sinh}^2 \zeta Y} (dX^2 + dY^2) ~,   
\ee
with conformal factor that satisfies the equation
\be
\partial_+ \partial_- \Phi - {\Lambda^2 \over 2} e^{\Phi} + 
2\pi (1- \zeta) \delta^{(2)}(z) = 0 ~.  
\ee
As such, it describes a space of constant negative curvature $-\Lambda^2$, but
there is also a curvature singularity at $z_+ = 0 =z_-$, where the metric becomes 
flat with a conical tip and deficit angle 
$2\pi(1 - \zeta)$. We may also restrict the range of $Y$ in the 
fundamental domain $-\infty < Y \leq 0$, as before, rather than taking values on the 
entire real line.  

It is convenient to pass from the conformal variable 
$Y$ to the proper coordinate $\tilde{Y}$ via $\tilde{Y} = 
-{\rm log}({\rm tanh} (-\zeta Y/2))$, in terms of which
the metric assumes the form
\be
ds^2 = {2 \over \Lambda^2} \left(d{\tilde{Y}}^2 + 
\zeta^2 {\rm sinh}^2 \tilde{Y}
dX^2 \right) ,   
\ee
with $\tilde{Y}$ ranging from $0$ to $+\infty$ as $Y$ ranges from 
$-\infty$ to $0$. 
The two-dimensional space exhibits rotational symmetry and it can be 
foliated by circles with variable size, starting from zero size at 
$\tilde{Y}=0$
and expanding exponentially as $\tilde{Y} \rightarrow +\infty$. 
The $X$-independent elliptic solution is depicted in figure 7 below.  

\vspace{-50pt}
\begin{figure}[h] \centering
\epsfxsize=8cm
\epsfbox{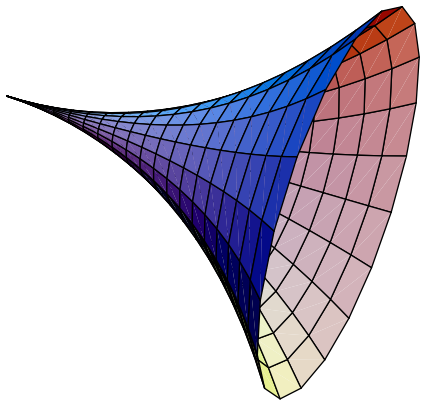}
\\[-25pt] \caption{An axially symmetric elliptic solution.}
\end{figure}

Concluding our general discussion of the constant negative curvature metrics in two
dimensions, we note that they come in three different classes depending on the
monodromy of the functions $F^{\pm}(z_{\pm})$. In fact, this is a simple example 
of a more general fact that metrics of constant negative curvature on a given
manifold ${\cal M}$ are all locally, but not globally equivalent. Then, it is
expected on general grounds that perturbative renormalization cannot reliably
distinguish among such metrics and non-perturbative effects enter significantly
into their renormalization \cite{fried}. 
As a result, the geometric deformations will 
look different in different local patches. In two dimensions, in particular, 
the infra-red limit of hyperbolic and parabolic metrics looks like a
cane with variable length and width, as we will see next, but the elliptic
metric exhibits a very different behavior within the given set of ansatz.    
The presence of a conical singularity in the latter case holds the key for this 
differentiation in the perturbative regime. 
    
\subsection{Deformation of hyperbolic solutions}

We first examine the geometric deformations of the hyperbolic solution 
induced by the renormalization group flows. We choose to work in the 
conformal frame with sigma model action 
\be
S_{\rm t} = {1 \over 2} 
\int e^{\Phi(Y; t)} \left((\partial_{\mu} X)^2 + (\partial_{\mu} 
Y)^2 \right) d^2 w ~. 
\ee
Here, the conformal factor $\Phi(Y;t)$ is given by the 
one-parameter family of functions
\be
e^{\Phi(Y;t)} = {2\zeta^2 \over a(t) + b(t) {\rm cos}(2 \zeta Y)} ~, 
\ee
which preserve the axial symmetry of the configuration for all $t$, and 
$X$ is the angular coordinate ranging between 0 and $2\pi$. As for the
variable $Y$, it takes values in the fundamental region 
$-\pi / \zeta \leq Y \leq 0$, as before, but it can also be extended
to the whole real axis by plumbing together the individual 
fundamental pieces. 
For $a(t) = 1 = -b(t)$ we have the usual hyperbolic solution 
with constant negative curvature $- \Lambda^2$, which is 
normalized to $-1$, 
\be
e^{\Phi(Y; t)} = {\zeta^2 \over {\rm sin}^2 (\zeta Y)} ~, 
\ee
whereas for $b(t)=0$ we have the geometry
of a flat cylinder with radius depending on the size of the 
parameter $a$, which has
to be positive. 

The metric is positive definite for all values of $t$ provided that 
$a(t) \geq -b(t) \geq 0$. This determines the physical wedge in the 
parameter space $(a(t), b(t))$ of the model, whose boundary 
line $a = -b$ consists by all 
constant negative curvature metrics with $\Lambda^2 = a$. 
Using this ansatz, we find that the 
renormalization group flows yield the following system of first 
order differential equations,
\be
a^{\prime}(t) = - 2b^2(t) ~, ~~~~~ b^{\prime}(t) = 
-2a(t)b(t) ~,  
\ee
where $a^2(t) - b^2(t) = \gamma^2 \geq 0$ is conserved on the 
physical trajectories.
This system is identical to the sausage model, 
provided that we flip the sign $t \rightarrow -t$. 
The sign difference
is also expected on general grounds, since negative 
curvature spaces flow to the infra-red, $t \rightarrow +\infty$, 
whereas positive curvature spaces start flowing from the ultra-violet
region, $t \rightarrow -\infty$. Thus, choosing 
$\gamma$ to be any real constant $0 \leq \gamma < +\infty$, we 
find the solution 
\be
a(t) = \gamma {\rm coth} 2\gamma (t-t_0) ~, ~~~~ 
b(t) = -{\gamma \over {\rm sinh} 2\gamma (t-t_0)} ~,  
\ee
which also takes into account the appropriate restriction to the physical
wedge of the parameter space\footnote{If we were deforming the 
hyperbolic metric with ${\rm exp}\Phi = \zeta^2 / {\rm cos}^2(\zeta Y)$,
which is obtained from above by a simple shift $Y \rightarrow 
Y + \pi / (2\zeta)$, the sign of $b(t)$ would be opposite, thus restricting 
the physical wedge in the same domain as in the sausage model. In this case,
the renormalization group equations, which are inert to the transformation
$b(t) \rightarrow -b(t)$, lead to the same evolution as in the 
sausage model, but with reverse direction in the flow.}.  

The trajectories are hyperbola, as in the 
sausage model, but they are now filling a different region in the  
parameter space $(a, b)$.
All trajectories start from the point $(+\infty , -\infty)$ 
at some arbitrary initial
time $t_0$, where they exhibit the universal 
behavior  
\be
a(t) \simeq -b(t) \simeq {1 \over 2(t-t_0)} ~, ~~~~
e^{\Phi(Y; t)} \simeq 2(t-t_0) {\zeta^2 \over {\rm sin}^2(\zeta Y)} ~, 
\ee
which is independent of $\gamma$ for $t \rightarrow t_0^+$. 
Thus, the geometry starts from  
constant negative curvature with $\Lambda^2 = 1/2(t-t_0) 
\rightarrow \infty$, but there are also higher order curvature terms
which can become important in this region and invalidate the 
lowest order approximation to the renormalization group
equations. 
Following the renormalization group equation for $t>t_0$, 
we find that the trajectories flow 
towards the fixed points
\be
a = \gamma~, ~~~~~ b=0 
\ee
as $t \rightarrow +\infty$, depending on $\gamma$. 
They are schematically depicted in figure 8 below. 

\vspace{20pt}
\begin{figure}[h] \centering
\epsfxsize=10cm \epsfysize=8cm \epsfbox{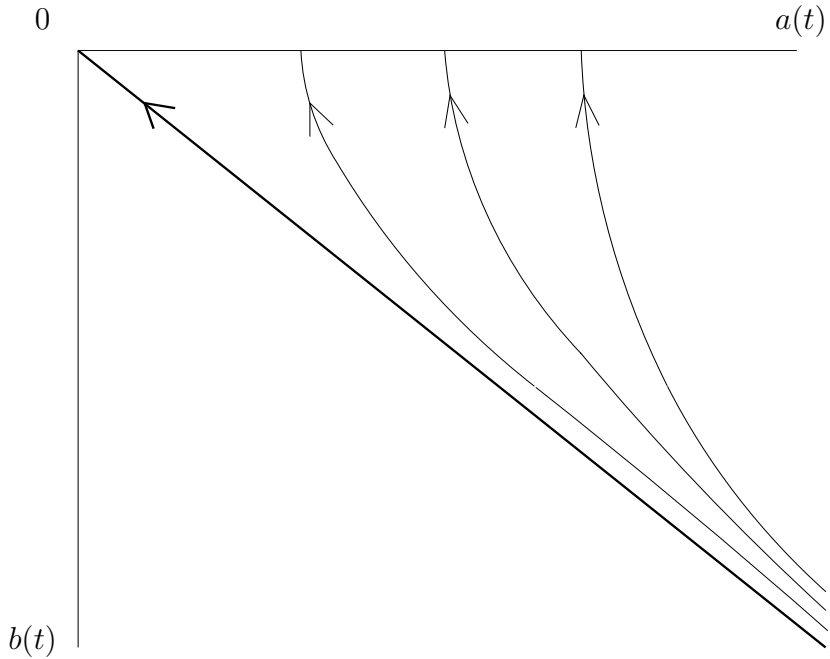}
\put(-20,235){$a(t)$} \put(-310,0){$b(t)$} 
\put(-300,235){$0$} 
\\[20pt]
\caption{Renormalization group trajectories of the hyperbolic model.}
\end{figure}

The limiting  
points $(\gamma, 0)$ on the semi-infinite axis $b=0$  
provide the set of all {\em infra-red points} of
the model. 
In the infra-red limit the geometry is described by a 
flat cylinder of radius $\zeta / \sqrt{\gamma}$, 
\be
ds^2 = {\zeta^2 \over \gamma}(dX^2 + dY^2) ~, 
\ee
which seemingly  
corresponds to the conformal field theory $RU(1)_{\gamma / \zeta^2}$, in  
close analogy with the ultra-violet fixed points of the sausage model. 
Actually, since the fundamental domain of $Y$ is taken between $- \pi / \zeta$
and $0$, the infra-red conformal field theory is only a segment of the 
$RU(1)_{\gamma / \zeta^2}$ model. It corresponds to a cylinder with finite length 
$\pi / \sqrt{\gamma}$ and radius $\zeta / \sqrt{\gamma}$, which 
describes the {\em orbifold conformal field theory} $(S^1 / Z_2) \times S^1$,
since any finite interval can be viewed as an orbifold $S^1 / Z_2$ with 
the two edges being its fixed points. Then, in this context,
the parameters $\zeta$ and $\gamma$ determine the size and the shape of the
limiting infra-red geometry.    

For $\gamma =0$, the radius as well as the length of the cylinder become 
infinite and the infra-red geometry is a two-dimensional plane,   
provided that $\zeta \neq 0$.  
For $\zeta = 0$, the infra-red fixed points  
become identical and they correspond to degenerate cylinders 
with vanishing radius, provided that $\gamma 
\neq 0$. This behavior can also be seen in the infra-red limit of the
parabolic model, since the hyperbolic conformal   
factor $\zeta^2 / {\rm sin}^2(\zeta Y)$ tends to the parabolic 
limit $1/Y^2$ when 
$\zeta \rightarrow 0$; in that case, however, the length of the 
degenerate cylinder is  
infinite, as the fundamental region of $Y$ also becomes  
infinite. We will say more about the deeper relation that exists between 
the geometric deformations of the hyperbolic and parabolic models in the 
next subsection. 

The boundary trajectory, $a(t) = -b(t)$, 
which corresponds to the value 
$\gamma = 0$, as in the sausage model,
describes the special flow
\be
e^{\Phi(Y; t)} = 2(t-t_0) {\zeta^2 \over {\rm sin}^2 (\zeta Y)}  
\label{toda93}
\ee
for all values of $t$. It provides the boundary Liouville line
of the hyperbolic model associated to the standard embedding of 
Liouville theory into the renormalization group equation. 
The geometry of the deformation is very easy to visualize in this case,
as the space remains hyperbolic for all $t_0 \leq t < +\infty$ 
with constant negative curvature
$-\Lambda^2 = 1/2(t_0 -t)$. It starts from 
infinite curvature at $t=t_0$ and flows towards zero curvature 
in the infra-red limit, where it becomes flat cylinder
stretched to a plane.    
This special flow can be intuitively understood  
using the membrane paradigm, 
where the stretching of an infinite curved hyperboloid  
is achieved by slowly broadening its
throat to infinite size, while reducing  
$\Lambda^2$ to zero at the same time. Clearly, the final 
configuration is independent of $\zeta$, which is the parameter that
determines the initial size of the throat.  
All other trajectories with $\gamma \neq 0$ describe 
geometric deformations of the initial hyperboloid 
to flat cylinders  
with radius $\zeta / \sqrt{\gamma}$.
   
We note at this point that in the vicinity of the infra-red limit 
the conformal factor behaves as
\be
e^{\Phi(Y; t)} \simeq {2\zeta^2 \over \gamma} \left(1 + 
2{\rm cos}(2\zeta Y) 
e^{-2\gamma (t-t_0)} + {\cal O}(e^{-4\gamma (t-t_0)}) 
\right) ~~~~~ {\rm for} ~~
t \rightarrow +\infty 
\ee
and describes small deviations from flatness. In this case, 
the metric perturbation is proportional to   
\be
\Theta(Y;t)= {\rm cos}(2\zeta Y) {\rm exp}\left(-2\gamma (t-t_0) \right)
\label{specia}
\ee
and it remains small for all 
$Y$. This perturbation should be compared with the 
analogous term ${\rm cosh}(2Y) 
{\rm exp}(2\gamma (t-t_0))$ that describes the deviation of the sausage metric 
away from flatness in the vicinity of the ultra-violet points 
$t \rightarrow -\infty$, which, however, can become arbitrary large at the
two ends of the sausage, $Y \rightarrow \pm \infty$.   
Then, the continual Toda equation linearizes in the infra-red region 
of the deformed hyperbolic model and  
reduces to the heat equation for $\Theta(Y;t)$,
\be
{2\zeta^2 \over \gamma} {\partial \over \partial t} \Theta(Y;t) = 
{\partial \over \partial Y^2} \Theta(Y;t) ~, 
\label{heaty}
\ee
which is defined over a periodic spatial variable $Y$. 
The factor $2\zeta^2 / \gamma$ that appears on the 
left-hand side of the linearized 
heat equation, arises from the normalization of 
the metric at the infra-red 
points ${\rm exp}\Phi |_{\rm IR} = 2 \zeta^2 / \gamma$.   
This periodic solution is a basic constituent of the Jacobi theta 
function that describes more general periodic solutions of the 
heat equation and it is smooth at $t=0$. 

We are now in position to describe the geometry of the deformed 
hyperbolic model by changing variables  
from $Y$ to the proper coordinate  
$\tilde{Y}$,
\be
\tilde{Y} = \sqrt{2} {\rm cosh} \gamma (t-t_0) \int 
{dY \over \sqrt{{\rm cosh}2\gamma (t-t_0) - 
{\rm cos}2\zeta Y}}  = - {1 \over \zeta} 
F(\psi, k) ~.    
\ee
Here, $F(\psi, k)$ is the incomplete elliptic integral of
the first kind with parameter 
\be
{\rm sin} \psi = \left\{ \begin{array}{ll} 
{\rm cosh}\gamma (t-t_0) 
\sqrt{{1 - {\rm cos}2\zeta Y
\over {\rm cosh}2\gamma (t-t_0) - 
{\rm cos}2\zeta Y}}  & ~~ {\rm for} ~~ -{\pi \over 2\zeta} 
\leq Y \leq 0 \\
 & \\
- {\rm cos}(\zeta Y)  & ~~ {\rm for} ~~ -{\pi \over \zeta} 
\leq Y \leq -{\pi \over 2\zeta} 
\end{array}  \right.    
\ee
and modulus $0 \leq k \leq 1$, which is equal to  
\be
k = {1 \over {\rm cosh}\gamma(t-t_0)} ~.  
\ee
Then, after some calculation, the metric assumes the form
\be
ds_{\rm t}^2 = \left\{ \begin{array}{l} 
{\zeta^2 \over \gamma} {k^{\prime}}^2  
\left(d{\tilde{Y}}^2 + {1 \over {k^{\prime}}^2}  
{\rm dn}^2(\zeta \tilde{Y} ; k) dX^2 \right) \\
 \\
{\zeta^2 \over \gamma} {k^{\prime}}^2  
\left(d{\tilde{Y}}^2 + {1 \over   
{\rm dn}^2(\zeta \tilde{Y} ; k)} dX^2 \right) 
\end{array} \right.  
\ee
respectively, 
where $\tilde{Y}$ ranges from $-K(k)/\zeta$ to $0$ in both cases 
and $k^{\prime} = \sqrt{1-k^2}$ denotes the complementary modulus.  
Thanks to the identity 
${\rm dn}(u - K(k); k) = k^{\prime} / {\rm dn}(u; k)$, which is 
obeyed by the delta-amplitude Jacobi elliptic function, we arrive
at the final form of the deformed metric
\be
ds_{\rm t}^2 =  
{\zeta^2 \over \gamma} {k^{\prime}}^2  
\left(d{\tilde{Y}}^2 + {1 \over {k^{\prime}}^2}  
{\rm dn}^2(\zeta \tilde{Y} ; k) dX^2 \right) \label{dnf} 
\ee
with $-2K(k)/\zeta \leq \tilde{Y} \leq 0$ as $Y$ ranges all over the 
fundamental domain $-\pi /\zeta \leq Y \leq 0$.  

The change of coordinates above is only valid when  
${\rm cosh}2\gamma (t-t_0)$ is strictly bigger than $1$, which implies
that $t>t_0$ or else $k<1$. This can also be seen 
from the resulting metric, which becomes ill-defined for 
$k^{\prime} =0$.  
The value $k=1$ corresponds to the initial
time $t=t_0$, where the change of variables 
is simply given by $d\tilde{Y} = - \zeta dY/{\rm sin} (\zeta Y)$ and 
yields the standard form of the hyperbolic metric in proper coordinates.    
The metric behaves well for all values $k<1$ 
and reaches the metric of a flat cylinder at $k=0$. The value  
$k=0$ corresponds to the infra-red limit of 
the deformations, where ${\rm dn}(u; k=0) 
= 1$ for all $u$.  
Also, since ${\rm dn}(0; k)= 1$ for all $k$, we see immediately
that the deformed geometry reduces to a very  
thin cylinder in the infra-red limit, when $\zeta \rightarrow 0$.  
We will say more about it later by studying separately the
deformations of the parabolic model.  

The general form of the interpolating geometry in the hyperbolic model 
is represented schematically in the figure 9 below, which is plotted
for $k$ close to $0$ and looks like a cane.

\begin{figure}[h] \centering
\epsfxsize=8cm
\epsfbox{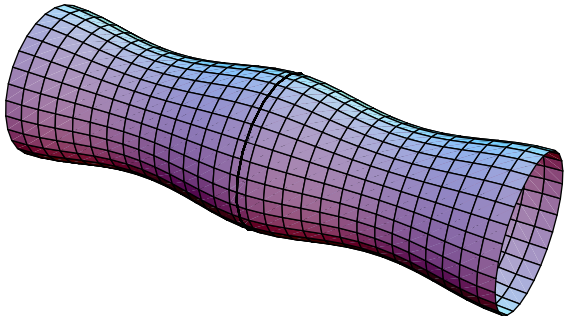} \caption{An axially symmetric deformed hyperbolic 
configuration.} 
\end{figure}

\noindent
The distance between the joins of the cane is equal to $2K(k)/\zeta$, 
which tends to infinity when $k \rightarrow 1$ and becomes 
$\pi / \zeta$ when $k=0$. The mid-point in the 
fundamental region $-\pi/ \zeta \leq Y \leq 0$, i.e., $Y=-\pi / (2\zeta)$, 
corresponds to the mid-point $\tilde{Y} = -K(k) / \zeta$ in the 
fundamental region of the proper variable $\tilde{Y}$, where the 
cane makes a bump. Then, in the infra-red limit the bump disappears 
and we arrive at the segment of a cylinder, $(S^1/Z_2) \times S^1$.
When $Y$ is taken on the whole real line, rather than being restricted 
in the fundamental domain, the deformed geometry looks like an 
infinite long cane, which is obtained by gluing together the 
individual pieces. Further details on the resulting periodic 
structure will also be given later in subsection 5.3.       
    
The metric of the cane model can be expressed 
in terms of Jacobi theta functions, as in the sausage model, 
but we should also identify the ``would be"
dilaton field associated with the change of variables to the
proper coordinates
$(X, \tilde{Y})$. Using the more general system of the renormalization 
group flows with a vector field $\xi$ that generates the relevant 
reparametrizations, we find in 
this case that the solution is given by  
\ba
ds_{\rm t}^2 & = & 
{\zeta^2 \over \gamma} k^{\prime} \left( k^{\prime}  
d{\tilde{Y}}^2 + \left({\Theta_1 (\zeta \tilde{Y}) \over 
\Theta (\zeta \tilde{Y})} \right)^2 dX^2 \right) , \nonumber\\ 
\tilde{\Phi}(\tilde{Y}) & = & {\rm log} \Theta (\zeta \tilde{Y}) + 
{1 \over 2} \left({E(k) \over K(k)} - 1 + {1 \over 2} k^2 
\left(1+{2 \over \gamma} k^{\prime}\right) \right) {\tilde{Y}}^2  ~,  
\ea
with $\xi_{X} = 0$ and $\xi_{\tilde{Y}} = 
\partial_{\tilde{Y}} \tilde{\Phi}$.    
One may also verify independently that this provides a solution of  
the renormalization group equations using the heat equation 
for the Jacobi theta functions for all times.

\underline{\em Free field realization}

We complete the discussion of the hyperbolic model with the free field
realization of the renormalization group trajectories. In analogy with
the sausage model, we select 
\be
\Phi_0 (Y; t) = 2i \zeta Y + {\rm log} \left(-{4 \zeta^2 \over \gamma} 
{\rm sinh} 2 \gamma (t-t_0) \right) 
\ee
as the corresponding free field configuration. It fits in the special 
class of free fields \eqn{toda56} with $c=2i \zeta$. 
Then, we may verify that the 
algebraic power series expansion \eqn{toda10} sums up to the conformal
factor $\Phi (Y; t)$ of the deformed hyperbolic solution.   

Here, we will demonstrate this for the special Liouville 
trajectory \eqn{toda93}, which corresponds to $\gamma = 0$ and 
has $\Phi_0 (Y; t) = 2i\zeta Y + {\rm log} (-8\zeta^2 (t-t_0))$.  
We have, in particular, the following expansion 
\ba
\Phi & = & \Phi_0 + {1 \over c^2}  \partial_t e^{\Phi_0} +  
{1 \over 4c^4} \partial_t 
\left(e^{\Phi_0} \partial_t e^{\Phi_0} \right) +  
{1 \over 36 c^6} \partial_t \left( 
3 e^{\Phi_0} \left(\partial_t e^{\Phi_0}\right)^2 
+ e^{2\Phi_0} \partial_t^2 e^{\Phi_0} \right) 
+ \cdots \nonumber\\
& = & {\rm log} \left(-8 \zeta^2 (t-t_0) e^{2i \zeta Y} \right) 
+ 2 e^{2i\zeta Y} + e^{4i\zeta Y} + {2 \over 3} e^{6i\zeta Y} 
+ \cdots \nonumber\\
& = & {\rm log} \left(-8 \zeta^2 (t-t_0) e^{2i \zeta Y} \right) 
-2 {\rm log} \left(1 - e^{2i \zeta Y} \right) = {\rm log} 
\left(2(t-t_0) {\zeta^2 \over {\rm sin}^2 \zeta Y} \right) 
\ea
that proves our assertion. The verification of the result for 
$\gamma \neq 0$ is left as exercise to the interested reader. 
 
\subsection{Deformation of parabolic solutions}

We consider the ansatz for the one-parameter family of two-dimensional metrics 
with
\be
e^{\Phi(Y; t)} = {1 \over a(t) + b(t) Y^2} ~, 
\ee
which describe axially symmetric deformations of the parabolic metric. 
For $a(t) = 0$ and $b(t) = 1$ the metric reduces to the standard Poincare metric 
on the upper (respectively lower) half-plane, which is defined for all 
$Y \geq 0$ (respectively $Y \leq 0$). 
For generic values of the parameters $a(t)$ and $b(t)$ the deformed
geometry remains conformally flat and obeys the continual Toda field equation, 
provided that the following system of 
first order differential equations is satisfied, 
\be
a^{\prime}(t) = 2 a(t)b(t) ~, ~~~~~ b^{\prime}(t) = -2b^2(t) ~. 
\ee
These equations can be easily integrated by noting the existence of the conserved
quantity
\be
a(t)b(t) = \gamma^2 ~, 
\ee
where $\gamma^2$ is taken to be a positive constant so that the metric is  
positive definite for all $Y$. Consequently, the general solution is given by
\be
a(t) = 2 \gamma^2 (t-t_0) ~, ~~~~~ b(t) = {1 \over 2(t-t_0)} ~, 
\ee
where $t_0$ is another integration constant, which sets the lower bound in the 
allowed range of t, $t_0 \leq t < +\infty$, so that $a(t)$ and $b(t)$ are 
both positive. The final result reads as follows, 
\be
e^{\Phi(Y; t)} = {2(t-t_0) \over 4\gamma^2 (t-t_0)^2 + Y^2}   
\ee
and provides a simple algebraic solution of the
continual Toda field equation. This solution was noted before 
\cite{misha3}, but without 
making reference to geometry as we do.  

The renormalization group trajectories are 
hyperbola in the $(a, b)$ parameter space 
using the integral of the flow,  
$a(t)b(t) = \gamma^2 \geq 0$, and they fill the entire physical region 
with $a(t) \geq 0$ and $b(t) \geq 0$.
The orbits start from the upper point $(0, \infty)$ at $t=t_0$ and they
flow towards the lower point $(\infty, 0)$, which is a universal 
{\em infra-red fixed point} for all $\gamma \neq 0$. For $\gamma =0$, however,
the trajectory is different and the model flows to the infra-red point
$(0, 0)$ located at the origin of parameter space. The physical trajectories are  
depicted in figure 10  below.

\vspace{15pt}
\begin{figure}[h] \centering
\epsfxsize=9cm
\epsfysize=8cm
\epsfbox
{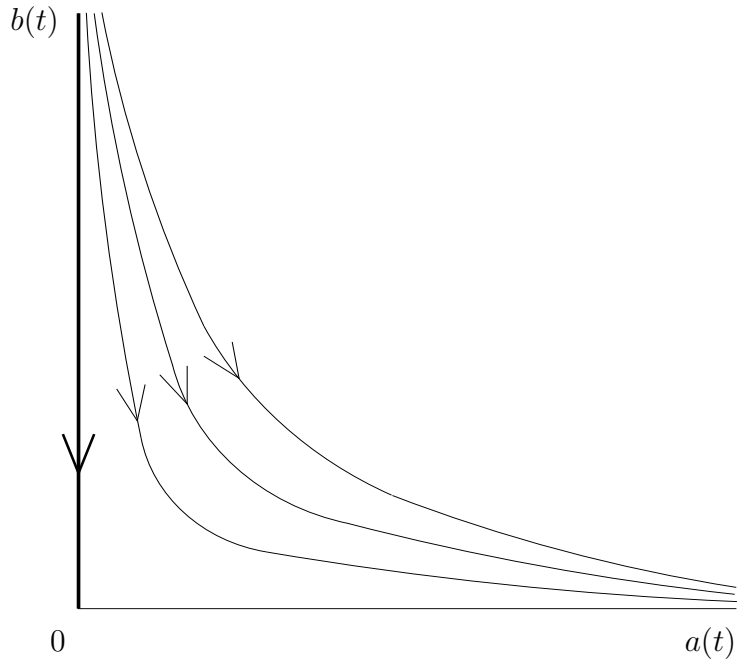} \put(-20,-15){$a(t)$} \put(-275,220){$b(t)$} 
\put(-260,-15){0}
\\[10pt]
\caption{Renormalization group trajectories of the parabolic 
model.} 
\end{figure}

The special trajectory with $\gamma = 0$ is a Liouville line, in which  
case the solution reads 
\be
e^{\Phi(Y;t)} = {2(t-t_0) \over Y^2}  
\ee
for all $t \geq t_0$. It corresponds to the standard embedding of Liouville 
theory in the continual Toda  
system and it is depicted by the dark line.  
It describes constant negative curvature spaces with 
$\Lambda^2 = 1/2(t-t_0)$ at all times, which start from infinity and flow
to zero at the infra-red point. 
All other trajectories with $\gamma \neq 0$ flow towards  
another infra-red limit,  
where the geometry looks like 
\be
ds_{\rm t}^2 \simeq {1 \over 4\gamma^2 (t-t_0)} (dX^2 + dY^2) ~, 
\ee
as $t \rightarrow +\infty$. The limiting metric describes a 
degenerate cylinder with vanishing radius for all $\gamma \neq 0$, as it can be 
easily seen by rescaling the coordinates.
Of course, this universal behavior is also expected   
following the general remarks on the infra-red limit of hyperbolic 
models for $\zeta \rightarrow 0$. 

The shape of the interpolating geometries for the 
parabolic model is best described by changing  
variables to the proper coordinate $\tilde{Y}$, 
\be
Y = 2\gamma (t-t_0) {\rm sinh}\left({\tilde{Y} \over \sqrt{t-t_0}}\right) ,   
\ee
which is appropriate for $t>t_0$ when $\gamma \neq 0$. 
Then, it is straightforward to cast the deformed metric in the form
\be
ds_{\rm t}^2 = d{\tilde{Y}}^2 + {dX^2 \over 4\gamma^2 (t-t_0) {\rm cosh}^2 
(\tilde{Y} / \sqrt{t-t_0})} ~, \label{coshf} 
\ee
and the geometry looks like a blob with two spikes that extend to
infinity, as depicted in 
figure 11 below.  

\vspace{-30pt}
\begin{figure}[h] \centering
\epsfxsize=8cm \epsfbox{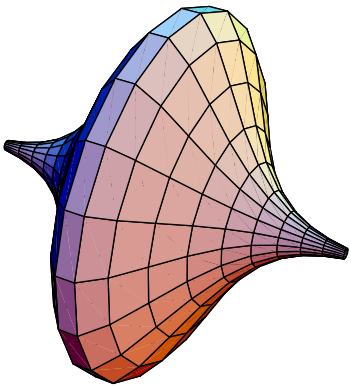} 
\\[-50pt] \caption{An axially symmetric deformed parabolic 
configuration.} 
\end{figure}

We actually drew the picture by letting 
$\tilde{Y}$ range from $-\infty$ to 
$+\infty$ as $Y$ ranges from $-\infty$ to $+\infty$. The restriction
to the fundamental domain $-\infty < Y \leq 0$ cuts the figure in
half, since it also restricts the range of $\tilde{Y}$ to 
$-\infty < \tilde{Y} \leq 0$.    
It is also interesting to note that the size of the blob diminishes 
as $t \rightarrow +\infty$, in which case the two-dimensional space 
degenerates to a 
line parametrized by $\tilde{Y}$; it is nothing else but a 
cylinder with zero radius, which describes the infra-red limit
of the model. 

The deformations of the parabolic model can be fully investigated   
in proper coordinates $(X, \tilde{Y})$ using the generalized 
system of renormalization group equations that incorporate the change of 
coordinates along the flow. In this
case, the solution is completely described by
introducing the ``would be" dilaton field
\be
\tilde{\Phi}(\tilde{Y}) = {\rm log} \left({\rm cosh} 
{\tilde{Y} \over \sqrt{t-t_0}}\right)  
-{{\tilde{Y}}^2 \over 4(t-t_0)}  
\ee
that serves to define   
the generating vector field of the reparametrizations, as usual,   
with $\xi_{X} = 0$ and $\xi_{\tilde{Y}} = 
\partial_{\tilde{Y}} \tilde{\Phi}$.    
The verification is very simple and boils to the 
equation $R_{\tilde{Y} \tilde{Y}} = 
2 \partial_{\tilde{Y}}^2 \tilde{\Phi}$.

Next, we append some comments regarding the interpretation of the hyperbolic solution
as an infinite long train of parabolic solutions, which 
are equally spaced on the axis of symmetry 
$\tilde{Y}$. This interpretation is useful for understanding the link between the
two models and puts on firm basis the
sporadic comments we made before for the behavior of the deformed
geometries as $\zeta \rightarrow 0$.   
We will present an appropriate superposition 
principle that is reminiscent of free field theories, and which is 
also applicable to the Toda field equations by their    
integrability. Such 
superposition rules are widely used in 
many other integrable field theories in two dimensions; for example, 
static periodic solutions of the sine-Gordon model can be described as 
an infinite sum of equally spaced kinks and/or anti-kinks on the line (see, 
for instance \cite{sourdis}, and references therein).  

We already know that the conformal factor of the constant curvature hyperbolic
space can be written as linear superposition of parabolic conformal factors, 
\be
{\zeta^2 \over {\rm sin}^2 (\zeta Y)} = \sum_{n \in Z} {1 \over 
(Y - \pi n/\zeta)^2} ~, 
\ee
which are equally spaced in $Y$. This is made possible by the integrability 
properties of the Liouville
equation and accounts for the period of the hyperbolic solution,
which is $\pi / \zeta$. 
Likewise, it is possible to decompose the metric \eqn{dnf} of the deformed
hyperbolic model into an infinite sum of equally spaced deformed parabolic
blocks, whose metric is given by equation \eqn{coshf} in proper coordinates. 
It is achieved by employing the remarkable identity \cite{watson}  
\be
{\rm dn}^2(u; k) = \left({\pi \over 2K(k^{\prime})}\right)^2 \sum_{n \in Z} 
{1 \over {\rm cosh}^2 \left({\pi \over 2K(k^{\prime})} (u+2nK(k)) \right)} 
+ 1 - {E(k^{\prime}) \over K(k^{\prime})} \label{swsix}  
\ee
that relates the two characteristic profiles for all $t>t_0$.  
This superposition is also consistent with the
fact that ${\rm dn}(u; k) = 1/{\rm cosh}u$ for $k=1$, in which case 
$K(k^{\prime}) = \pi / 2 = E(k^{\prime})$ and the spacing of
the lattice, $2K(k)$, tends to infinity. 

Thus, it becomes clear that as we flow towards the infra-red limit, 
$k \rightarrow 0$, the blobs come closer   
without touching each other. At the same time, the  
inter-connecting tubes are becoming thicker by superposition of the
narrow spikes that join two neighboring blobs. The resulting picture
looks like a cane, as shown in figure 9, and one arrives at an 
alternative (though more fundamental)  
description of the cane model in terms of its constituents. 
When $\zeta \rightarrow 0$ 
the blobs diminish in size and their separation 
becomes large at the same time, thus forming a very long but thin 
cane in the infra-red region.
Thus, we obtain an alternative, yet intuitive, derivation 
of the infra-red limit of the 
deformed parabolic model, which is described by a degenerate cylinder 
for all $\gamma \neq 0$.  
The geometry of the general superposition \eqn{swsix}  
is depicted schematically in figure 12 below.  

\begin{figure}[h] \centering
\epsfxsize=8cm \epsfbox{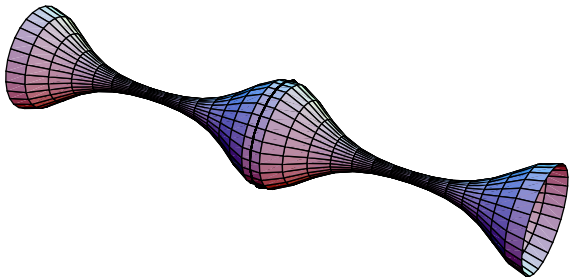} \caption{Superposition of deformed 
parabolic solutions.} 
\end{figure}

\subsection{Deformation of elliptic solutions} 

Let us assume that 
the conformal factor of the axially symmetric deformed elliptic 
model can be written in the form  
\be
e^{\Phi(Y; t)} = {2\zeta^2 \over a(t) + b(t) {\rm cosh}
(2\zeta Y)} ~, \label{ellipa}  
\ee
which looks like the metric of the sausage model, but it has a different
domain of validity. 
For $a(t) = -1$ and $b(t) = 1$, we 
obtain the standard elliptic metric with constant negative curvature and 
$\Lambda^2$ normalized to $1$. 
Then, we impose the restriction  
$b(t) \geq -a(t) \geq 0$ in order to ensure that we are 
describing deformations of the elliptic model; this determines
the physical wedge in the parameter
space $(a(t), b(t))$, which is different from the sausage model. 

The first order differential equations for the evolution of  
the coefficients $a(t)$ and $b(t)$ are
identical to the sausage model and they read as follows,  
\be
a^{\prime} (t) = 2 b^2(t) ~, ~~~~~ b^{\prime}(t) = 2a(t) b(t) ~, 
\ee
for all values of $\zeta$. The trajectories are again hyperbola in 
the parameter space, but in this case the conserved quantity is given by  
\be
b^2(t) - a^2(t) = \gamma^2 
\ee
and the general solution is obtained by analytic continuation in time,  
\be
a(t) = - \gamma {\rm cot} 2\gamma (t-t_0) ~, ~~~~~ b(t) = {\gamma \over 
{\rm sin} 2\gamma (t-t_0)} ~.  
\ee

The orbits of the elliptic model with $\gamma \neq 0$ are   
periodic in time with period equal to $\pi / \gamma$. They all  
start from the same point $(-\infty , +\infty )$ in the parameter
space $(a, b)$ at some initial time $t_0$, but after some time  
$t= t_0 + \pi / 4\gamma$ the coefficient $a(t)$ becomes zero.
For later times it  
turns positive by approaching the Liouville line of the sausage model, 
as $t$ goes to $t_0 + \pi / 2\gamma$. This is
already an indication that the model behaves badly by exiting the
domain of negative curvature and entering into another domain. 
Continuing the validity of our solution beyond it, we find that  
both metric coefficients $a(t)$ and $b(t)$ turn
negative by following trajectories in the lower mirror image of the upper
wedge, which is clearly not physical. 
After a full period, the trajectories repeat 
themselves oscillating between the two regions of parameter space. 
Thus, the geometric deformations of the elliptic model do not
possess an infra-red limit within the ansatz \eqn{ellipa}. 
This negative result may also be viewed as an indication that 
the particular mini-superspace model we are using is not physically correct
when $\gamma \neq 0$.
The origin of the problem we are
facing here is technical and seems to be intimately 
related to the presence of a conical
singularity, which, hence, deserves  separate study.  
The orbits of the model are depicted schematically in figure 13 below.

\vspace{15pt}
\begin{figure}[h] \centering
\epsfxsize=12cm \epsfysize=8cm \epsfbox{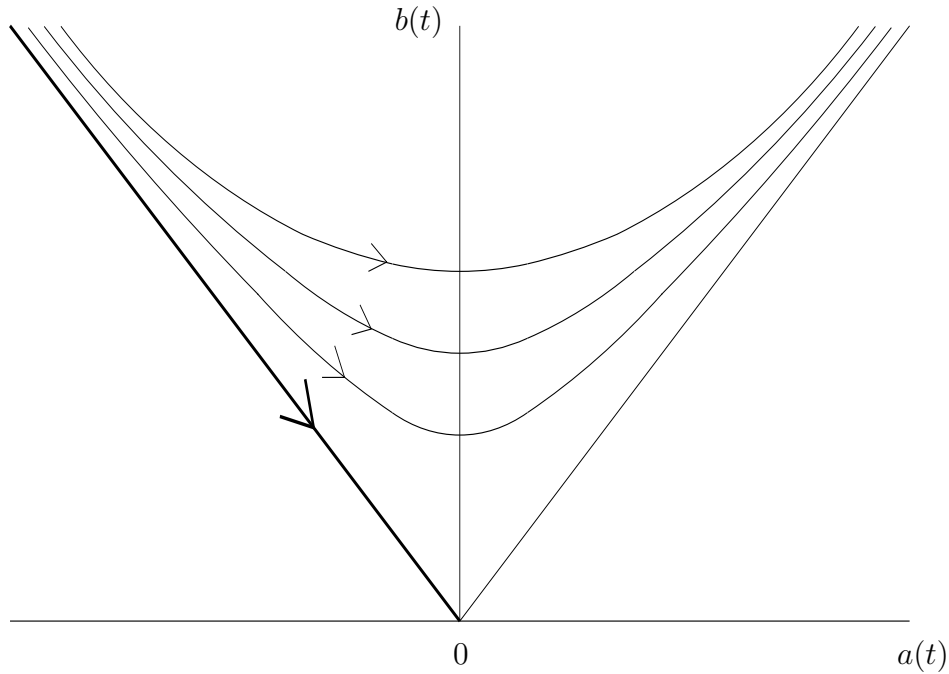}
\put(-5,-15){$a(t)$} \put(-195,225){$b(t)$} \put(-173,-15){$0$}
\caption{Renormalization group trajectories of the elliptic model.} 
\end{figure}

For $\gamma = 0$ we obtain the exact solution 
\be
b(t) = - a(t) = {1 \over 2(t-t_0)} ~, ~~~~~ 
e^{\Phi(Y;t)} = 2(t-t_0) {\zeta^2 \over {\rm sinh}^2 (\zeta Y)} ~,  
\ee
which is valid for all $t > t_0$ up to infinity. 
This is the Liouville
trajectory sitting at the boundary of the physical wedge, 
which describes a special deformation 
of the elliptic model via constant negative curvature metrics with 
$\Lambda^2 = 1/2(t-t_0)$ for all $t$. In 
this case, the geometry starts from 
infinite negative curvature and flows towards zero curvature in the 
infra-red region, $t \rightarrow +\infty$. Passing to proper coordinates
$(X, \tilde{Y})$ with $\tilde{Y} = -{\rm log} 
({\rm tanh}( -\zeta Y/2))$, 
as in section 5.1, and absorbing $t$ into a further
redefinition of the coordinates, 
\be
Y^{\prime} = 2\sqrt{t-t_0} ~ \tilde{Y} ~,   
\ee
we note that the metric becomes
\be
ds^2 \simeq d{Y^{\prime}}^2 + {\zeta}^2 
{Y^{\prime}}^2 dX^2  
\ee
as $t \rightarrow +\infty$. Thus, the end-point of this 
special trajectory is a two-dimensional flat cone with opening 
angle $2\pi \zeta$. The evolution turns a negatively curved cone into
a flat cone, but the opening angle of the cone remains the same 
through out the process.  

Summarizing, we note that there is only one physical trajectory
which corresponds to the Liouville line and exhibits a 
well-defined  infra-red
behavior. This configuration can be deformed further
and decay into a two-dimensional plane following another renormalization
group trajectory that corresponds to different ansatz for 
the metric, as outlined in section 6. 
 
\section{The decay of conical singularities}
\setcounter{equation}{0}

We are turning now to other exact solutions of the renormalization group
equations that describe the decay of conical singularities in non-compact
spaces. These processes provide a non-linear realization of the
dissipative properties of the heat equation, where it is found that
the geometry exhibits a smooth limit in the infra-red region.
The main example is the transition of a two-dimensional cone 
$C/Z_n$, with $n \geq 2$, to flat space, but it can also be 
generalized to describe the transition of $C/Z_n$ to another
cone $C/Z_m$ with $m<n$. The cone is the simplest singular configuration 
of this kind, where the geometry is everywhere flat apart from a point. 
Its curvature has delta function singularity at the tip of the cone, 
which is similar in that regard 
to the initial profile of the Gaussian solution 
of the heat equation. The evolution under the renormalization group
equation can be described in closed form, but the algebraic structure
of the solution turns out to be more complicated when compared
to the geometric deformations of the sausage and cane models. 
It can also be used as a guide to explore the behavior of the 
renormalization group flows in the vicinity of singular geometries
in superspace. Moreover, 
this solution is very important for addressing the problem of
tachyon condensation in closed string theory, since orbifold models 
(like the cone) exhibit tachyonic states in their string spectrum. 
These issues will be discussed next by expanding on earlier work
by other authors \cite{polch, minwa}, 
using the framework of Toda field equations. Thus,
tachyon condensation in closed string theory is found to exhibit a hidden 
relation to the infinite dimensional algebra ${\cal G}(d/dt; 1)$, 
which hopefully can be used further 
to address many conceptual and technical 
questions in the future. 

\subsection{Decay of a flat cone}  

It is convenient to describe the geometry of a two-dimensional cone 
$C/Z_n$ using polar coordinates,  
\be
ds^2 = dr^2 + r^2  d\phi^2 ~,  
\ee
where $\phi$ ranges from 0 to $2\pi/n$; the opening angle of the
cone is $2\pi / n$ and its deficit angle is $2\pi(1-1/n)$. 
This metric assumes the equivalent conformally flat form
\be
ds^2 = 2 e^{\Phi(\rho)} \left(d\rho^2 + \rho^2 d\theta^2\right) , 
\ee
with
\be
\Phi(\rho) = 2\left({1 \over n} -1 \right) {\rm log} \rho 
~ + ~ {\rm constant} ~,
\ee
using the change of coordinates 
\be
\rho = r^n ~, ~~~~~ \theta = n \phi ~. 
\ee
Here, $\theta$ runs from $0$ to $2\pi$ covering the whole 
two-dimensional plane instead of a single wedge with opening 
angle $2\pi /n$.  
Using the complex coordinates $z_{\pm} = \rho {\rm exp}(\pm i\theta)$,  
the metric of a cone assumes the conformally flat form 
\be
ds^2 = {2 \over (z_+ z_-)^{1-1/n}} dz_+ dz_- 
\ee
up to an overall normalization constant. 

This metric is also  
obtained by taking the $n$-th root of the complex coordinates of 
the plane, which are multi-valued functions that arise naturally by 
moding with the $Z_n$-rotations   
$z_{\pm} \rightarrow {\rm exp}(\pm 2\pi i/n) z_{\pm}$ 
acting on the plane.
The resulting space, $C/Z_n$, is an orbifold whose fixed point 
is the tip of the 
cone located at $z=0$, where the curvature is infinite  
as $R \sim \delta(z)$. Thus, 
it is natural to expect that the renormalization group flow 
of the metric towards the infra-red region will have the
tendency to smooth out the curvature singularity by diffusing it
all over the place and induce a transition to the two-dimensional plane.
The intuitive picture will be
made explicit in the following by analyzing the exact solution that 
describes this very transition. Furthermore, by simple 
generalization, one can
show that any cone with deficit angle $2\pi(1-1/n)$ flows towards another
cone with smaller deficit angle $2\pi(1-1/m)$ with $m<n$.

We limit ourselves to the renormalization group flow of purely 
gravitational backgrounds by also allowing the freedom to perform continuous
changes of the target space coordinates. Thus, we seek solutions of the 
lowest order equations  
\be
{\partial G_{\mu \nu} \over \partial t} = - R_{\mu \nu} + 
\nabla_{\mu} \xi_{\nu} + \nabla_{\nu} \xi_{\mu} ~, \label{naniep}   
\ee
where $\xi_{\mu}$ denote the components of the vector field that generate
diffeomorphisms of the target space coordinates, and we make the ansatz
\cite{minwa}  
\be
ds^2 = t\left(f^2(r) dr^2 + r^2 d\phi^2 \right) ~, ~~~~ \xi_r = {1 \over 2} 
rf(r) ~, ~~~~ \xi_{\phi} = 0 ~. \label{metansz} 
\ee
This frame turns out to be very convenient for the description of the 
decay process in closed form, but later we will also transform the 
solution in conformally flat coordinates. Here, we are assuming  
a factorized linear dependence of the two-dimensional target 
space metric on time. The radial coordinate 
$r$ ranges from 
$0$ to $\infty$, as usual, whereas the angular variable $\phi$ 
is taken to be periodic 
with arbitrary period $2\pi /n$, thus allowing for geometries 
with deficit angle. 

Since the components of the Ricci curvature tensor are in this 
frame 
\be
R_{rr} = {f^{\prime}(r) \over rf(r)} ~,  ~~~~ R_{\phi \phi} = 
r{f^{\prime}(r) \over f^3(r)} ~, ~~~~  R_{r \phi} = 0 ~, 
\ee
the renormalization
group flows reduce to the following differential equation for the
unknown function $f(r)$:  
\be
f^{\prime}(r) = r f^2(r) \left(1 - f(r) \right) . 
\ee
It is convenient to define new variables, as
\be
x= {1 \over 2} r^2 ~, ~~~~~ y(x) = {1 \over f(x)} - 1 ~, 
\ee
since the differential equation simplifies to
\be
y^{\prime}(x) = -{y(x) \over 1 + y(x)} ~. 
\ee
The latter is easily solved as follows,
\be
\left({1 \over f(x)} -1 \right) {\rm exp} \left({1 \over f(x)} -1 \right) 
= Ce^{-x} ~, \label{rockyk1} 
\ee
where $C$ is an arbitrary positive integration constant. 
Then, parametrizing $C$ 
in terms of $n \geq 1$, as follows,  
\be
C= \left(n - 1 \right) {\rm exp}
\left(n -1\right) , \label{rockyk2}
\ee
we observe that the function $f$ interpolates smoothly between $f=1/n$ at 
$r=0$ and $f=1$ at $r = \infty$, \cite{minwa}. 

The solution we have described above is the most general axi-symmetric 
deformation of a conical space with factorized time dependence. To 
demonstrate our claim, let us consider the generalized ansatz
\ba
ds^2 & = & t \left(f^2 (r) dr^2 + g^2 (r) d \phi^2 \right) , \nonumber\\
\xi_r & = & \xi (r) ~, ~~~~ \xi_{\phi} = 0 ~, \label{turta1}  
\ea
which depends on three function $f(r)$, $g(r)$ and $\xi (r)$. 
Explicit calculation shows in this case
that the components of the Ricci curvature tensor are
\be
R_{rr} = -{g^{\prime \prime}(r) \over g(r)} + 
{f^{\prime}(r) g^{\prime}(r) \over f(r) g(r)} ~, ~~~~ 
R_{\phi \phi} = -{g(r) g^{\prime \prime}(r) \over f^2(r)} + 
{g(r) g^{\prime}(r) f^{\prime}(r) \over f^3(r)} ~, ~~~~ 
R_{r \phi} = 0 ~, 
\ee
and the renormalization group equations \eqn{naniep} reduce to the 
following system, 
\ba
f(r) & = & {1 \over g(r)} \left({g^{\prime} (r) \over f(r)} 
\right)^{\prime} + 2 \left({\xi (r) \over f(r)} 
\right)^{\prime} ~, \nonumber\\
g(r) & = & {1 \over f(r)} \left({g^{\prime} (r) \over f(r)} 
\right)^{\prime} + 2 {g^{\prime} (r) \over f^2 (r)} \xi (r) ~.
\ea

Consistency of the equations implies that $\xi / (fg)$ is independent of 
$r$. Without loss of generality, we choose
\be
\xi (r) = {1 \over 2} f(r) g(r) ~, \label{turta2}  
\ee
thus setting the integration constant equal to $1/2$. This is precisely 
the form of $\xi (r)$ in the ansatz \eqn{metansz} when $g(r) = r$. 
Then, the remaining equation simplifies to 
\be
f(r) g(r) = \left( {g^{\prime} (r) \over f(r)} + {1 \over 2} 
g^2 (r) \right)^{\prime} ~, 
\ee
and its general solution is given by  
\be
\left( {g^{\prime} (r) \over f (r)} - 1 \right) {\rm exp} \left( 
{g^{\prime} (r) \over f (r)} - 1 \right) = C {\rm exp} \left(-{1 \over 2} 
g^2 (r) \right) ~, \label{rockyk11}  
\ee
where $C \geq 0$ is an arbitrary integration constant. Since $g(r)$ can always 
be set equal to $r$ by appropriate redefinition of the radial variable, we 
conclude that the solution described above is indeed the most general 
axi-symmetric deformation of the conical space.   

The explicit form of the solution \eqn{rockyk1}, or equivalently 
\eqn{rockyk11} with $g(r) = r$ and $C$ given by equation \eqn{rockyk2},  
shows that the renormalization group flow proceeds
by linear expansion of the metric, which is a global Weyl transformation.  
Its physical interpretation is best described by introducing
a $t$-dependent change of coordinates \cite{minwa}, namely,
\be
\tilde{r} = r \sqrt{t} ~,  
\ee
which absorbs the linear expansion scale of the metric into the radial coordinate.
Then, keeping $\tilde{r}$ fixed, we see that $t \rightarrow 0$ corresponds to 
$r \rightarrow \infty$, and hence $f=1$, which describes a cone with opening 
angle $2\pi /n$ at the initial time. 
On the other hand, keeping $\tilde{r}$ fixed, 
we see that $t \rightarrow 
+ \infty$ corresponds to $r \rightarrow 0$, and hence $f= 1/n$, which 
describes the metric of a two-dimensional plane, up to an overall scale 
$1/n^2$, 
at the infra-red point of the renormalization group flow. 
When $0 \leq t < + \infty$, the function $f(r)=f(\tilde{r}; t)$ interpolates
smoothly between the two geometries by diffusing the curvature all over the 
space, as it was also anticipated above on intuitive grounds. In fact, for 
generic values of $t$, the two-dimensional space is conical at infinity and 
smooth at the origin, as the curvature equals to 
\be
R = {2f^{\prime}(r) \over r f^3(r)} \equiv {2 \over t} 
\left({1 \over f(r)} - 1\right) .  
\ee
The shape of this transient space is depicted schematically in figure 
14 below:     

\vspace{-35pt}
\begin{figure}[h] \centering
\epsfxsize=8cm \epsfysize=10cm \epsfbox{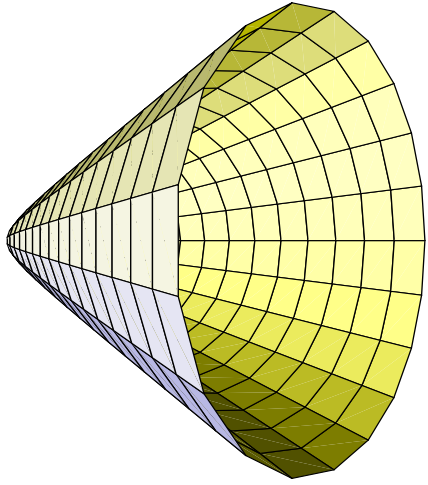} 
\mbox{\hspace{1.5cm}}  
\\[-45pt] \caption{An asymptotically conical space with smooth tip.} 
\end{figure}

The exact solution \eqn{rockyk1}   
can be used further to describe the transition of a cone 
$C/Z_n$ to another cone $C/Z_m$ by choosing the
integration constant $C$ as  
\be
C= \left({n \over m} - 1 \right) {\rm exp}\left({n \over m} - 
1\right) , \label{rockyk3}
\ee
using two positive integers with $n > m$, \cite{minwa}. 
In this case the
angular variable $\phi$ is also taken to range between 0 and $2\pi/n$, which 
equals to the opening of the cone $C/Z_n$ at the initial time $t=0$. 
Keeping $\tilde{r}$ fixed, as before, we obtain the metric of a cone 
with opening angle $2\pi / m$, up to an overall scale $m^2 / n^2$, 
as $t \rightarrow + \infty$. Successive transitions can also be considered, 
e.g., $n \rightarrow m \rightarrow m^{\prime} 
\rightarrow \cdots \rightarrow 1$,
which eventually  
lower the value of $n$ to 1 and yield a trivial conformal field theory at the
end-point of the process.  

The solution that describes the transition of a cone $C/Z_n$ to 
$C/Z_m$ can be brought
in a conformal frame. First, we consider the change of coordinates
$\rho (r)$, $\tilde{\phi}(\phi)$ given by 
\be
\rho^2 (r) = {\rm exp} \left({n \over m} 
\int^{r^2 / 2} 
{dx \over x} f(x) \right) , ~~~~~ \tilde{\phi} = {n \over m} 
\phi ~, \label{imptra1}  
\ee
which casts the metric \eqn{metansz} in the conformal form
\be
ds^2 =  2t e^{\Phi (\rho)} \left( d{\rho}^2 + {\rho}^2 
d{\tilde{\phi}}^2 \right) \label{confla}   
\ee
with   
\be
e^{\Phi (\rho)} = {m^2 \over n^2}  
{r^2 \over 2 \rho^2} \equiv {\rm exp} \left( 
\int_0^{r^2 / 2} {dx \over x} \left(1 - {n \over m}  
f(x) \right) \right) . \label{imptra2}   
\ee
Here, $\tilde{\phi}$ runs from $0$ to 
$2\pi /m$ and the metric \eqn{confla} is conformally related to 
the metric of the cone $C/Z_m$, $d{\rho}^2 + {\rho}^2 d{\tilde{\phi}}^2$, 
that corresponds to the end-point of the general process $C/Z_n 
\rightarrow C/Z_m$. Also, the lower value of the integration range 
in \eqn{imptra1} is left arbitrary to account for the integration 
constant of the transformation $\rho (r)$.   
Note that this change of coordinates alters the vector field $\xi$,   
which now becomes 
\be 
\xi_{\rho} = \rho ~ {\rm exp}(\Phi / 2) ~, 
\ee
according to the 
generalized ansatz \eqn{turta1}, \eqn{turta2}. The fact that 
$\xi_{\rho} \neq 0$ implies that $\Phi (\rho)$ above does not 
satisfy the Liouville equation, since, otherwise, 
the metric \eqn{confla} would have been a Liouville line with 
factorized linear time dependence.  
 
Next, we absorb the time dependence by the introducing the 
rescaling
\be
\tilde{\rho} = \sqrt{t} \rho ~, \label{imptra3} 
\ee
which yields the solution of the continual Toda field equation 
for the transition of conical geometries in the 
conformal frame with $\xi_{\tilde{\rho}} = 0$.     
The resulting Toda field $\Phi (\tilde{\rho}; t)$ is an  
implicit function of $(\tilde{\rho} , t)$, which follows   
by inverting the transformation $\rho (r)$ in the defining 
relations \eqn{imptra1}, \eqn{imptra2}, \eqn{imptra3} 
with 
\be
ds^2 = 2 e^{\Phi (\tilde{\rho} ; t)} \left(d{\tilde{\rho}}^2 
+ {\tilde{\rho}}^2 d {\tilde{\phi}}^2 \right) . 
\ee
For this reason,   
the solution can only be written  
as series expansion in the variable $\tilde{\rho}$, 
which, despite its limitations, 
will prove useful for studying the infra-red asymptotic behavior as    
well as the free field realization of the corresponding solution.  

Before we proceed with the analysis of the 
general solution, note that the transformation
$\rho (r)$ can be easily inverted  when   
both $n$ and $m$ are very large, in which case the 
renormalization group equation linearizes and becomes the heat flow equation.
Choosing $m=n-k$ for definiteness, where $k$ is arbitrary but finite, 
we note that $C = k/n + {\cal O}(1/n^2)$ 
and $f(x)$ can be approximated by the function 
\be
f(x) \simeq 1 - {k \over n} e^{-x} ~,  
\ee
which is valid for all $x$ to lowest order in $1/n$. 
Then, using equation \eqn{imptra1} we obtain 
\be
\rho^{2} \simeq {r^2 \over 2} + {\cal O} \left({1 \over n}
\right) ,  
\ee
whereas from equation \eqn{imptra2} we have 
\be
\Phi (\rho) \simeq -{k \over n-k} 
\int_0^{r^2 / 2} {dx \over x} 
\left(1-e^{-x} \right)   
\ee
to lowest order in $1/n$. 
Thus, combining these equations, we may eliminate $r$ and 
write $\Phi$ as a function of $\tilde{\rho}$, 
\be
\Phi(\tilde{\rho}; t) \simeq -{k \over n} 
\int_0^{{\tilde{\rho}}^2 / t} {dx \over x} 
\left(1 - e^{-x} \right) + {\cal O} \left({1 \over n^2} 
\right) . \label{imptra4} 
\ee

In this case, $\Phi(\tilde{\rho}; t)$ remains 
small for all times, and therefore
the continual Toda field equation can be approximated by the heat equation.
Furthermore, using the expansion  
\be
\int_0^{z} {dx \over x} \left(1 - e^{-x} \right) \simeq {\cal C} + 
{\rm log} z + {e^{-z} \over z} \left(1 - {1 \over z} + {2! \over z^2} 
- {3! \over z^3} + \cdots \right) ,   
\ee
which is valid for large $z= {\tilde{\rho}}^2 / t$ 
with appropriately chosen integration 
constant ${\cal C}$, we find that  
the solution \eqn{imptra4} interpolates smoothly between 
the function $\Phi (\tilde{\rho}; t= 0) = -2 (k/n) 
{\rm log} \tilde{\rho}$, 
which describes the conformal factor of the initial cone 
$C/Z_n$ in the reference frame of $C/Z_{n-k}$, and the 
constant function  
$\Phi (\tilde{\rho}; t =+\infty)=0$, which corresponds to the end-point
of the transition in the same reference frame.    
This particular expression has been derived before by considering the 
transition $C/Z_n \rightarrow C/Z_{n-2}$ with $k=2$ and $n$ 
large, \cite{polch}.  

The infra-red behavior of the geometric transition  
$C/Z_n \rightarrow C/Z_m$ is easily obtained when $n$ and $m$ 
are both finite by expanding $f(x)$ around $x=0$. 
The power series expansion turns out to be  
\be
f(x) = {m \over n} + {m^2 \over n^2} \left(1- {m \over n} \right) x + 
{1 \over 2} {m^3 \over n^3} \left(1 - {m \over n} \right) 
\left(2-3 {m \over n} \right) x^2 + \cdots 
\ee
following equation \eqn{rockyk1}. 
Then, using the defining relation \eqn{imptra1}, 
we find that $\rho^2$ also admits the power series expansion 
\be
\rho^2(r) = {r^2 \over 2} \left(1 +  
{m \over n} \left(1 - {m \over n} \right) {r^2 \over 2} + 
{1 \over 4} {m^2 \over n^2}  
\left(1- {m \over n} \right) \left(4-5 {m \over n} \right) 
{r^4 \over 4} + \cdots \right) ,  
\ee
which can be inverted to yield the following power series expansion of 
$r (\rho)$,       
\be
{r^2 \over 2 \rho^2} = 1 - {m \over n} 
\left(1- {m \over n} \right) \rho^2 + 
{1 \over 4} {m^2 \over n^2} \left(1 - {m \over n} \right) 
\left(4-3 {m \over n} \right) \rho^4 + \cdots ~ .  
\ee
Thus, using the rescaling of the coordinates \eqn{imptra3}, we 
obtain the general infra-red behavior of the conformal factor
\be
e^{\Phi (\tilde{\rho} ; t)} = 1 - {m \over n} \left(1 - 
{m \over n} \right) {{\tilde{\rho}}^2 \over t} + 
{\cal O} \left({1 \over t^2} \right)  
\ee
up to an overall (but irrelevant) normalization constant $m^2 / n^2$.  

\underline{\em Free field realization} 

Following \eqn{rockyk1}, we may also expand $f(x)$ in an infinite series  
\be
f(x) = 1 - C e^{-x} + {1 \over 2!} (2C)^2 e^{-2x} - 
{1 \over 3!} (3C)^3 e^{-3x} + {1 \over 4!} (4C)^4 e^{-4x} + \cdots ~,  
\ee
which is valid for all $x$ with $C$ given by \eqn{rockyk3}. This 
expansion is particularly useful when $x \rightarrow + \infty$ 
and only the first few terms are retained. It can be used to show
that 
\be
\Phi_0 (z_+ , z_-; t) = - \left(1 - {m \over n} \right) {\rm log} 
{z_+ z_- \over t} ~, ~~~~~~ {\rm with} ~~ {\tilde{\rho}}^2 = 
z_+ z_-  
\ee
is the appropriate free field configuration that reproduces the 
solution of the Toda field equation given by \eqn{imptra1}, 
\eqn{imptra2}, \eqn{imptra3}, according to        
the perturbative series expansion following from the general formulae   
\eqn{toda133} and \eqn{toda9}. The details of the calculation are 
left to the reader.

\subsection{On the decay of a negatively curved cone}

In section 5.4 we examined deformations of the elliptic constant 
negative curvature metric and found that, generically, 
the renormalization group  
flow does not possess a well defined infra-red limit in the naive 
mini-superspace approximation. It rather exhibits an 
oscillatory behavior, which is 
not physically acceptable since the metric changes signature 
beyond a certain length scale, and shows that the given 
mini-superspace approximation is not physically correct.
This behavior was attributed  
to the presence of a conical singularity in the class of elliptic  
metrics, which in proper coordinates take the form
\be
ds^2 = dr^2 + {\zeta}^2 {\rm sinh}^2 r d {\theta}^2 ~. 
\ee
Indeed, close to $r = 0$, the metric becomes  
\be
ds^2 \simeq dr^2 + r^2 d {\phi}^2 
\ee
with $\phi = \zeta \theta$ ranging from $0$ to $2\pi \zeta$, and the 
space looks like a flat cone with opening angle $2\pi \zeta$; 
everywhere else, away from $r=0$, it has 
constant negative curvature as it was discussed in   
section 5.1. 

One may think that there are other trajectories of 
the renormalization group flow with  
well defined infra-red limit that describe the decay of a  
negatively curved cone (as depicted in figure 7) into a two-dimensional 
plane, in close analogy with the decay of a flat cone to the plane. 
Unfortunately, the dissipation of the singularity can not be 
technically achieved in this case using the 
axi-symmetric ansatz \eqn{turta1} with factorized linear time 
dependence. We have already seen that this ansatz is only 
applicable to the decay of a flat cone and, therefore, 
generalized axi-symmetric ansatz with more complicated time 
dependence should be devised for this purpose. So far, we 
have not been able to construct a solution of the renormalization 
group equation that describes the
decay of the conical singularity in negatively curved spaces     
following a single trajectory in 
superspace. The best we can do at this moment is to view the 
decay of a negatively curved cone to the plane as a composite 
process. It proceeds by first deforming it into a flat cone
using the special Liouville trajectory found in section 5.4, and then 
deforming the resulting cone with opening angle $2\pi \zeta$ 
to the plane, along the previous lines.  

It will be interesting to construct other trajectories in 
superspace that are capable to resolve curvature singularities 
in the infra-red limit, since the behavior of the renormalization 
group flows close to singular geometries holds the key for
the systematic understanding of the dissipative properties of
the continual Toda field equation. It may also help to analyze  
the stability properties of the flows in superspace and the 
physical relevance of various mini-superspace approximations.

\subsection{Tachyon condensation in closed string theory}

The renormalization group approach to the problem of tachyon 
condensation in closed string theory is based on the assumption 
that the full dynamical evolution in string theory can be 
well approximated in the world-sheet theory by flowing from 
one critical point to another more stable conformal solution. 
Instabilities in target space are indicated by the presence of
tachyonic states in the string spectrum, which in turn 
correspond to relevant operators on the world-sheet theory. 
Consider ten-dimensional string backgrounds of the simple form
\be
{\cal C} = {\rm CFT}_{\rm d} + R^{9-d, 1} ~, \label{enpt}  
\ee
where ${\rm CFT}_{\rm d}$ is any conformal field theory block with
central charge $c=d$.  One may use operators ${\cal O}$ 
of the $d$-dimensional block to construct marginal 
operators of the full conformal background by 
applying momentum dressing, 
i.e.,
\be
\hat{\cal O} = e^{iP \cdot X} {\cal O} ~~~~~ {\rm with} ~~ 
{1 \over 2} P^2 + \delta = 1 ~,   
\ee
where $(\delta, \delta)$ is the conformal dimension of ${\cal O}$ 
and $P$ is the momentum in $R^{9-d, 1}$ with coordinates $X$.   
Thus, the world-sheet operators $\hat{\cal O}$ correspond to 
space-time fluctuations with mass-squared $M^2 = 2(\delta -1)$,
which are tachyonic in space-time when ${\cal O}$ are relevant 
operators on the world-sheet with
$\delta < 1$. Put it differently, the operator spectrum of the 
conformal field theory block ${\rm CFT}_{\rm d}$ determines 
the stability of the full string background under time evolution. 
Furthermore, the infra-red limit of the world-sheet renormalization 
group flow driven by ${\cal O}$ 
should coincide with the end-point of the dynamical 
transition from the background \eqn{enpt} to another solution 
of the same type, but with no tachyons in the spectrum.    

It is quite difficult to provide a systematic description of this 
problem in all generality, but for tachyons that are localized
in target space the investigation becomes more tractable. The most 
characteristic example, where there is an exact conformal field 
theory description of the spectrum, is provided by the 
orbifold backgrounds
\be
{\cal C} = C/Z_n + R^{7, 1} ~, 
\ee
where $C/Z_n$ is a cone with central charge $c=2$. In this case, 
there are tachyonic states coming from the twisted sectors of the 
orbifold labeled by $k=0, 1, 2, \cdots , n-1$. The ground state 
in each twisted sector is known to be tachyonic 
\cite{orbif} (but see also \cite{atish}) with mass-squared equal to    
\be
M^2 = \left\{ \begin{array}{ll}
-2k / n~, & {\rm for} ~ k ~ {\rm even} \\
  &  \\
-2\left(1 - k / n \right) , & {\rm for} ~ k ~ {\rm odd} \\
\end{array} \right. 
\ee
which then lead to instabilities according to the general 
framework above. 
There are also excited state tachyons in many sectors, but the 
details are not essential for the purposes of the present work. 
These instabilities trigger the decay of
the conformal field theory $C/Z_n$ to another cone $C/Z_m$ with 
$m<n$, and eventually to two-dimensional flat space $R^2$, which
has no tachyons. 

The renormalization group approach to the metric 
deformation that describes this transition was summarized in section 
6.1 above, but recently there has also been a systematic description 
of the transition in real time with the same qualitative 
features; it should be mentioned, however, that the dynamical 
equations that describe this transition are not the same as the 
renormalization group equations, \cite{harv2}.  
Furthermore, tachyon condensation
in this conical space is closely related to the computation of 
black-hole entropy \cite{atish}, 
which is also very interesting to consider 
on physical grounds.  

The results we described so far show that the infinite dimensional 
algebra ${\cal G}(d/dt; 1)$ arises as hidden symmetry of the 
tachyon condensation process in the regime of gravity. Different 
geometric transitions correspond to different solutions of the 
continual Toda field equation, and they should all have an infinite
set of charges in target space, which 
are associated to the conserved currents of the system.
Thus, the problem of vacuum selection in string theory appears to
be part of an integrable 
model in superspace that consists of all target space metrics. 
The notion of integrability refers to the zero curvature formulation
in two-dimensional target space, whereas the renormalization 
group time serves as a label on the root system of the underlying
continual Lie algebra and it should be able to determine the 
form of the integrals. Unfortunately, the explicit construction of
these charges, as well as their systematic interpretation in 
string theory remain unknown. It is natural to expect, however, 
that they are associated to other modes of the string which 
participate in the decay process.

Since tachyons become massless at the end-point of the condensation
process, one is lead to consider the beta function equations of all 
``would be massless" fields on equal footing. We do not yet know 
how to extend the use of the algebra ${\cal G}(d/dt; 1)$ to such 
generalized systems of beta function equations, which remain 
unexplored to a large extend, but it is certain that the inclusion
of other string states will shed new light on the decay of 
singularities in string theory. Also, the breakdown of the lowest
order approximation to the metric beta functions, which shows  
at some finite world-sheet length scale due to the 
appearance of high curvature regions  
in target space, is indicative of the influence of other 
string states to the general problem. After all, string effects 
always manifest as paradoxes in the lowest order effective theory
of the massless modes and break down its validity. We hope to 
be able to report elsewhere on these interesting possibilities.   

\section{Conclusions and discussion}
\setcounter{equation}{0}

We have shown that the beta function equations of two-dimensional 
non-linear sigma models admit a zero curvature formulation with the 
aid of the infinite dimensional continual Lie algebra 
${\cal G}(d/dt; 1)$. 
The Cartan-Weyl generators of 
the algebra are sufficient to formulate the geometric deformations 
of two-dimensional target space metrics as an integrable Toda
field equation in target space using the system of conformally 
flat coordinates. This equation, which arises to lowest order 
in $\alpha^{\prime}$, provides a non-linear generalization
of the heat equation and shares the same dissipative properties 
in renormalization group time by resolving curvature singularities 
in the infra-red limit of the geometries. We have constructed 
the general solution in terms of two-dimensional free fields,
via B\"acklund transformations, and described a formal power
series expansion of the non-linear field configurations,   
using group theoretic techniques as in finite dimensional Toda 
field theory. We have also examined several 
examples of axially symmetric renormalization group flows for 
compact and non-compact constant curvature metrics, using 
a mini-superspace approximation in the space of all target space 
metrics, and demonstrated the validity of the formal power 
series expansion of the general solution for appropriate choices
of free field configurations.   

The sausage model arises by introducing axially symmetric 
deformations of the standard $O(3)$ sigma model, and describes 
the transition of an infinite long cylinder from the ultra-violet
region towards a round sphere with diminishing size at some 
finite world-sheet length scale. Likewise, the cane model, 
which was introduced here by considering axially symmetric 
deformations of constant negative curvature spaces, describes 
the transition from space with infinite negative curvature 
towards the infra-red region where the geometry looks like a 
cane with variable length and width. Both cases were also 
described in proper coordinates, which are more suitable  
for visualizing the corresponding geometric deformations, 
where it was found that they are parametrized by 
Jacobi theta functions. The latter satisfy the heat equation    
and, therefore, we have obtained 
a direct embedding of the heat equation into
the non-linear system of renormalization group equations which
is valid for all times. It should be noted, however, that 
these trajectories are physically sensible only in regions 
with small curvature, as all higher order curvature terms have
been suppressed by considering the one-loop beta function 
equations. Finally, we have revisited the transition from 
a two-dimensional cone $C/Z_n$ to the plane, which is 
indicative of the dissipative nature of the non-linear 
Toda field equation to resolve conical singularities  
after infinitely long time. At the same time, this 
transition takes place between two exact conformal field 
theories and provides a concrete framework for studying 
the problem of tachyon condensation in closed string theory 
in the gravity regime.    
   
The algebra ${\cal G}(d/dt; 1)$ exhibits exponential growth, 
thus making it difficult to write the complete system of 
commutation relations among all generators, and, 
furthermore, it has no 
clear geometric interpretation which could be presently used for 
understanding the dynamics of all other string modes under the 
renormalization group flow. Fortunately, the complete structure of
the algebra is not needed for the formulation of the metric 
beta functions as Toda system, but it is certainly 
required for making further progress in the algebraic 
description of the renormalization group equations. 
First of all, the construction of conserved currents in target 
space and their systematic interpretation in string theory
relies on this mathematical problem. Integrability of the
equations implies their existence according to the 
standard lore, but their construction is not straightforward
as in other 
Toda field theories based on simple Lie algebras. The main 
difficulty originates from the anti-symmetric character of the
Cartan kernel $\delta^{\prime} (t-t^{\prime})$, which is 
also responsible for the lack of Lagrangian description of 
the corresponding Toda field equation. 
It is quite natural to expect, however, 
that the conserved currents in target space should
be related to higher spin fields, in close analogy with 
the higher spin interpretation of the conserved currents 
in ordinary Toda field theory. As such, they should be 
associated to higher spin modes of string theory, but we
have not succeeded so far to construct even the simplest 
representatives using local expressions in target space. There
are many reasons to believe that such conservation laws 
will also involve non-local expressions in target space. 

From a mathematical point of view, it remains to 
develop the structural theory of continual Lie algebras and 
the theory of their representations beyond the formal 
definition of highest weight modules that has been used in 
this paper. Also, the definition 
of bilinear forms, such as trace formulae for the 
Lie algebra elements of ${\cal G}(d/dt; 1)$ and their 
products, is lacking    
at this moment and prevents the construction of conserved
currents. Finally, it will be interesting to construct
systems of roots and their reflections in a systematic 
way, as mirror symmetry might have a natural manifestation
in this algebraic context.    

The complete structure of the Lie algebra ${\cal G}(d/dt; 1)$ 
is also important for extending the use of our methods to 
higher string modes beyond gravity. There has been no systematic 
description of the beta functions for higher 
string modes to this day,
apart from some general results \cite{tsey5} that are 
awaiting better understanding. Also, results have been derived for 
some special states in the framework of
two-dimensional string theory \cite{ed3}. We have suggested that 
generalized systems of multi-component Toda field equations, 
which can be naturally introduced by considering the zero 
curvature conditions $[\partial_+ + A_+ , \partial_- + A_-] = 0$ 
with $A_{\pm}$ taking appropriate values in the subspaces 
${\cal G}_N \oplus \cdots \oplus {\cal G}_0 \oplus \cdots 
{\cal G}_N$ with $N>1$ might be relevant for the inclusion 
of such states in the system of renormalization group 
equations. If this expectation materializes, it will lead
to a systematic realization of all Lie algebra elements 
in terms of off-shell string theory. In any case, the 
algebraic 
description of the renormalization group flows for all other 
additional operators, which can be added to the world-sheet 
action on top of the metric, pose a very interesting but 
difficult problem that is worth studying further.        
 
Another important question that remains unsolved, and 
most likely the answer relies on the exponential growth of  
the Lie algebra ${\cal G}(d/dt; 1)$, is the general 
proof of the dissipative 
behavior of the renormalization group flows in time.  
This issue is very important for the problem of tachyon 
condensation in closed string theory, where unstable vacua 
decay to more stable ones and curvature singularities 
dissipate away. Also, it is well known for some time that 
the renormalization group flows in the space of 
two-dimensional quantum field theories have the tendency to 
decrease the central charge as one moves from the 
ultra-violet to the infra-red fixed points. Of course, there
can be some differences on the properties of Zamolodchikov's
$c$-function for sigma models with compact or non-compact 
target spaces \cite{joe2}, but the essential idea is that generically 
the number of degrees of freedom is thinning towards the 
infra-red region; likewise the space-time energy 
decreases under the world-sheet renormalization group flow
for appropriate asymptotic conditions. It will be very 
interesting to provide an algebraic derivation of these
results based on the structure of the underlying 
continual Lie algebra by defining an entropy function
based on its exponential growth.  
  
The problem of tachyon condensation in closed string theory 
and the transitions among vacua that 
describe different conformal field theories is a very 
important issue. We believe that the algebraic methods 
discussed in this paper will help to understand the 
mechanism of vacuum selection in string theory, at least
from the world-sheet point of view. It will be also 
interesting to study the meaning of our algebraic methods
in real time string dynamics, as well as in other physical 
problems that involve time dependent gravitational 
backgrounds, like cosmological backgrounds and transitions
that describe gravitational collapse. Finally, it will 
be technically useful to provide a deeper understanding 
of the geometric transitions based on the gauged linear sigma model 
approach to supersymmetric theories \cite{ed4} and the 
powerful techniques of mirror symmetry \cite{vafa, hori1, hori2}, 
using the infinite
dimensional algebra ${\cal G}(d/dt; 1)$. All these 
questions are left open for future work.       

\vspace{.5in}

\centerline{\bf Note added}
\noindent 
After the completion of this work I became aware of several  
important developments in the theory of Ricci flows, which 
were introduced in the mathematics literature twenty years 
ago. Since then they have become a major tool for addressing
a variety of problems in geometry in diverse dimensions;
for a recent account of the main results    
see, for instance, \cite{chow} and 
references therein. Ricci flows are the same as the renormalization
group flows of sigma models and, therefore, it will be interesting 
to compare notes in view of their physical applications to 
string theory. 

\vspace{.5in}
 
\centerline{\bf Acknowledgments}
\noindent
This work was supported in part by the European Research and Training Networks
``Superstring Theory" (HPRN-CT-2000-00122) and ``Quantum Structure of 
Space-time" (HPRN-CT-2000-00131), as well as the Greek State Foundation 
Award ``Quantum Fields and Strings" (IKYDA-2001-22) and NATO Collaborative
Linkage Grant ``Algebraic and Geometric Aspects of Conformal Field Theories 
and Superstrings" (PST.CLG.978785). A first account of the main results 
was presented  
in the annual meeting of the Hellenic society for the study of high energy physics
held in the National Technical University of 
Athens, 17--20 April 2003; I thank the organizers for their kind invitation
to participate to this enjoyable event. I also wish to
thank George Savvidy for useful exchanges, as well as Brett
Taylor for his help with the figures.  
Finally, I thank the theory division at CERN for hospitality during 
the final stages of this work. 

\vspace{.5in}

\centerline{\bf Dedication}
\noindent
The present paper is  
a long due dedication to the memory of my friend  
Misha Saveliev with whom I had several enlightening conversations 
about continual Lie algebras, Toda theories, and their physical applications.

\end{document}